\documentclass[aps,prd,preprintnumbers,showpacs,nofootinbib,amssymb]{revtex4}
\usepackage{graphicx}
\usepackage{amsmath}
\usepackage{amssymb}
\usepackage{amsfonts}
\usepackage{bm}
\usepackage{color}

\def\be{\begin{equation}}
\def\ee{\end{equation}}
\def\bea{\begin{eqnarray}}
\def\eea{\end{eqnarray}}

\begin{document}

\title{Kinetic theory of two-dimensional point vortices and fluctuation-dissipation theorem}
\author{Pierre-Henri Chavanis}
\affiliation{Laboratoire de
Physique Th\'eorique, Universit\'e de Toulouse, CNRS, UPS, France}

\begin{abstract}

We complete the  kinetic theory of two-dimensional (2D) point vortices
initiated in previous works. We use a simpler and more physical formalism. We
consider a system of 2D point vortices submitted to a small external stochastic
perturbation and determine the response of the system to the perturbation. We
derive the diffusion coefficient and the drift by polarization of a test vortex.
We introduce a general Fokker-Planck equation involving a diffusion term and a
drift term. When the drift by polarization can be neglected, we obtain a secular
dressed diffusion (SDD) equation sourced by the external noise. When the
external
perturbation is created by a discrete collection of $N$ point vortices, we
obtain a Lenard-Balescu-like kinetic equation reducing to a Landau-like kinetic
equation when collective effects are neglected. We consider a multi-species
system of point vortices. We discuss the process of kinetic blocking in the
single and multi-species cases. When the field vortices are at statistical
equilibrium (thermal bath), we establish the proper expression of the
fluctuation-dissipation theorem for 2D point vortices relating the power
spectrum of the fluctuations to the response
function of the system.  In
that case, the drift coefficient and the diffusion coefficient satisfy an
Einstein-like relation and the Fokker-Planck equation reduces to a
Smoluchowski-like equation. We mention the analogy between 2D point vortices and
stellar systems. In particular, the drift of a point vortex in 2D hydrodynamics
[P.H. Chavanis, Phys. Rev. E {\bf 58}, R1199 (1998)] is the counterpart of the
Chandrasekhar dynamical friction in astrophysics. We also
consider a gas of 2D Brownian point vortices described by $N$ coupled stochastic
Langevin
equations and determine its mean and mesoscopic evolution. In the present
paper, we treat the case of unidirectional flows but our results can be
straightforwardly generalized to axisymmetric flows.

\end{abstract}


\maketitle

\section{Introduction}
\label{sec_intro}

There exist remarkable analogies 
between 2D point vortices  and stellar systems
\cite{csr,houchesPH,tcfd}.
This is basically due to the fact that these systems have long-range
interactions \cite{campabook}. As a result, they self-organize into coherent
structures such as large-scale vortices (e.g. Jupiter's Great Red spot)
\cite{bouchetvenaille,houchesPH} or
globular clusters and galaxies \cite{bt}. However, the relaxation towards these
organized states is nontrivial. Systems with long-range interactions
experience two successive types of relaxation. There is first a
violent collisionless relaxation to a metaequilibrium state on a very
short timescale of the order of the dynamical time $t_D$. The 
collisionless evolution of stellar systems  is governed by the Vlasov-Poisson
equations
\cite{jeans,vlasov} and the metaequilibrium state resulting from violent
relaxation can be predicted by the
statistical theory of Lynden-Bell \cite{lb}. Similarly, the
collisionless evolution of 2D
point vortices is governed by the Euler-Poisson equations
\cite{newton} and the metaequilibrium state can be predicted by the
Miller-Robert-Sommeria (MRS) statistical theory
\cite{miller,rs}, which is the hydrodynamic analogue of the Lynden-Bell
theory \cite{csr}.\footnote{The MRS theory applies either to the 2D point vortex
gas in the
collisionless regime (when $N\rightarrow +\infty$ with $\gamma\sim 1/N$) 
 or to continuous 2D
incompressible flows in the inviscid regime (when $\nu\rightarrow 0$). The late
time evolution of
continuous 2D incompressible flows is dominated by viscous decay and does not
relax towards an equilibrium state.} These theories rely on an assumption
of ergodicity which is not always fulfilled in practice. This is the
difficult problem of incomplete violent relaxation \cite{incomplete}. Then, on a
much longer timescale, there is
a slow (secular) collisional relaxation due to finite $N$ effects
(granularities) leading, for $t\rightarrow +\infty$, to the Boltzmann
equilibrium
distribution predicted by conventional statistical
mechanics. For stellar systems,
this equilibrium state has been considered by
Ogorodnikov \cite{ogo1,ogo2}, Antonov \cite{antonov}, and Lynden-Bell and Wood
\cite{lbw}. They showed that an equilibrium state does not always exist,
even when the system is artificially enclosed within a box in order to prevent
its
evaporation. Indeed, self-gravitating systems may experience a gravothermal
catastrophe \cite{lbw}. For 2D point
vortices, the Boltzmann equilibrium state has been considered by Joyce and
Montgomery \cite{jm,mj}, Kida \cite{kida}, and Pointin and Lundgren \cite{pl,lp}
following the pioneering work of Onsager \cite{onsager}.\footnote{In his seminal
paper on the statistical
mechanics of 2D point vortices, Onsager
\cite{onsager} related the formation of large-scale vortices to the existence of
negative temperature states. Later on, in unpublished notes \cite{esree}, he
developed a mean field theory of 2D point vortices and derived the
Boltzmann-Poisson equation several years before the authors of Refs.
\cite{jm,mj,kida,pl,lp}.}
The relaxation time towards the Boltzmann distribution diverges
algebraically with $N$ so that, when $N\rightarrow +\infty$, the
system
is always in the collisionless regime. Here, we focus on the secular evolution
of the system for large but finite values of $N$.\footnote{The kinetic
theory
of collisionless relaxation for systems with long-range interactions is
discussed in \cite{lbnew} and references therein.} We briefly review the
case of systems of material particles with long-range interactions (stellar
systems, plasmas, HMF
model...) before considering the case of 2D point vortices. 

Let us first consider the kinetic theory of spatially homogeneous systems 
with long-range interactions in dimension $d=3$. The evolution of a test
particle in a thermal bath  is
described by a Fokker-Planck equation of the Kramers form involving a diffusion
term and a friction term. The friction and the diffusion coefficients satisfy
the Einstein relation. At statistical equilibrium, the diffusion and the
friction balance each other and the Boltzmann distribution is established. This
approach was developed by Chandrasekhar
\cite{chandra,chandra1,chandra2,chandra3,chandrany,nice}
for stellar systems by analogy with the theory of Brownian motion
\cite{chandrabrown}. The theory of stochastic gravitational fluctuations was
studied in \cite{cvn0,cvn1,cvn2,cvn3,cvn4,cvn5,kandruprep}. If we consider the
evolution of the system as a whole, the
kinetic evolution is described by the Landau
\cite{landau} or Lenard-Balescu \cite{lenard,balescu} equation
introduced in
plasma physics.
These equations describe the collisional evolution of the system
at the order $1/N$ due to the development of two-body correlations.  They
conserve mass and energy,
satisfy an $H$-theorem for the Boltzmann entropy, and relax towards the
Boltzmann distribution. The relaxation time scales as $N t_D$.\footnote{In
plasma physics and stellar dynamics the relaxation time scales as
$(N/\ln N) t_D$ because of logarithmic corrections. In plasma physics, $N$
represents the number of charges in the Debye sphere (usually denoted
$\Lambda$). In stellar dynamics, $N$
represents the number of stars in the Jeans sphere which corresponds to the
typical size of the cluster.}
The Lenard-Balescu equation takes into account collective
effects \cite{lenard,balescu,rostrosen,th,hubbard1,hubbard2} that 
are neglected in the Landau approach. The Landau equation has also been applied
to stellar systems by making a local approximation (which amounts to considering
that the system is spatially  homogeneous). In the thermal bath approximation,
we recover the original Fokker-Planck (Kramers) equation of
Chandrasekhar.\footnote{Chandrasekhar
\cite{chandra,chandra1,chandra2,nice} also
considered the evolution of a test star experiencing gravitational encounters
with field stars that are
not necessarily  at statistical equilibrium. His work was further developed by
Rosenbluth {\it et al.} \cite{rosenbluth}. If we assume that 
the system evolves self-consistently \cite{kingL,henon}, the corresponding
Fokker-Planck equation is equivalent to the Landau equation although it appears
in a different form (see \cite{aa} for the correspondence between the
Chandrasekhar and Landau equations).} However,
self-gravitating systems are
spatially inhomogeneous and the local approximation leads to a logarithmic
divergence at large scales. Furthermore, the Landau equation does not take into
account collective effects. Recently, the Landau and Lenard-Balescu equations
have been generalized to the case of spatially inhomogeneous systems by
Heyvaerts \cite{heyvaerts} and Chavanis \cite{angleaction2} using angle-action
variables.  For gravitational systems, the proper treatment
of spatial inhomogeneity removes the logarithmic divergence at large scales. The
inhomogeneous  Landau and Lenard-Balescu equations conserve mass and energy and
satisfy an $H$-theorem for the Boltzmann entropy. They usually relax towards the
Boltzmann distribution except in
the case of unconfined stellar systems where the relaxation is hampered by
the phenomena  of  evaporation and gravothermal catastrophe (core collapse). The
Landau and Lenard-Balescu equations are valid for all systems with long-range
interactions in any dimension of space $d$. However, for spatially homogeneous
systems in $d=1$ (like the  HMF model or spins with long-range interactions
moving on a sphere), the Landau and Lenard-Balescu collision terms vanish
identically \cite{epjp,fbc}. In that case, there is no kinetic evolution at the 
order $1/N$. This is a situation of kinetic blocking due to the absence of
resonances. As a result, the system does not reach the Boltzmann distribution
on a timescale $N t_D$. We thus have to take into account three-body
correlations and
develop the kinetic theory at the order $1/N^2$. An explicit kinetic equation
that is valid at the order $1/N^2$ has been obtained recently by Fouvry {\it et
al.} \cite{fbcn2,fcpn2}  for arbitrary homogeneous 1D systems with long-range
interactions in the approximation where collective effects can be neglected.
Remarkably, this equation satisfies an $H$-theorem and relaxes towards the
Boltzmann distribution. This implies that the relaxation time scales as $N^2
t_D$ for homogeneous 1D systems with long-range interactions.

The evolution of a test vortex in a thermal bath is described by a 
Fokker-Planck equation of the Smoluchowski form involving a diffusion term and a
drift term. The drift and the diffusion coefficients satisfy an Einstein-like
relation. At statistical equilibrium, the diffusion and the drift balance each
other and the Boltzmann distribution is established. This approach was developed
by Chavanis \cite{preR,pre} by analogy with the theory of Chandrasekhar
\cite{chandra,chandra1,chandra2,nice} for stellar systems and the theory of
Brownian motion \cite{chandrabrown}. The theory of stochastic fluctuations in
the point vortex gas was studied in \cite{fluc1,fluc2,fluc3}. If we
consider the evolution of the system
as a whole, the kinetic evolution is more complicated.
It is described by a
Landau-like \cite{pre,cl,bbgky,kindetail,fcp} or a Lenard-Balescu-like
\cite{dubin,dubin2,klim,onsagerkin} equation. These equations describe the
collisional evolution of the system at the order $1/N$ due to two-body
correlations. They conserve circulation and energy and satisfy an $H$-theorem
for the Boltzmann entropy.  The Lenard-Balescu-like
equation takes into account collective effects that are
ignored in the Landau-like equation.
In the thermal bath
approximation,
we recover the original Fokker-Planck (Smoluchowski-like) equation of Chavanis
\cite{preR,pre}. For general flows that are neither unidirectional nor
axisymmetric, there is a collisional evolution at the order $1/N$. The
corresponding kinetic equation [see Eq.
(128) or Eq. (137) of \cite{pre}] approaches the Boltzmann
distribution on a timescale
$Nt_D$ if there are sufficient ``resonances'' between
the point vortices \cite{pre}.
However, for unidirectional flows and for axisymmetric flows with a monotonic
profile of angular velocity, the Landau and Lenard-Balescu-like collision terms
vanish identically. Therefore, there is no kinetic evolution for unidirectional
flows at the order  $1/N$. For axisymmetric flows, the system
evolves until the profile of angular velocity becomes monotonic and then stops
evolving at the order $1/N$. This is a situation of kinetic
blocking \cite{cl} due to the absence of resonances. As a result, the system
does
not reach the Boltzmann distribution on a timescale $N t_D$. In order to
describe the relaxation of the system towards the Boltzmann distribution it is
necessary to take into account three-body correlations and develop the kinetic
theory of 2D point vortices at the order $1/N^2$ like in the work of Fouvry {\it
et al.} \cite{fbcn2,fcpn2}.

In the present paper, we focus on the Landau and Lenard-Balescu equations
for 2D
point vortices at the order $1/N$. In previous works, these kinetic equations
have been derived in different manners using the linear response theory
\cite{preR}, the projection operator formalism \cite{pre}, the BBGKY hierarchy
\cite{bbgky,onsagerkin}, the Klimontovich equation \cite{dubin,bbgky,klim}, the
Fokker-Planck equation \cite{preR,pre,cl,klim,onsagerkin}, 
and the functional approach \cite{fcp}. These derivations are rather formal and
technical. In the present paper, we complement the kinetic theory of point
vortices in the following manner:

(i) We provide a simpler and more physical derivation of the kinetic 
equation of 2D point vortices. We compute the diffusion coefficient $D$ and the
drift by polarization $V_{\rm pol}$ by a direct approach and substitute these
expressions into the Fokker-Planck equation written in a suitable form in which
the diffusion coefficient is ``sandwiched'' between the two gradients in
position so that the drift by polarization appears naturally.

(ii) We derive the proper expression of
the
fluctuation-dissipation 
theorem for 2D point vortices.

(iii) We consider a multi-species system of point vortices  while 
previous works were mostly restricted to point vortices with the same
circulation.

(iv) We consider unidirectional flows\footnote{A kinetic theory of Stewart
point vortices moving on the background of a shear flow with uniform vorticity
has been developed in \cite{nz}. It is valid in the case of short-range
interactions between point vortices making the problem similar to the kinetic
theory of gases. This is substantially different from the problem that we
consider here where the evolution of the point vortices is due to long-range
collisions and collective effects.} while previous works 
were developed for axisymmetric flows. We show that in the thermal bath
approximation the kinetic equation becomes similar to the Smoluchowski equation
of Brownian theory with a Rosen-Morse \cite{rosenmorse} (or P\"oschl-Teller
\cite{poschlteller}) potential and a
constant diffusion coefficient. This equation can be
transformed into a Schr\"odinger-like equation (in imaginary time) which can be
solved analytically \cite{prep1}. 

(v) We consider a system of collisionless 2D point
vortices (or a continuous vorticity field) submitted to a small external
stochastic
perturbation of arbitrary origin and derive a secular dressed diffusion
(SDD) equation sourced by the
external noise. 

(vi) The Landau and Lenard-Balescu equations are associated with the
microcanonical ensemble where the system of point vortices is isolated. In that
case, the point vortices are fundamentally described by $N$-body Hamiltonian
equations \cite{newton}. We compare these results with those obtained for a gas
of 2D Brownian point vortices \cite{mp,bv} described by  $N$-body stochastic
Langevin equations. We
establish a drift-diffusion equation governing their mean evolution as well
as a stochastic partial differential equation governing their mesoscopic
evolution. Similar equations are obtained from the stochastic damped 2D Euler
equations.

(vii) Throughout the paper, we mention the numerous analogies 
between the kinetic theory of 2D point vortices and the kinetic theory of
stellar systems (and other systems with long-range interactions).

The paper is organized as follows. In Sec. \ref{sec_inhos}, we
present the
basic equations describing a system of 2D point vortices
submitted to a small external stochastic perturbation and introduce the
quasilinear approximation. In Sec. \ref{sec_bog}, we
explain how the linearized equation for the perturbation can be
analytically solved with Fourier transforms by
making the Bogoliubov ansatz. In Sec. \ref{sec_inhosrf},  we determine the
linear response of the flow to a small external perturbation. In Sec.
\ref{sec_cf}, we relate the
dressed power
spectrum of the total fluctuating  stream function to the correlation function
of the external perturbation and consider
the case where the external perturbation  is due to a random distribution of
$N$ field vortices. In Sec. \ref{sec_fdi}, we derive the fluctuation-dissipation
theorem satisfied by an isolated system of point vortices at statistical
equilibrium. In
Sec. \ref{sec_fp}, we introduce the general Fokker-Planck equation adapted to a
gas of
point vortices. In Sec. \ref{sec_diffco}, we derive the diffusion coefficient
of a test vortex experiencing an external stochastic perturbation and consider
the case where the external perturbation  is due to a random distribution of
$N$ field vortices. In Sec. \ref{sec_ifpol}, we derive the drift by
polarization experienced by a test vortex traveling in a background flow
possibly created by a smooth distribution of field vortices. In
Sec. \ref{sec_ein}, we consider the evolution of a test vortex is a sea of field
vortices at statistical equilibrium and establish the appropriate form of
Einstein relation between the drift and the diffusion. In Sec.
\ref{sec_avp}, we
derive the kinetic equation of 2D point vortices with an arbitrary
velocity profile. In Sec. \ref{sec_mono}, we show how this kinetic
equation
simplifies itself when the velocity profile is monotonic.  In Sec.
\ref{sec_tdbv}, we
contrast the kinetic theory of an isolated Hamiltonian system of point vortices
to the
kinetic theory of a gas of 2D Brownian point vortices.
In Sec. \ref{sec_sdd}, we  derive the SDD equation
describing the mean evolution of a continuous vorticity field submitted to
a small external stochastic perturbation of arbitrary origin. In Sec.
\ref{sec_fd}, we study
the mean evolution and the mesoscopic evolution of a continuous vorticity
field described by the stochastic damped 2D Euler equations. We recover by
this approach the
power spectrum and the diffusion coefficient of a gas
of point vortices.   In Sec.
\ref{sec_spv}, we apply the same approach to the case of stochastically forced
2D point
vortices. In Sec. \ref{sec_diff}, we summarise our results and compare the
different kinetic equations obtained in this paper. The Appendices provide
useful complements to the results established in the main text.

\section{Basic equations}
\label{sec_inhos}

We consider a system of 2D point vortices of individual
circulation
$\gamma$ (see Appendix \ref{sec_pvmod}). We assume that the point
vortices move under
their own interactions and under the effect of an external stochastic
incompressible velocity field ${\bf u}_e({\bf r},t)$ (exterior
perturbation) of zero mean.
The equations of motion of the
point vortices   are  
\begin{eqnarray}
\frac{d{\bf
r}_{i}}{dt}=-{\bf z}\times \nabla\psi_d({\bf r}_i)-{\bf
z}\times \nabla\psi_e({\bf r}_i,t),
\label{n1zero}
\end{eqnarray}
where $\psi_d({\bf r})=-(1/2\pi)\sum_j \gamma \ln|{\bf r}-{\bf r}_j|$ is the
exact stream function
produced by the point vortices. They can be written in
Hamiltonian form as $\gamma d{\bf r}_i/dt=-{\bf z}\times
\nabla(H_d+H_e)$, where $H_d=-(1/2\pi)\sum_{i<j} \gamma^2 \ln|{\bf
r}_i-{\bf r}_j|$ is the Hamiltonian of the point vortices and
$H_e=\sum_i\gamma \psi_e({\bf r}_i,t)$ is the Hamiltonian associated with the
external flow. 
The discrete
vorticity field
$\omega_d({\bf r},t)=\sum_i \gamma\delta({\bf r}-{\bf r}_i(t))$ of the point
vortex gas
satisfies the equations
\begin{eqnarray}
\frac{\partial\omega_d}{\partial t}+({\bf u}_d+{\bf u}_e)\cdot \nabla\omega_d=0,
\label{ham1}
\end{eqnarray}
\begin{eqnarray}
{\bf u}_d=-{\bf z}\times\nabla\psi_d,\qquad \omega_d=-\Delta\psi_d,
\label{ham2}
\end{eqnarray}
\begin{eqnarray}
{\bf u}_e=-{\bf z}\times\nabla\psi_e,\qquad \omega_e=-\Delta\psi_e,
\label{ham2b}
\end{eqnarray}
where $\psi_d({\bf r},t)$ is the 
stream function produced by the point vortices and $\psi_e({\bf r},t)$ is the
external stochastic stream function. These equations are similar to the 2D
Euler-Poisson equations for an incompressible continuous flow but they apply
here to a singular vorticity field which is a sum of Dirac distributions. The
2D Euler-Poisson equations for an incompressible continuous flow are the
counterparts of the Vlasov-Poisson equations  in  stellar dynamics
\cite{jeans} and plasma physics \cite{vlasov} and the 2D Euler-Poisson equations
for a
singular system of point vortices are the counterparts of the Klimontovich
equations \cite{klimontovich} in plasma physics.

We introduce the mean vorticity $\omega({\bf r},t)=\langle
\omega_{d}({\bf r},t)\rangle$ corresponding to an ensemble
average of
$\omega_{d}({\bf r},t)$. We then write
$\omega_d({\bf r},t)=\omega({\bf r},t)+\delta \omega({\bf r},t)$ where
$\delta \omega({\bf r},t)$ denotes the fluctuations about the mean vorticity.
Similarly, we write $\psi_d({\bf r},t)=\psi({\bf
r},t)+\delta\psi({\bf r},t)$ where $\delta\psi({\bf
r},t)$ denotes the fluctuations about the mean stream function $\psi({\bf
r},t)=\langle \psi_d({\bf
r},t)\rangle$. Substituting this decomposition into Eq. (\ref{ham1}), we get
\begin{eqnarray}
\frac{\partial\omega}{\partial t}+\frac{\partial\delta\omega}{\partial t}+({\bf u}+\delta{\bf u}+{\bf u}_e)\cdot \nabla(\omega+\delta\omega)=0,
\label{ham3}
\end{eqnarray}
\begin{eqnarray}
{\bf u}=-{\bf z}\times\nabla\psi,\qquad \omega=-\Delta\psi,
\label{ham2bb}
\end{eqnarray}\begin{eqnarray}
\delta{\bf u}=-{\bf z}\times\nabla\delta\psi,\qquad \delta\omega=-\Delta\delta\psi.
\label{ham4}
\end{eqnarray}
If we introduce the total fluctuations $\delta {\bf u}_{\rm tot}=\delta{\bf
u}+{\bf u}_e$, $\delta\psi_{\rm tot}=\delta\psi+\psi_e$ and $\delta\omega_{\rm
tot}=\delta\omega+\omega_e$, which include the contribution of
the external perturbation, we can rewrite the foregoing
equations as
\begin{eqnarray}
\frac{\partial\omega}{\partial t}+\frac{\partial\delta\omega}{\partial t}+({\bf u}+\delta{\bf u}_{\rm tot})\cdot \nabla(\omega+\delta\omega)=0,
\label{ham5}
\end{eqnarray}
\begin{eqnarray}
{\delta{\bf u}}_{\rm tot}=-{\bf z}\times\nabla\delta\psi_{\rm tot},\qquad \delta\omega_{\rm tot}=-\Delta\delta\psi_{\rm tot}.
\label{ham5b}
\end{eqnarray}
Expanding the advection term in Eq. (\ref{ham5}) we obtain
\begin{eqnarray}
\frac{\partial\omega}{\partial t}+\frac{\partial\delta\omega}{\partial t}+{\bf u}\cdot \nabla \omega+{\bf u}\cdot\nabla\delta\omega+\delta{\bf u}_{\rm tot}\cdot \nabla\omega+\delta{\bf u}_{\rm tot}\cdot \nabla\delta\omega=0.
\label{ham6}
\end{eqnarray}
Taking the ensemble average of Eq. (\ref{ham6}) and subtracting the resulting equation from Eq. (\ref{ham6}) we obtain the two coupled equations
\begin{eqnarray}
\frac{\partial\omega}{\partial t}+{\bf u}\cdot \nabla \omega=-\nabla\cdot \langle \delta\omega \delta{\bf u}_{\rm tot}\rangle,
\label{ham7}
\end{eqnarray}
\begin{eqnarray}
\frac{\partial\delta\omega}{\partial t}+{\bf u}\cdot\nabla\delta\omega+\delta{\bf u}_{\rm tot}\cdot \nabla\omega=-\nabla\cdot (\delta\omega \delta{\bf u}_{\rm tot})+\nabla\cdot \langle \delta\omega \delta{\bf u}_{\rm tot}\rangle,
\label{ham8}
\end{eqnarray}
which govern the evolution of the mean flow  and the fluctuations. 
To get the right hand side of Eqs. (\ref{ham7}) and (\ref{ham8}) we have used
the incompressibility of the flow $\nabla\cdot \delta{\bf u}_{\rm tot}=0$ (see
Appendix \ref{sec_pvmod}). Equations (\ref{ham7}) and (\ref{ham8})  are exact
in the sense that no approximation has been made for the moment. The right hand
side of Eq.
(\ref{ham7}) can be interpreted as
a ``collision'' term arising from the granularity of the
system
(finite $N$ effects) and the correlations of the fluctuations due to
the external stochastic perturbation (forcing).\footnote{We
generically 
call it the ``collision'' term  although it may have a more general meaning due
to the
contribution of the external perturbation. A more
proper name could be the ``correlational'' term.}

We now assume that the external velocity
is weak and treat the stochastic stream function
$\psi_e({\bf r},t)$ as a small perturbation to the mean field dynamics.
We also assume that  the
fluctuation of the stream function $\delta\psi({\bf r},t)$ created by the point
vortices is weak. Since the circulation of the point vortices scales as
$\gamma\sim 1/N$ this approximation is valid when $N\gg
1$. If we ignore the external stochastic perturbation and the fluctuations of
the stream function due to finite $N$ effects altogether, the
collision term vanishes and Eq. (\ref{ham7}) reduces to the 2D Euler
equation
\begin{eqnarray}
\frac{\partial\omega}{\partial t}+{\bf u}\cdot \nabla \omega=0.
\label{ham7ep}
\end{eqnarray}
The 2D Euler-Poisson equations (\ref{ham7ep}) and  (\ref{ham2bb})  describe a
self-consistent mean field dynamics. It is valid in the limit
$\psi_e\rightarrow 0$ and  in a proper thermodynamic limit
$N\rightarrow +\infty$ with $\gamma\sim 1/N$. It is also valid for
sufficiently short times.

We now take into account a small correction to the 2D Euler equation obtained by
keeping the 
collision term on the right hand side of Eq. (\ref{ham7}) but neglecting the
quadratic terms on the right hand side of Eq. (\ref{ham8}). We therefore obtain
a set of two coupled equations
\begin{eqnarray}
\frac{\partial\omega}{\partial t}+{\bf u}\cdot \nabla \omega=-\nabla\cdot \langle \delta\omega \delta{\bf u}_{\rm tot}\rangle,
\label{ham7b}
\end{eqnarray}
\begin{eqnarray}
\frac{\partial\delta\omega}{\partial t}+{\bf u}\cdot\nabla\delta\omega+\delta{\bf u}_{\rm tot}\cdot \nabla\omega=0.
\label{ham8b}
\end{eqnarray}
These equations form the starting point of the quasilinear theory of 2D
point vortices which is
valid in a weak
coupling  approximation ($\gamma\sim 1/N\ll 1$) and for a weak external
stochastic perturbation ($\psi_e\ll 1$). Equation (\ref{ham7b})
describes the evolution of the mean vorticity sourced by
the correlations of the fluctuations and Eq. (\ref{ham8b}) describes the evolution
of the fluctuations due to  the
granularities of the system (finite $N$ effects) and the external noise. These
equations are valid at the order $1/N$ and to leading order in  
$\psi_e$.

If we restrict ourselves to unidirectional mean
flows\footnote{In
the following, we
assume
that the system remains unidirectional during the whole  evolution. This
may not always be the case. Even if we start from a unidirectional flow 
$\omega_0(y)$, 
the ``collision'' term (r.h.s. in Eq. (\ref{ham7b})) will change it and induce a
temporal 
evolution of the vorticity field $\omega(y,t)$. The system may
become
dynamically (Euler) unstable and undergo a dynamical phase
transition from a unidirectional flow to a more complicated flow (e.g., a large
scale vortex). 
We assume here that this transition does not take place or we consider a period
of time preceding this transition.} and introduce a cartesian system of
coordinates, we have
\begin{eqnarray}
{\bf u}=U(y,t){\bf x},\qquad \psi=\psi(y,t),\qquad \omega=\omega(y,t),
\label{ham9}
\end{eqnarray}
\begin{eqnarray}
U=\frac{\partial\psi}{\partial y}, \qquad \omega=-\frac{\partial^2\psi}{\partial y^2},\qquad  \omega=-U'(y,t).
\label{ham10}
\end{eqnarray}
On the other hand,  the two components of the  fluctuating velocity field read
\begin{eqnarray}
(\delta{\bf u}_{\rm tot})_x=\frac{\partial\delta\psi_{\rm tot}}{\partial y},\qquad (\delta{\bf u}_{\rm tot})_y=-\frac{\partial\delta\psi_{\rm tot}}{\partial x}.
\label{ham11}
\end{eqnarray}
As a result, Eqs. (\ref{ham7b}) and (\ref{ham8b}) become 
\begin{eqnarray}
\frac{\partial\omega}{\partial t}=\frac{\partial}{\partial y} \left\langle \delta\omega \frac{\partial\delta\psi_{\rm tot}}{\partial x}\right\rangle,
\label{ham12}
\end{eqnarray}
\begin{eqnarray}
\frac{\partial\delta\omega}{\partial t}+U\frac{\partial\delta\omega}{\partial x}- \frac{\partial\delta\psi_{\rm tot}}{\partial x}
\frac{\partial\omega}{\partial y}=0.
\label{ham13}
\end{eqnarray}
For the simplicity of the presentation, we have assumed that
the external velocity field ${\bf u}_e$ is of zero mean. If there is an external
(unidirectional) mean flow $U_e$,
it can be included in $U$ by making the substitution $U\rightarrow U+U_e$. In
other words, $U$ represents the total mean flow
including the mean flow produced by the system of point vortices and by the
external perturbation.

{\it Remark:} Although we have introduced the above equations for a system of
2D point vortices, they are  also valid for a continuous 2D incompressible flow
forced by an external velocity field.  In that case, Eqs.
(\ref{ham1})-(\ref{ham2b}) are  the 2D Euler equations for a continuous
vorticity field $\omega_c$ and a continuous velocity field ${\bf u}_c$ replacing
 the discrete vorticity field $\omega_d$ and the discrete velocity field ${\bf
u}_d$. If the
continuous flow is submitted to an external  stochastic perturbation ${\bf
u}_e$, we can decompose the vorticity and the velocity into  a mean component
plus  a fluctuation, writing  $\omega_c=\omega+\delta\omega$ and ${\bf u}_c={\bf
u}+\delta{\bf u}$, and obtain the same equations as above with a different
interpretation (see Sec. \ref{sec_sdd}).

\section{Bogoliubov ansatz}
\label{sec_bog}

In order to solve Eq. (\ref{ham13})  for the fluctuations, we resort to
the Bogoliubov ansatz. We  assume that there exist a timescale separation
between a slow and a fast dynamics and we regard $U(y)$ and $\omega(y)$ in Eq.
(\ref{ham13}) as ``frozen'' (independent of time). This amounts to neglecting
the temporal variation of the mean flow when we consider the evolution of the
fluctuations. This is possible when the mean
vorticity field evolves on
a
secular timescale that is long compared to the timescale over which the
correlations of the fluctuations have their essential
support.  We can then
introduce Fourier transforms in $x$ and $t$ for the vorticity fluctuations,
writing
\begin{eqnarray}
\delta{\omega}(x,y,t)=\int dk\int\frac{d\sigma}{2\pi}\, e^{i(kx-\sigma
t)}\delta{\hat \omega}(k,y,\sigma),
\label{ham15}
\end{eqnarray}
\begin{eqnarray}
\delta{\hat\omega}(k,y,\sigma)=\int\frac{dx}{2\pi}\int dt\, e^{-i(kx-\sigma
t)}\delta\omega(x,y,t).
\label{ham14}
\end{eqnarray}
Similar expressions hold  for the stream functions  $\delta\psi({\bf r},t)$
and $\psi_e({\bf r},t)$.  For future reference, we recall the Fourier
representation of the Dirac $\delta$-function
\begin{equation}
\delta(\sigma)=\int_{-\infty}^{+\infty} e^{i \sigma t}\, \frac{dt}{2\pi}.
\label{delta}
\end{equation} 

Before going further, some comments about our procedure of derivation are
required.
In order to derive the
Lenard-Balescu equation describing the mean evolution of a system of
point vortices under discreteness effects (``collisions'') at the order
$1/N$  we usually take
$\omega_e=0$ in Eq. (\ref{ham1}) and consider an initial value problem as
described in Sec. 3 of \cite{klim} (see also Appendix \ref{sec_j}). In that
case, Eq. (\ref{ham13}) has
to be solved by
introducing
a Fourier transform in space and a Laplace transform in time.
This brings a term $\delta{\hat\omega}(k,y,0)$ related to the initial condition
in the equation for the
fluctuations [see Eq. (\ref{j3})]. This is how discreteness
effects (granularities) are taken into account in this approach. Calculating the
correlation
function and
substituting the result into Eq.
(\ref{ham12}), one obtains a Lenard-Balescu-like equation in which the 
diffusion
and the drift terms appear simultaneously.   This derivation involves, however,
rather technical calculations. In the
present paper, we shall
derive the Lenard-Balescu equation differently by using a simpler and more
physical (or more pedagogical) approach based on the
Fokker-Planck
equation (see Sec. \ref{sec_fp}). In this approach, 
discreteness effects are taken into account  in the  external perturbation
$\omega_e$.  Indeed, we can regard $\omega_e$ either as having an arbitrary 
origin
(see
Sec. \ref{sec_inhosgr}) or as being generated by a collection
of $N$ point vortices -- the so-called field vortices (see Sec.
\ref{sec_inhoscf}). In the presence of an
external perturbation, Eq.
(\ref{ham13}) can be solved
by introducing
Fourier transforms in space and time.\footnote{In the
absence of external perturbation, we need to introduce a
Laplace transform in time yielding a term related to the initial condition.
In our approach, the initial condition is rejected to the infinite past but we
have to add a small imaginary term $i0^+$ in the
pulsation $\sigma$ of the Fourier transform to make the fluctuations vanish for
$t\rightarrow -\infty$. In a sense, this procedure amouts to using a Laplace
transform in time but neglecting the initial condition.} We
can then derive the 
dressed power spectrum and the diffusion coefficient of a test vortex by taking
collective effects
into
account (see Sec. \ref{sec_diffco}). On the other hand, the drift by
polarization of a test vortex can be
obtained by determining the response of the background flow to the perturbation
that it has caused (see Sec. \ref{sec_ifpol}). Substituting these coefficients
into
the Fokker-Planck equation, we obtain the Lenard-Balescu equation (see
Secs. \ref{sec_avp} and \ref{sec_mono}). This formalism also
allows us to treat situations in which the external perturbation $\omega_e$ is
not necessarily due to a discrete distribution of point vortices. In that case,
we can derive more
general kinetic equations. When $N\rightarrow +\infty$, i.e. when the collisions
between the point vortices are negligible, we obtain the
secular
dressed diffusion (SDD) equation 
involving a diffusion term due to the external perturbation (see
Sec. \ref{sec_sdd}). For 
finite $N$, we obtain a mixed kinetic equation involving a SDD term and a
Lenard-Balescu term (see
Sec. \ref{sec_diff}).

\section{Response function}
\label{sec_inhosrf}

Let us determine the linear response of a 2D incompressible flow to a small
external
perturbation
$\omega_e(x,y,t)$. 
In the present case, the perturbation is not necessarily stochastic. Since the
perturbation is small, we can use the linearized 2D
Euler equation (\ref{ham13}). Taking the Fourier transform of this equation in
$x$ and $t$, we obtain
\begin{eqnarray}
\delta{\hat\omega}(k,y,\sigma)=\frac{k\frac{\partial\omega}{\partial y}}{kU(y)-\sigma}\delta{\hat\psi}_{\rm tot}(k,y,\sigma).
\label{ham17}
\end{eqnarray}
On the other hand, according to  Eqs. (\ref{ham2b}) and (\ref{ham4}), we have 
\begin{eqnarray}
\Delta\delta\psi_{\rm tot}=-\delta\omega_{\rm tot}=-\delta\omega-\omega_e.
\label{ham18}
\end{eqnarray}
Writing this equation  in Fourier space and combining the result with Eq.
(\ref{ham17}), we get
\begin{eqnarray}
\left\lbrack \frac{d^2}{dy^2}-k^2+\frac{k\frac{\partial\omega}{\partial y}}{kU(y)-\sigma}\right\rbrack \delta{\hat\psi}_{\rm tot}=-{\hat\omega}_e.
\label{ham19}
\end{eqnarray}
The formal solution of this differential equation is
\begin{eqnarray}
\delta{\hat\psi}_{\rm tot}(k,y,\sigma)=\int G(k,y,y',\sigma) {\hat\omega}_e(k,y',\sigma)\, dy',
\label{ham20}
\end{eqnarray}
where  $G(k,y,y',\sigma)$ is the Green function defined by
\begin{eqnarray}
\left\lbrack \frac{d^2}{dy^2}-k^2+\frac{k\frac{\partial\omega}{\partial y}}{kU(y)-\sigma}\right\rbrack G(k,y,y',\sigma)=-\delta(y-y').
\label{ham21}
\end{eqnarray}
Although not explicitly written, we must use
the Landau
prescription $\sigma\rightarrow \sigma+i0^{+}$ in Eq. (\ref{ham21}). As
a result, $G(k,y,y',\sigma)$ is a complex function which plays the role of the
response function (or dielectric function) in plasma physics and stellar
dynamics \cite{bt,nicholson}.
It determines the
response of the system $\delta{\hat\psi}_{\rm tot}(k,y,\sigma)$ to an external
perturbation $ {\hat\omega}_e(k,y,\sigma)$ through Eq. (\ref{ham20}).  Assuming
that the perturbation $\omega_e(x,y)$ is time-independent and taking $\sigma=0$,
Eq. (\ref{ham21}) reduces to
\begin{eqnarray}
\left\lbrack \frac{d^2}{dy^2}-k^2+\frac{\frac{\partial\omega}{\partial
y}}{U(y)}\right\rbrack G(k,y,y')=-\delta(y-y').
\label{ham22}
\end{eqnarray}
This equation determines  the static response function $G(k,y,y')\equiv
G(k,y,y',\sigma=0)$ 
of the system to a time-independent perturbation.

Without external perturbation, the flow would be (by assumption) purely
unidirectional, described by $\omega(y)$. The external perturbation
$\omega_{e}(x,y,t)$ creates a weak
flow $\psi_{e}(x,y,t)$. This flow polarizes the system and induces through
Eq. (\ref{ham13}) a small change in the vorticity field $\delta\omega(x,y,t)$
producing in turn a weak flow $\delta\psi(x,y,t)$ through Eq. (\ref{ham4}).
As a result, the total stream function acting on a point vortex, sometimes
called the dressed or
effective stream function, is $\delta\psi_{\rm
tot}(x,y,t)=\psi_{e}(x,y,t)+\delta\psi(x,y,t)$. This  is the sum of the stream
function due to the external perturbation plus the stream
function induced by
the system itself (i.e. the system's own
response). Since $\delta\psi$
occurs in Eqs. (\ref{ham4}) and (\ref{ham13}), we have to
solve a loop. The total stream function is related to the external perturbation
$\omega_{e}(x,y,t)$ by Eq. (\ref{ham20}). The dressed Green function
$G(k,y,y',\sigma)$ takes into account the polarization of the system due to its
self-interaction. This corresponds to the so-called ``collective effects''. 
The polarization cloud surrounding a point vortex may amplify
or shield the action of the imposed external perturbation.  
Therefore,  the stream function is modified by collective
effects: $\delta\psi_{\rm tot}=\psi_{e}+\delta\psi\neq \psi_e$.
This leads to the notion of ``dressed'' point
vortices.\footnote{The notion of
``dressed''
particles (or quasiparticles) 
has been introduced and developed by Hubbard \cite{th,hubbard1,hubbard2} and
Rostoker \cite{rosenrost,rost1,rost2} in plasma physics.} If we
neglect collective effects ($\delta\omega=\delta\psi=0$), we have
$\delta\psi_{\rm tot}=\psi_e$ with
\begin{eqnarray}
{\hat\psi}_{e}(k,y,\sigma)=\int G_{\rm bare}(k,y,y') {\hat\omega}_e(k,y',\sigma)\, dy',
\label{ham20bare}
\end{eqnarray}
where $G_{\rm bare}(k,y,y')$ satisfies the equation 
\begin{eqnarray}
\left (\frac{d^2}{dy^2}-k^2\right ) G_{\rm
bare}(k,y,y')=-\delta(y-y').
\label{ham21bare}
\end{eqnarray}
Equations (\ref{ham20bare}) and (\ref{ham21bare}) define the bare stream 
function and the bare Green function. The bare Green function $G_{\rm
bare}(k,y,y')$ is just the Fourier transform in $x$ of the
potential of interaction between the vortices (see Appendix \ref{sec_gfn}). 
Similarly, the Green function $G(k,y,y',\sigma)$ can be interpreted as a dressed
potential of interaction between the vortices taking into account
collective effects. Neglecting collective
effects amounts to replacing the
dressed stream function (or dressed Green function) by the bare stream function
(or bare Green function).

{\it Remark:} When $\omega_e=0$, Eq. (\ref{ham19}) reduces to 
\begin{eqnarray}
\left\lbrack \frac{d^2}{dy^2}-k^2-\frac{U''(y)}{U(y)-\sigma/k}\right\rbrack
\delta{\hat\psi}=0,
\label{ham23}
\end{eqnarray}
where we have used Eq. (\ref{ham10}).  This is the celebrated 
Rayleigh equation \cite{rayleigh,drazin} which determines the proper complex
pulsations $\sigma$ of the flow associated with the velocity field $U(y)$. It
plays the role of the dispersion relation in plasma physics and stellar
dynamics \cite{nicholson,bt}. It can be used to study
the linear dynamical stability of unidirectional flows.

\section{Correlation function}
\label{sec_cf}

We now assume that 
the external perturbation $\omega_e(x,y,t)$ is a stochastic process and we
determine the dressed correlation function of the total stream function
$\delta\psi_{\rm tot}(x,y,t)$ that it
induces. We first give general results and then consider the case
where the external perturbation is produced by a random distribution of $N$
point
vortices.

\subsection{General results}
\label{sec_inhosgr}

We assume that the time evolution of the external vorticity field is
a stationary stochastic  process and write its auto-correlation function as
\begin{eqnarray}
\langle \omega_e (x,y,t)\omega_e(x',y',t')\rangle=\delta(y-y')C(x-x',y,t-t').
\label{ham24i}
\end{eqnarray}
We assume that the fluctuations are $\delta$-correlated in $y$ but not
necessarily in $x$ and $t$. The function $C(x-x',y,t-t')$ describes a
possibly colored
noise. The
Fourier transform in $x$ and $t$ of
the correlation function of the
external vorticity 
field is therefore 
\begin{eqnarray}
\langle {\hat \omega}_e (k,y,\sigma){\hat
\omega}_e(k',y',\sigma')\rangle=2\pi\delta(k+k')
\delta(\sigma+\sigma')\delta(y-y'){\hat C}(k,y,\sigma),
\label{ham24}
\end{eqnarray}
where ${\hat C}$ depends on $k$ and $\sigma$ (for a white noise
it would be constant). Similarly, we define the
correlation function $P(k,y,\sigma)$ of the
total fluctuating stream function acting on the point vortices by
\begin{eqnarray}
\langle \delta{\hat \psi}_{\rm tot} (k,y,\sigma)\delta{\hat \psi}_{\rm tot} (k',y,\sigma') \rangle=2\pi\delta(k+k')\delta(\sigma+\sigma')P(k,y,\sigma).
\label{ham25}
\end{eqnarray}
We call it the power spectrum by analogy with plasma
physics \cite{hubbard1}.\footnote{More precisely, the power spectrum is $k^2
P(k,y,\sigma)$.} Using Eqs.
(\ref{ham20}) and (\ref{ham24}), we get
\begin{eqnarray}
\langle \delta{\hat\psi}_{\rm tot}(k,y,\sigma)\delta{\hat\psi}_{\rm tot}(k',y,\sigma')\rangle&=&\int dy' dy''\, G(k,y,y',\sigma) G(k',y,y'',\sigma') \langle {\hat\omega}_e(k,y',\sigma) {\hat\omega}_e(k',y'',\sigma')\rangle\nonumber\\
&=&2\pi\delta(k+k')\delta(\sigma+\sigma')\int dy'\, G(k,y,y',\sigma) G(-k,y,y',-\sigma) {\hat C}(k,y',\sigma)\nonumber\\
&=&2\pi\delta(k+k')\delta(\sigma+\sigma')\int dy'\, |G(k,y,y',\sigma)|^2 {\hat C}(k,y',\sigma),
\label{ham26}
\end{eqnarray}
where we have used Eq. (\ref{obv}).  Comparing Eq. (\ref{ham26})
with Eq. (\ref{ham25}), we obtain
\begin{eqnarray}
P(k,y,\sigma)=\int dy'\, |G(k,y,y',\sigma)|^2 {\hat C}(k,y',\sigma).
\label{ham28}
\end{eqnarray}
This equation relates the power spectrum $P(k,y,\sigma)$ of the total
fluctuating stream function acting on the point vortices
to the correlation function ${\hat C}(k,y,\sigma)$ of the external
stochastic perturbation.
The ``dressed'' power spectrum $P(k,y,\sigma)$  takes into account collective
effects. If we
neglect collective effects, we just have to replace $G(k,y,y',\sigma)$ by
$G_{\rm bare}(k,y,y')$ in Eq. (\ref{ham28}). In that case, we find that the
``bare'' power spectrum $P_{\rm bare}(k,y,\sigma)$ defined by
\begin{eqnarray}
\langle {\hat \psi}_{e} (k,y,\sigma){\hat \psi}_{e}
(k',y,\sigma') \rangle=2\pi\delta(k+k')\delta(\sigma+\sigma')P_{\rm
bare}(k,y,\sigma)
\end{eqnarray}
is given by
\begin{eqnarray}
P_{\rm bare}(k,y,\sigma)=\int dy'\,
G_{\rm bare}(k,y,y')^2 {\hat C}(k,y',\sigma).
\end{eqnarray}
We note that ${\hat
C}(k,y,\sigma)$,
$P(k,y,\sigma)$ and  $P_{\rm bare}(k,y,\sigma)$ are real and positive.

{\it Remark:} From Eq. (\ref{ham25}) we easily obtain
\begin{equation}
\label{sta1}
\langle \delta{\hat\psi}_{\rm tot}(k,y,t)\delta{\hat\psi}_{\rm
tot}(k',y,t')\rangle=\delta(k+k'){\cal P}(k,y,t-t'),
\end{equation}
where ${\cal P}(k,y,t)$ is the inverse Fourier transform in time of
$P(k,y,\sigma)$. Therefore, the static power spectrum
$P({k},{y})={\cal P}({k},{y},0)$ is
\begin{eqnarray}
P({k},{y})=\int P({k},{y},\sigma)\,
\frac{d\sigma}{2\pi}.
\label{sta2}
\end{eqnarray}

\subsection{Correlation function created by a random distribution of $N$ field
vortices}
\label{sec_inhoscf}

We now assume that the external vorticity is created by a random distribution of
$N$ point vortices. We allow for different
species of point vortices with circulations $\lbrace \gamma_b\rbrace$. The
discrete vorticity of the field vortices is
\begin{eqnarray}
\omega_e(x,y,t)= \sum_i \gamma_i \delta(x-x_i(t))\delta(y-y_i(t)).
\label{ham29}
\end{eqnarray}
The initial positions $(x_i,y_i)$ of the point vortices are 
assumed to be uncorrelated and randomly distributed.  The  distribution
function of
point vortices of species $b$, with circulation $\gamma_b$,  is $P_1^{(b)}(y)$.
The mean vorticity  of species $b$ is therefore $\omega_b(y)=N_b \gamma_b
P_1^{(b)}(y)$,
where $N_b$ is the total number of point vortices of species $b$. When
$N\rightarrow +\infty$ with $\gamma\sim 1/N$, the point vortices are
advected by the mean flow $U(y)$  produced by the total vorticity
$\omega(y)=\sum_b \omega_b(y)$. Their mean field
trajectories are straight lines at constant $y$:
\begin{eqnarray}
x_i(t)=x_i+U(y_i)t,\qquad y_i(t)=y_i.
\label{ham10b}
\end{eqnarray}
As a result, the external vorticity can be written as
\begin{eqnarray}
\omega_e(x,y,t)=\sum_i \gamma_i \delta(x-x_i-U(y)t)\delta(y-y_i).
\label{ham29b}
\end{eqnarray}
Passing in Fourier space, we obtain
\begin{eqnarray}
{\hat\omega}_e(k,y,\sigma)&=& \sum_i\gamma_i \int\frac{dx}{2\pi}\int dt\, e^{-i(kx-\sigma t)} \delta(x-x_i-U(y)t)\delta(y-y_i)\nonumber\\
&=&\frac{1}{2\pi} \sum_i \gamma_i \int dt\, e^{-i k x_i} e^{-i(kU(y)-\sigma) t} \delta(y-y_i)\nonumber\\
&=& \sum_i  \gamma_i e^{-i k x_i} \delta(kU(y)-\sigma) \delta(y-y_i).
\label{ham30}
\end{eqnarray}
The correlation function of the external vorticity field in Fourier space is therefore  
\begin{eqnarray}
\langle {\hat\omega}_e(k,y,\sigma) {\hat\omega}_e(k',y',\sigma')\rangle&=& \left\langle \sum_{ij}  \gamma_i\gamma_j e^{-i k x_i} e^{-i k' x_j} \delta(kU(y)-\sigma)  \delta(k'U(y')-\sigma') \delta(y-y_i) \delta(y'-y_j)\right\rangle\nonumber\\
&=& \left\langle \sum_{i}  \gamma_i^2 e^{-i (k+k') x_i} \delta(kU(y)-\sigma)  \delta(k'U(y')-\sigma') \delta(y-y_i) \delta(y'-y_i)\right\rangle\nonumber\\
&=&\sum_b N_b \gamma_b^2 \int dx_1 dy_1\,  e^{-i (k+k') x_1} \delta(kU(y)-\sigma)  \delta(k'U(y')-\sigma') \delta(y-y_1) \delta(y'-y_1)P_1^{(b)}(y_1) \nonumber\\
&=&\sum_b \gamma_b \int dx_1 dy_1\, e^{-i (k+k') x_1} \delta(kU(y)-\sigma)  \delta(k'U(y')-\sigma') \delta(y-y_1) \delta(y'-y_1) \omega_b(y_1) \nonumber\\
&=&\sum_b 2\pi \gamma_b  \delta(k+k') \delta(\sigma+\sigma') \delta(y-y')  \delta(kU(y)-\sigma)\omega_b(y).\label{ham31}
\end{eqnarray}
To get the second line, we have used the 
decomposition $\sum_{ij}=\sum_i+\sum_{i\neq j}$ and the fact that the terms
involving different vortices  ($i\neq j$) vanish in average since the
point vortices
are initially uncorrelated. To get the third
line, we have
used the fact that the point vortices of the same species are
identical. Comparing Eqs. (\ref{ham24}) and
(\ref{ham31}), we find that 
\begin{eqnarray}
{\hat C}(k,y,\sigma)= \sum_b \gamma_b   \delta(kU(y)-\sigma)\omega_b(y).
\label{ham32}
\end{eqnarray}
This is the bare correlation function of the
vorticity field created by
the background point vortices. Using Eqs. (\ref{ham28}) and (\ref{ham32}), we
obtain the
dressed power
spectrum of the total
fluctuating stream function created by a random distribution of point vortices
\begin{eqnarray}
P(k,y,\sigma)=\sum_b \gamma_b\int dy'\, |G(k,y,y',\sigma)|^2  \delta(kU(y')-\sigma) \omega_b(y').
\label{ham33}
\end{eqnarray}
This returns the expression  (34) of the power spectrum
given in Ref. \cite{klim} which was obtained from the Klimontovich
formalism. The present approach provides an alternative, more physical,
manner to derive this result. Using Eqs. (\ref{sta2}) and  (\ref{ham33}), we
obtain the static power spectrum
\begin{eqnarray}
P(k,y)=\frac{1}{2\pi}\sum_b \gamma_b\int dy'\, |G(k,y,y',kU(y'))|^2
\omega_b(y').
\label{ham33m}
\end{eqnarray}

If we neglect collective effects, we find that the
bare correlation function
of the total
fluctuating stream function produced by a random
distribution of field vortices is
\begin{eqnarray}
P_{\rm bare}(k,y,\sigma)=\sum_b \gamma_b\int dy'\, G_{\rm bare}(k,y,y')^2 
\delta(kU(y')-\sigma) \omega_b(y').
\end{eqnarray}
It can be obtained from Eq. (\ref{ham33}) by replacing the dressed Green
function (\ref{ham21}) by the bare Green function (\ref{ham21bare}). The same
presciption can be used to obtain the bare static spectrum from Eq.
(\ref{ham33m}).

\subsection{Energy of fluctuations}
\label{sec_ieof}

The energy of fluctuations 
\begin{eqnarray}
{\cal E}=\frac{1}{2}\int \langle \delta\omega_{\rm tot} \delta\psi_{\rm
tot}\rangle\, dy=\frac{1}{2}\int \langle (\nabla\delta\psi_{\rm
tot})^2\rangle\, dy,
\label{nrj1}
\end{eqnarray}
where $\delta\omega_{\rm tot}=\delta\omega+\omega_e$ and $\delta\psi_{\rm
tot}=\delta\psi+\psi_e$ are the total fluctuations of vorticity and stream
function, can be calculated as follows.
Decompositing the fluctuations of stream function in Fourier modes, we get
\begin{eqnarray}
{\cal E}=-\frac{1}{2}\int dy\int dk\int
dk'\int\frac{d\sigma}{2\pi}\int\frac{d\sigma'}{2\pi}kk'e^{i(kx-\sigma
t)}e^{i(k'x-\sigma' t)}\langle \delta{\hat\psi}_{\rm
tot}(k,y,\sigma)\delta{\hat\psi}_{\rm tot}(k',y,\sigma')\rangle.
\label{nrj2}
\end{eqnarray}
Introducing the power spectrum from Eq. (\ref{ham25}) and integrating over
$k'$ and $\sigma'$, we obtain 
\begin{eqnarray}
{\cal E}=\frac{1}{2}\int dy\int
dk\int\frac{d\sigma}{2\pi}\,
k^2P(k,y,\sigma)=\frac{1}{2}\int dy\int
dk\, k^2P(k,y).
\label{nrj3}
\end{eqnarray}
The energy of fluctuations is equal to the integral of $k^2P(k,y)$ over
the wavenumber $k$ and over the position $y$. Using Eq. (\ref{ham28}), we can
rewrite the energy of fluctuations as 
\begin{eqnarray}
{\cal E}=\frac{1}{2}\int dy\int
dk\int\frac{d\sigma}{2\pi}\int dy'\, k^2 |G(k,y,y',\sigma)|^2 {\hat
C}(k,y',\sigma).
\label{nrj4}
\end{eqnarray}
When the perturbation is due to a distribution of $N$ point vortices, using
Eq. (\ref{ham32}), we obtain 
\begin{eqnarray}
{\cal E}=\frac{1}{4\pi}\sum_b\int dy\int
dk\int dy'\, k^2 |G(k,y,y',kU(y'))|^2\gamma_b \omega_b(y').
\label{nrj5}
\end{eqnarray}
We note that ${\cal E}$ is constant in time.

\section{Fluctuation-dissipation theorem for an isolated system of point vortices at
statistical equilibrium}
\label{sec_fdi}

\subsection{Fluctuation-dissipation theorem}
\label{sec_fdib}

The fluctuation-dissipation theorem for a gas of point vortices can be written as
\begin{eqnarray}
P(k,y,\sigma)=-\frac{1}{\pi\beta\sigma}{\rm Im}\,  G(k,y,y,\sigma).
\label{ham50}
\end{eqnarray}
It relates the power spectrum $P(k,y,\sigma)$ of the fluctuations 
 to the response function $G(k,y,y,\sigma)$ of a system of point vortices at
statistical
equilibrium with an inverse temperature $\beta$.\footnote{Note that Eq.
(\ref{ham50}) is
valid at positive and negative temperatures.} Although the
fluctuation-dissipation theorem can be established from very general arguments
\cite{nyquist,welton,greene1,greene2,kubo0,kubo},
we shall derive Eq. (\ref{ham50}) directly from the results obtained in the
preceding sections. 

We start from the general identity (see Appendix \ref{sec_id})
\begin{eqnarray}
{\rm Im}\,  G(k,y,y,\sigma)= \pi\int dy'\,  |G(k,y,y',\sigma)|^2 \delta(\sigma-kU(y')) k\frac{\partial\omega'}{\partial y'}.
\label{ham51i}
\end{eqnarray}
If the vorticity field is created by different species of point vortices, it
can be rewritten as
\begin{eqnarray}
{\rm Im}\,  G(k,y,y,\sigma)=\pi\sum_b\int dy'\,  |G(k,y,y',\sigma)|^2 \delta(\sigma-kU(y')) k\frac{\partial\omega'_b}{\partial y'}.
\label{ham51}
\end{eqnarray}
If the point vortices are at statistical equilibrium with the Boltzmann distribution
\begin{eqnarray}
\omega_b(y)=A_b\, e^{-\beta \gamma_b \psi(y)},
\label{ham52}
\end{eqnarray}
we have the identity
\begin{eqnarray}
\frac{\partial \omega_b}{\partial y}=-\beta \gamma_b \psi'(y)\omega_b(y).
\label{ham53}
\end{eqnarray}
Substituting Eq. (\ref{ham53}) into Eq. (\ref{ham51}), we obtain
\begin{eqnarray}
{\rm Im}\,  G(k,y,y,\sigma)&=&-\beta \pi\sum_b \gamma_b\int dy'\,  |G(k,y,y',\sigma)|^2 \delta(\sigma-kU(y')) k \frac{\partial \psi'}{\partial y'}\omega_b(y')\nonumber\\
&=&-\beta \pi\sum_b\gamma_b\int dy'\,  |G(k,y,y',\sigma)|^2 \delta(\sigma-kU(y')) k U(y') \omega_b(y')\nonumber\\
&=&-\beta\sigma \pi\sum_b\gamma_b\int dy'\,  |G(k,y,y',\sigma)|^2 \delta(\sigma-kU(y')) \omega_b(y')\nonumber\\
&=&-\beta\sigma \pi P(k,y,\sigma).
\label{ham54}
\end{eqnarray}
where we have used Eq. (\ref{ham10}) to get the second line, the property of the
$\delta$-function to get
the third line,
and the expression  (\ref{ham33}) of the power spectrum to get the fourth
line. This establishes Eq. (\ref{ham50}).

\subsection{Static power spectrum}
\label{sec_sps}

At statistical equilibrium, we have the following relation
\begin{eqnarray}
P(k,y)=-\frac{1}{\beta}G(k,y,y)
\label{ham55}
\end{eqnarray}
between the static  power spectrum $P(k,y)=\frac{1}{2\pi}\int P(k,y,\sigma)\,
d\sigma$ of
the fluctuations (see Sec.
\ref{sec_cf}) and
the static response
function $G(k,y,y)=G(k,y,y,0)$ (see Sec. \ref{sec_inhosrf}). This relation can
be derived directly 
from the microcanonical distribution of point vortices at statistical
equilibrium. It can also be recovered from  the fluctuation-dissipation theorem
(\ref{ham50}) as follows. 

Integrating  Eq. (\ref{ham50}) between
$-\infty$ and $+\infty$, we get
\begin{eqnarray}
P(k,y)=-\frac{1}{\beta}\int_{-\infty}^{+\infty} \frac{{\rm Im}\, 
G(k,y,y,\sigma)}{\pi\sigma}\, d\sigma.
\label{ham56}
\end{eqnarray}
On the other hand, applying at $\sigma=0$ the Kramers-Kronig relation
\cite{kram,kron}
\begin{eqnarray}
\label{ham57}
G(k,y,y,\sigma)={\rm P} \frac{1}{\pi}\int_{-\infty}^{+\infty}
\frac{{\rm
Im}\left\lbrack G(k,y,y,\sigma')\right\rbrack}{\sigma'-\sigma}\, d\sigma',
\end{eqnarray}
where P is the principal value,  we get
\begin{eqnarray}
\label{ham58}
G(k,y,y,0)=\frac{1}{\pi}\int_{-\infty}^{+\infty}
\frac{{\rm
Im}\left\lbrack G(k,y,y,\sigma)\right\rbrack}{\sigma}\, d\sigma.
\end{eqnarray}
Comparing Eqs. (\ref{ham56}) and (\ref{ham58}) we obtain Eq. (\ref{ham55}).

\section{Fokker-Planck equation}
\label{sec_fp}

Here, we consider the evolution of a test vortex of circulation $\gamma$
submitted to an external stochastic stream function $\psi_e(x,y,t)$. The
equations of motion of the test vortex  are (see Appendix
\ref{sec_pvmod} and
Sec. \ref{sec_inhos})
\begin{equation}
\label{fp1}
\frac{dx}{dt}=U(y,t)+\frac{\partial
\delta\psi_{\rm tot}}{\partial y}(x,y,t),\qquad \frac{dy}{dt}=-\frac{\partial
\delta\psi_{\rm tot}}{\partial x}(x,y,t),
\end{equation}
where $\delta\psi_{\rm tot}(x,y,t)$ is the total fluctuation of the stream
function. They can be written in Hamiltonian form as
$\gamma d{\bf
r}/dt=-{\bf z}\times \nabla (H+\delta H_{\rm tot})$, where
$H=\gamma\int^yU(y')\, dy'$
is the mean
Hamiltonian and $\delta H_{\rm tot}$ is the fluctuating Hamiltonian. The test
vortex follows a rectilinear trajectory 
at constant velocity $U(y)$ on the line level $y$ but it also experiences
a small stochastic perturbation $\delta\psi_{\rm tot}=\psi_e+\delta\psi$ which
is equal
to the external stream function $\psi_e$ plus the fluctuating stream function
$\delta\psi$ produced by the
system itself (collective effects). Eq. (\ref{fp1})
can be formally
integrated
into
\begin{equation}
\label{fp2}
 \quad x(t)=x+\int_0^t U(y(t'),t')\, dt'+\int_0^t \frac{\partial
\delta\psi_{\rm tot}}{\partial y}(x(t'),y(t'),t')\, dt', \qquad \quad y(t)=y-\int_0^t \frac{\partial
\delta\psi_{\rm tot}}{\partial x}(x(t'),y(t'),t')\, dt',
\end{equation}
where we have assumed that, initially, the test vortex is at position $(x,y)$. Since the fluctuations
$\delta\psi_{\rm tot}$ of the stream function are small, the changes in the position of the test vortex in the $y$-direction are also small, and the dynamics of the test vortex can
be represented by a stochastic process governed by a Fokker-Planck equation
\cite{risken}. The Fokker-Planck equation can be derived from the Master
equation by using the
Kramers-Moyal expansion truncated at the level of the second moments of the position increment. If we denote by $P(y,t)$ the  probability density that the
test vortex is at $y$ at time $t$,  the general form of the Fokker-Planck 
equation is 
\begin{equation}
\label{fp5}
\frac{\partial P}{\partial t}=\frac{\partial^{2}}{\partial y^2}\left (D P\right
)
-\frac{\partial}{\partial y}\left (P V_{\rm tot}\right
).
\end{equation}
The diffusion and drift coefficients are defined by
\begin{equation}
\label{fp6}
D(y)=\lim_{t\rightarrow +\infty}\frac{1}{2t} \langle
(y(t)-y)^2\rangle=\frac{\langle(\Delta y)^2 \rangle}{2\Delta t},
\end{equation}
\begin{equation}
\label{fp7}
V_{\rm tot}(y)=\lim_{t\rightarrow +\infty}\frac{1}{t} \langle
y(t)-y\rangle=\frac{\langle\Delta y\rangle}{\Delta t}.
\end{equation}
In writing these limits, we have implicitly assumed that the time $t$ is long
compared to the fluctuation time but short compared to the evolution
time. As shown in our previous papers
\cite{bbgky,klim}, it is relevant to rewrite the Fokker-Planck equation
in the alternative form
\begin{equation}
\label{fp8}
\frac{\partial P}{\partial t}=\frac{\partial}{\partial y} \left
(D\frac{\partial P}{\partial y}- P V_{\rm pol}\right ).
\end{equation}
The total drift can be written as
\begin{equation}
\label{fp9}
V_{\rm tot}=V_{\rm pol}+\frac{\partial D}{\partial y},
\end{equation}
where $V_{\rm pol}$ is the drift by  polarization (see Sec. \ref{sec_ifpol} of
this paper  and Sec.
4.3 of
\cite{klim}) while the
second term is due to the variation of the diffusion coefficient  with $y$ (see
Sec. 4.4 of \cite{epjp}).   The drift by
polarization ${V}_{\rm pol}$ arises from the retroaction (response) of the
perturbation caused by the test vortex on the mean flow. It
represents, however, only one
component of the total drift $V_{\rm tot}$ experienced by the test vortex, the
other
component being $\partial D/\partial y$. 

The two expressions (\ref{fp5}) and (\ref{fp8}) of the
Fokker-Planck equation have their own interest. The expression (\ref{fp5}),
where the diffusion coefficient is placed after the second derivative
$\partial^2(DP)$, involves the total drift $V_{\rm tot}$ and the expression
(\ref{fp8}), where the diffusion coefficient is placed between the derivatives
$\partial D\partial P$, isolates the drift by polarization $V_{\rm pol}$. We
shall see in Sec. \ref{sec_avp} that this second form is directly related to the
Lenard-Balescu equation. It has therefore a clear physical meaning.

\section{Diffusion coefficient}
\label{sec_diffco}

\subsection{General expression of the diffusion coefficient}
\label{sec_gei}

We now calculate the diffusion coefficient of the test vortex from Eq.
(\ref{fp6}) following the
approach 
developed in \cite{klim}. The increment in position of the test vortex in the
$y$-direction  is
\begin{equation}
\Delta y=-\int_0^t\frac{\partial\delta\psi_{\rm tot}}{\partial x}(x(t'),y(t'),t')\, dt'.
\label{delty}
\end{equation}
Substituting Eq. (\ref{delty}) into Eq. (\ref{fp6}) and assuming that the 
correlations of the fluctuating velocity persist for a time less than the
time for the trajectory of the test vortex to be much altered, we can make a
linear trajectory approximation
\begin{equation}
x(t')=x+U(y) t', \qquad y(t')=y,
\label{lin}
\end{equation}
and write
\begin{equation}
D=\lim_{t\rightarrow +\infty}\frac{1}{2t}\int_{0}^{t}dt'\int_{0}^{t}dt'' \,
\left\langle
\frac{\partial\delta\psi_{\rm tot}}{\partial x}(x+U(y)t',y,t')\frac{\partial\delta\psi_{\rm tot}}{\partial x}(x+U(y)t'',y,t'')\right\rangle.
\label{diff1}
\end{equation}
Introducing
the Fourier transform of the total fluctuating stream function, we obtain
\begin{eqnarray}
\left\langle \frac{\partial\delta\psi_{\rm tot}}{\partial x}(x+U(y)t',y, t')
\frac{\partial\delta\psi_{\rm tot}}{\partial x}(x+U(y)
t'',y,t'')\right\rangle=\int d{k}\int\frac{d\sigma}{2\pi}\int
d{{k}'}\int\frac{d\sigma'}{2\pi}\nonumber\\
\times k k' e^{i{k} ({x}+U(y) t')}e^{-i\sigma
t'}e^{i {k}' (x+U(y)
t'')}e^{-i\sigma' t''} \langle \delta\hat\psi_{\rm
tot}({k},y,\sigma)\delta\hat\psi_{\rm
tot}({k}',y,\sigma')\rangle.
\label{diff2}
\end{eqnarray}
Introducing the power spectrum from Eq. (\ref{ham25}) and carrying out the
integrals over $k'$ and $\sigma'$, we end up with the result
\begin{eqnarray}
\left\langle \frac{\partial\delta\psi_{\rm tot}}{\partial x}(x+U(y)t',y, t')
\frac{\partial\delta\psi_{\rm tot}}{\partial x}(x+U(y)
t'',y,t'')\right\rangle=\int d{k}\int\frac{d\sigma}{2\pi} \, k^2
e^{i({k}U(y)-\sigma)(t'-t'')}P({k},y,\sigma).
\label{diff3}
\end{eqnarray}
This expression shows that the auto-correlation function of the total
fluctuating velocity appearing in Eq.
(\ref{diff1}) depends only on the difference of times $s=t''-t'$. Using
the
identity \cite{klim}
\begin{eqnarray}
\int_{0}^{t}dt'\int_{0}^{t}dt''\,
f(t'-t'')&=&2\int_{0}^{t}dt'\int_{t'}^{t}dt''\,
f(t'-t'')=2\int_{0}^{t}dt'\int_0^{t-t'} ds\, f(s)\nonumber\\
&=&2\int_{0}^{t}ds\int_0^{t-s} dt'\, f(s)=  2\int_{0}^{t}ds\, (t-s)f(s),
\label{diff4}
\end{eqnarray}
and assuming that the autocorrelation function of the total fluctuating velocity
$f(s)$ decreases more rapidly
than $s^{-1}$, we find for $t\rightarrow +\infty$
that\footnote{This formula can also be obtained by using the 
identity
\begin{eqnarray}
D=\frac{1}{2}\frac{d}{dt}\langle (\Delta y)^2\rangle=\langle \dot y \Delta y \rangle= \int_0^t  \left\langle V_y(x,y,0)V_y({x}(t'),y(t'),t')\right\rangle\, dt',
\label{th1}
\end{eqnarray}
where $V_y=(\delta u_{\rm tot})_y=-\partial\delta\psi_{\rm tot}/\partial x$.
Making the approximations discussed above and taking
the limit $t\rightarrow +\infty$, we obtain
\begin{eqnarray}
D= \int_0^{+\infty} \left\langle V_y(x,y,0)V_y(x+U(y)s,y,s) \right\rangle\, ds,
\label{th2}
\end{eqnarray}
which coincides with Eq. (\ref{gei1}).}
\begin{eqnarray}
D=\int_0^{+\infty}\left\langle \frac{\partial\delta\psi_{\rm tot}}{\partial
x}(x,y,0)\frac{\partial\delta\psi_{\rm tot}}{\partial
x}(x+U(y)s,y,s)\right\rangle\, ds.
\label{gei1}
\end{eqnarray}
Therefore, as in Brownian theory \cite{uo,chandrabrown,risken,kubo0,kubo} and
in fluid turbulence \cite{taylor}, the
diffusion coefficient of a point vortex
is equal to the integral of the temporal auto-correlation function $\langle
V_y(0)V_y(t)\rangle$ of the fluctuating velocity felt by the point vortex
\cite{pre}:
\begin{eqnarray}
D=\int_0^{+\infty}\langle V_y(0)V_y(t)\rangle\, dt.
\label{dcorr}
\end{eqnarray}
This is
similar to the diffusion tensor of a star in a globular cluster which is
given by $D_{ij}=\int_0^{+\infty}\langle F_i(0)F_j(t)\rangle\, dt$ where ${\bf
F}(t)$ is the gravitational force by unit of mass experienced by the star
\cite{cvn0,lee,cohenamad,kandruprep,kandrup1,
kandrupfriction,hb4,aa} (see also \cite{cohen,gabor,th,hubbard1,hubbard2} for
plasmas). 
Replacing
the velocity auto-correlation function by
its expression from Eq. (\ref{diff3}), which can be written as
\begin{equation}
\langle {V}_y(0){V}_y(t)\rangle=\int d{k}\int\frac{d\sigma}{2\pi} \,
k^2
e^{i({k}U(y)-\sigma)t}P({k},y,\sigma),
\end{equation}
we obtain 
\begin{eqnarray}
D=\int_0^{+\infty}dt\int dk \int\frac{d\sigma}{2\pi}
 k^2  e^{i(\sigma-kU(y)) t}P(k,y,\sigma).
 \label{gei2}
\end{eqnarray}
Making the change of variables $t\rightarrow -t$, ${k}\rightarrow -{k}$ 
and $\sigma\rightarrow -\sigma$,
and using the fact
that $P(-{k},{y},-\sigma)=P({k},{y},\sigma)$,
we see that we can replace $\int_{0}^{+\infty}dt$ by
$(1/2)\int_{-\infty}^{+\infty}dt$. Therefore, we get
\begin{eqnarray}
D=\frac{1}{2}\int_{-\infty}^{+\infty}dt\int dk \int\frac{d\sigma}{2\pi}k^2 e^{i(\sigma-kU(y)) t}P(k,y,\sigma).
\label{gei3}
\end{eqnarray}
Using the identity (\ref{delta}), we find that
\begin{eqnarray}
D=\pi \int dk \int\frac{d\sigma}{2\pi}k^2 \delta(\sigma-kU(y)) P(k,y,\sigma).
\label{gei3b}
\end{eqnarray}
The time integration has given a $\delta$-function which creates a
resonance condition for interaction. Integrating over the $\delta$-function
(resonance), we arrive at the following equation
\begin{eqnarray}
D=\frac{1}{2} \int dk \, k^2 P(k,y,kU(y)).
\label{gei4}
\end{eqnarray}
This equation expresses the diffusion coefficient of the test vortex in terms of
the power spectrum of the fluctuations at the resonance
$\sigma=kU(y)$. This is the general
expression of the diffusion coefficient of a test vortex submitted to a
stochastic perturbation. When collective effects are neglected,
we get
\begin{eqnarray}
D_{\rm bare}=\frac{1}{2} \int dk \, k^2 P_{\rm bare}(k,y,kU(y)).
\label{gei4nc}
\end{eqnarray}

Using the relation between the power spectrum 
and the correlation function of the external perturbation [see Eq.
(\ref{ham28})], we obtain
\begin{eqnarray}
D=\frac{1}{2} \int dy'\int dk  \, k^2 |G(k,y,y',kU(y))|^2 {\hat
C}(k,y',kU(y)).
\label{gei5}
\end{eqnarray}
This expression shows that the diffusion coefficient of the test vortex depends
on the correlation function of the
external perturbation ${\hat C}(k,y',\sigma)$ and 
on the response function of the flow $G(k,y,y',\sigma)$ both
evaluated at the resonance
frequencies $\sigma=kU(y)$. As a result, the
diffusion coefficient $D(y)$ depends on the position $y$ of the test vortex, on
the mean vorticity field $\omega(y)$ through the Green function
$G(k,y,y',kU(y))$ defined by Eq. (\ref{ham21}), and on the mean velocity $U(y)$.
When collective effects
are neglected, i.e., when we replace $G(k,y,y',kU(y))$ by   $G_{\rm bare}(k,y,y')$ in Eq. (\ref{gei5}), the
diffusion coefficient reduces to 
\begin{eqnarray}
D_{\rm bare}=\frac{1}{2} \int dy'\int dk  \, k^2 G_{\rm bare}(k,y,y')^2 {\hat
C}(k,y',kU(y)).
\label{gei6}
\end{eqnarray}

{\it Remark:} Alternative derivations of the general expression of the 
diffusion coefficient of a test vortex are given in Appendices \ref{sec_geia}
and \ref{sec_geiab} (see also Appendix \ref{sec_geic} where we
compute the velocity auto-correlation function of the test vortex).

\subsection{Expression of the diffusion coefficient due to $N$ point vortices}
\label{sec_geib}

We now assume that the external noise is due to a discrete collection of $N$
point vortices. 
In that case ${\hat C}(k,y,\sigma)$ is given by Eq. (\ref{ham32}) and we obtain
\begin{eqnarray}
D=\frac{1}{2} \sum_b  \gamma_b\int dy' \int dk \, k^2 |G(k,y,y',kU(y))|^2    \delta(kU(y')-kU(y))\omega_b(y').
\label{gei7}
\end{eqnarray}
Using  the identity
\begin{eqnarray}
\delta(\lambda x)=\frac{1}{|\lambda|}\delta(x),
\label{gei7b}
\end{eqnarray}
we can rewrite the foregoing equation as
\begin{eqnarray}
D=\frac{1}{2} \sum_b \gamma_b \int dy'\int dk  \, |k| |G(k,y,y',kU(y))|^2   \delta(U(y')-U(y)) \omega_b(y').
\label{gei7diff}
\end{eqnarray}
This is the general 
expression of the diffusion coefficient of a test vortex of circulation
$\gamma$ produced by $N$ field vortices of circulation $\lbrace
\gamma_b\rbrace$. It returns Eq. (68) of \cite{klim}. The diffusion of the
test vortex is due to the fluctuations of the field vortices induced by finite
$N$ effects (granularities). This is why it
depends on $\lbrace
\gamma_b\rbrace$ but not on $\gamma$.\footnote{Note, however, that some field
vortices may have the circulation $\gamma_b=\gamma$, i.e., they belong to the
same species as the test vortex.}

Introducing the function
\begin{eqnarray}
\chi(y,y',U(y))=\frac{1}{2}\int |k| |G(k,y,y',kU(y))|^2 \, dk,
\label{gei10}
\end{eqnarray}
the diffusion coefficient (\ref{gei7diff}) can be written in the more compact
form
\begin{eqnarray}
D=\sum_b \gamma_b \int dy' \chi(y,y',U(y))   \delta(U(y')-U(y)) \omega_b(y').
\label{gei7c}
\end{eqnarray}
Using the identity
\begin{eqnarray}
\delta[g(x)]=\sum_j \frac{1}{|g'(x_j)|}\delta(x-x_j),
\label{gei7d}
\end{eqnarray}
where the $x_j$ are the simple zeros of the function $g(x)$ (i.e. $g(x_j)=0$ and
$g'(x_j)\neq 0$), we can write the diffusion coefficient as
\begin{eqnarray}
D=\sum_b \sum_r \gamma_b  \frac{\chi(y,y_r,U(y))}{|U'(y_r)|} \omega_b(y_r),
\label{gei7e}
\end{eqnarray}
where the $y_r$ are the points that resonate with $y$, i.e., the points that satisfy $U(y_r)=U(y)$.

If the velocity profile is monotonic,\footnote{When the vorticity $\omega(y)$
is always of the same
sign, the relation $\omega(y)=-U'(y)$
from Eq. (\ref{ham10}) implies that the velocity profile $U(y)$ is
monotonic. This is the case, in
particular, when the circulations $\lbrace
\gamma_b\rbrace$  of the point vortices have the same sign.} 
using the identity
\begin{eqnarray}
\delta(U(y')-U(y))=\frac{1}{|U'(y)|}\delta(y-y'),
\label{gei8}
\end{eqnarray}
we find that 
\begin{eqnarray}
D=\sum_b \gamma_b \frac{\chi(y,y,U(y))}{|U'(y)|} \omega_b(y).
\label{gei9}
\end{eqnarray}
For a multispecies gas of field vortices the diffusion
coefficient $D\propto {1}/{|U'(y)|}$ decreases when the shear
increases.  If we
consider a single species gas of field vortices 
with circulation $\gamma_b$,  the foregoing expression reduces to
\begin{eqnarray}
D=\gamma_b  \frac{\chi(y,y,U(y)) }{|U'(y)|} \omega_b(y)= |\gamma_b| \chi(y,y,U(y)).
\label{gei11}
\end{eqnarray}
To obtain the second equality,  we have used the relation  $\omega_b(y)=-U'(y)$
from Eq.
(\ref{ham10}). 

{\it Remark:} An alternative derivation of the 
diffusion coefficient of a test vortex produced by $N$ field vortices is given
in Appendix \ref{sec_dn} (see also Appendix \ref{sec_geic}).

\subsection{Expression of the diffusion coefficient due to $N$ point vortices without collective effects}
\label{sec_dco}

If we neglect collective effects, the previous results remain valid provided
that $G(k,y,y',kU(y))$ is 
replaced by $G_{\rm bare}(k,y,y')$. The bare diffusion coefficient of a test
vortex is 
\begin{eqnarray}
D_{\rm bare}=\frac{1}{2}\sum_b  \gamma_b  \int dy'\int dk \, |k| G_{\rm bare}(k,y,y')^2   \delta(U(y')-U(y))\omega_b(y').
\label{gei12}
\end{eqnarray}
Introducing the
function 
\begin{eqnarray}
\chi_{\rm bare}(y,y')=\frac{1}{2}\int |k| G_{\rm bare}(k,y,y')^2 \, dk,
\label{gei10b}
\end{eqnarray}
it can be written as
\begin{eqnarray}
D_{\rm bare}&=&\sum_b \gamma_b \int dy' \chi_{\rm bare}(y,y')   \delta(U(y')-U(y)) \omega_b(y')\nonumber\\
&=&\sum_b \sum_r \gamma_b  \frac{\chi_{\rm bare}(y,y_r)}{|U'(y_r)|} \omega_b(y_r).
\label{gei7du}
\end{eqnarray}
Explicit expressions of the function $\chi_{\rm bare}(y,y')$ are given in
Appendix \ref{sec_gfn}. In the
dominant approximation, we have $\chi_{\rm bare}(y,y')\simeq
(1/4)\ln\Lambda$.

If the velocity profile is monotonic, we obtain 
\begin{eqnarray}
D_{\rm bare}=\sum_b \gamma_b  \frac{\chi_{\rm bare}(y,y)}{|U'(y)|} \omega_b(y)
=\frac{1}{4}\sum_b \gamma_b  \frac{\ln\Lambda}{|U'(y)|} \omega_b(y),
\label{gei13}
\end{eqnarray}
where we have used Eq. (\ref{hamg10}).
If the field vortices have the same circulation, Eq. (\ref{gei13}) becomes
\begin{eqnarray}
D_{\rm bare}= \frac{1}{4}\gamma_b  \frac{\ln\Lambda}{|U'(y)|}
\omega_b(y)=\frac{1}{4}|\gamma_b| \ln\Lambda,
\label{gei15}
\end{eqnarray}
where we have used Eq. (\ref{ham10}). In that case, the bare diffusion
coefficient is constant.

\section{Drift by polarization}
\label{sec_ifpol}

Let us consider a 2D incompressible flow with a continuous vorticity profile
$\omega(y)$ and let us introduce a test vortex with a small circulation
$\gamma$ in
that
flow. The vorticity profile $\omega(y)$ may be due to a collection
of field vortices with circulations $\lbrace \gamma_b\rbrace$, in which case it
represent
their mean vorticity in the limit $N_b\rightarrow +\infty$ with $\gamma_b\sim
1/N_b$, but it can also have a more general origin. 
We want to determine the drift by polarization experienced by the test vortex
due to the perturbation that it causes to the flow. We use the formalism of
linear response theory developed in the previous sections and treat the
perturbation induced by the test vortex as a
small external perturbation $\omega_e(x,y,t)$ to the flow.

\subsection{Drift by polarization with collective effects}

The Fourier transform of the vorticity of the test vortex is [see Eq. (\ref{ham30})]
\begin{eqnarray}
{\hat\omega}_e(k,y,\sigma)=\gamma  e^{-i k x_0} \delta(kU(y)-\sigma) \delta(y-y_0),
\label{ham59}
\end{eqnarray}
where $(x_0,y_0)$ denotes the initial position of the test vortex. According to Eqs. (\ref{ham20}) and (\ref{ham59}), the Fourier transform of the total stream function (including collective effects) created by the test vortex is given by
\begin{eqnarray}
\delta{\hat\psi}_{\rm tot}(k,y,\sigma)&=&\int G(k,y,y',\sigma) {\hat\omega}_e(k,y',\sigma)\, dy'\nonumber\\
&=& \gamma\int G(k,y,y',\sigma)  e^{-i k x_0} \delta(kU(y')-\sigma) \delta(y'-y_0)\, dy'\nonumber\\
&=& \gamma G(k,y,y_0,\sigma)  e^{-i k x_0} \delta(kU(y_0)-\sigma).
\label{ham60}
\end{eqnarray}
The Fourier transform of the corresponding  velocity field  in the
$y$-direction $V_y=(\delta u_{\rm tot})_y=-{\partial\delta\psi_{\rm
tot}}/{\partial
x}$ is 
\begin{eqnarray}
{\hat V}_y(k,y,\sigma)&=&-ik\delta{\hat\psi}_{\rm tot}(k,y,\sigma)\nonumber\\
&=&-ik \gamma G(k,y,y_0,\sigma)  e^{-i k x_0} \delta(kU(y_0)-\sigma).
\label{ham62}
\end{eqnarray}
Returning to physical space, we get
\begin{eqnarray}
V_y(x,y,t)&=&-i\gamma\int dk\int\frac{d\sigma}{2\pi}\, e^{i(kx-\sigma t)}k  G(k,y,y_0,\sigma)  e^{-i k x_0} \delta(kU(y_0)-\sigma)\nonumber\\
&=&-i\frac{\gamma}{2\pi}\int dk \, e^{ik(x-x_0-U(y_0) t)}k \, G(k,y,y_0,kU(y_0)).
\label{ham63}
\end{eqnarray}
The test vortex is submitted to the velocity field resulting from the
perturbation that
it has caused and, as
a result,  it experiences a systematic drift \cite{preR}.  Applying Eq.
(\ref{ham63}) at the position of the test vortex at time $t$ ($x=x_0+U(y_0)t$,
$y=y_0$),  we obtain the drift by polarization
\begin{eqnarray}
V_{\rm pol}=-i\frac{\gamma}{2\pi}\int dk \, k \, G(k,y,y,kU(y)).
\label{ham64}
\end{eqnarray}
Since $V_{\rm pol}$ is real, we can write
\begin{eqnarray}
V_{\rm pol}=\frac{\gamma}{2\pi}\int dk \, k \,  {\rm Im} \, G(k,y,y,kU(y)).
\label{ham65}
\end{eqnarray}
Using the identity from Eq. (\ref{hami4})  we can rewrite the foregoing
equation as
\begin{eqnarray}
V_{\rm pol}=\frac{\gamma}{2}\int dk \int dy'\,  k^2 |G(k,y,y',kU(y))|^2 \delta(kU(y')-kU(y))\frac{\partial\omega'}{\partial y'}.
\label{ham66i}
\end{eqnarray}
Finally, using the identity from Eq. (\ref{gei7b}), we obtain
\begin{eqnarray}
V_{\rm pol}=\frac{\gamma}{2}\int dk \int dy'\,  |k| |G(k,y,y',kU(y))|^2 \delta(U(y')-U(y))\frac{\partial\omega'}{\partial y'}.
\label{ham66}
\end{eqnarray}
This is the general expression of the drift by polarization of the test
vortex. It returns Eq. (81) of \cite{klim}.\footnote{The fact
that the drift velocity calculated in this section corresponds to ${V}_{\rm
pol}$ in Eq. (\ref{fp9}) is
justified in Ref. \cite{klim} by calculating $V_{\rm tot}$ directly from Eq.
(\ref{fp7}).} The drift by
polarization of the test vortex is due to the retroaction (response) of the
perturbation
that it has caused to the mean flow. This is why it is proportional to
$\gamma$. The calculation of the polarization cloud created by the test
vortex is detailed in Appendix \ref{sec_pc}.

Introducing the function from Eq. (\ref{gei10}), the drift by polarization can
be written in the more compact form as
\begin{eqnarray}
V_{\rm pol}=\gamma \int dy'\,  \chi(y,y',U(y)) \delta(U(y')-U(y))\frac{\partial\omega'}{\partial y'}.
\label{ham66qb}
\end{eqnarray}
Using the identity from Eq. (\ref{gei7d}) we get
\begin{eqnarray}
V_{\rm pol}=\gamma \sum_r    \frac{\chi(y,y_r,U(y))}{|U'(y_r)|}\frac{\partial\omega}{\partial y}(y_r).
\label{ham66qc}
\end{eqnarray}
If the velocity profile is monotonic, we obtain 
\begin{eqnarray}
V_{\rm pol}=\gamma  \frac{\chi(y,y,U(y))}{|U'(y)|} \frac{\partial\omega}{\partial y}.
\label{ham69}
\end{eqnarray}
The drift by polarization is proportional to the vorticity 
gradient ${\bf V}_{\rm pol}\propto \gamma\nabla\omega$. If $\gamma>0$ the test
vortex
ascends the vorticity gradient. If $\gamma<0$ the test vortex descends the
vorticity gradient \cite{pre}. 

If the vorticity $\omega$  is due to a collection of $N$ field vortices with
circulation $\lbrace\gamma_b\rbrace$, the drift by polarization takes the form
\begin{eqnarray}
V_{\rm pol}=\frac{\gamma}{2}\sum_b \int dk \int dy'\,  |k| |G(k,y,y',kU(y))|^2 \delta(U(y')-U(y))\frac{\partial\omega'_b}{\partial y'}.
\label{ham66q}
\end{eqnarray}
The other equations remain valid with $\omega=\sum_b\omega_b$. If
the field vortices have the same circulation $\gamma_b$, the velocity field is
monotonic (see footnote 15), and Eq.
(\ref{ham69}) reduces to
\begin{eqnarray}
V_{\rm pol}=\gamma \frac{\chi(y,y,U(y))}{|U'(y)|}\frac{\partial\omega_b}{\partial y}.
\label{ger1}
\end{eqnarray}
We note the similarity between Eq. (\ref{ham66q}) and the expression
(\ref{gei7diff})
 of the diffusion coefficient created by a collection of point vortices.  The
main difference is that the drift by polarization involves the gradient of the
vorticity instead of the vorticity itself. In addition the drift by polarization
is proportional to the circulation $\gamma$ of the test vortex while the
diffusion involves the circulations $\lbrace\gamma_b\rbrace$ of the   field
vortices.

{\it Remark:} The drift by polarization $V_{\rm pol}$ is just one component of
the total drift $V_{\rm tot}$ of the test vortex which is given by Eq.
(\ref{fp9}).
Substituting Eqs. (\ref{gei7diff}) and (\ref{ham66}) into Eq. (\ref{fp9}) and
making an integration by parts,
we find that the total drift is
\begin{eqnarray}
V_{\rm tot}=\sum_b  \int dy'\, \omega_b(y') \left
(\gamma_b\frac{\partial}{\partial
y}-\gamma \frac{\partial}{\partial
y'}\right ) \chi(y,y',U(y)) \delta(U(y')-U(y)).
\label{totdrift}
\end{eqnarray}

\subsection{Drift by polarization without collective effects}
\label{sec_ifpolwo}

It is instructive to redo the calculation 
of the drift by polarization by neglecting collective effects from the
start.\footnote{We note that we cannot simply replace $G$ by $G_{\rm bare}$ in
Eq. (\ref{ham65}) otherwise we would find $V_{\rm pol}=0$ since $G_{\rm bare}$
is real. We first have to use Eq. (\ref{hami4}), then replace $G$ by
$G_{\rm bare}$ in
Eq. (\ref{ham66}).}
In that case, the change of the vorticity caused by the external
perturbation is determined by the equation
\begin{eqnarray}
\frac{\partial\delta\omega}{\partial t}+U\frac{\partial\delta\omega}{\partial
x}- \frac{\partial\psi_{e}}{\partial x}
\frac{\partial\omega}{\partial y}=0,
\label{ham13bare}
\end{eqnarray}
where we have neglected the term $\delta\psi$ in Eq. (\ref{ham13}). Written  in
Fourier space, we get
\begin{eqnarray}
\delta{\hat\omega}(k,y,\sigma)=\frac{k\frac{\partial\omega}{\partial y}}{kU(y)-\sigma}{\hat\psi}_{e}(k,y,\sigma).
\label{ham70}
\end{eqnarray}
Using Eqs. (\ref{ham20bare}) and (\ref{ham59}), the stream function created by the test vortex is
\begin{eqnarray}
{\hat\psi}_e(k,y,\sigma)=\gamma G_{\rm bare}(k,y,y_0)  e^{-i k x_0} \delta(kU(y_0)-\sigma).
\label{ham74}
\end{eqnarray}
Therefore, according to Eq. (\ref{ham70}), the perturbed vorticity field is
\begin{eqnarray}
\delta{\hat\omega}(k,y,\sigma)=\gamma\frac{k\frac{\partial\omega}{\partial
y}}{kU(y)-\sigma} G_{\rm bare}(k,y,y_0)  e^{-i k x_0}
\delta(kU(y_0)-\sigma).
\label{ham70f}
\end{eqnarray}
The Fourier transform of the stream function associated with the perturbed
vorticity field is  [see Eqs. (\ref{ham4}) and (\ref{hamg3})]
\begin{eqnarray}
\delta{\hat\psi}(k,y,\sigma)=\int G_{\rm bare}(k,y,y') \delta{\hat\omega}(k,y',\sigma)\, dy'.
\label{ham71}
\end{eqnarray}
Combining Eqs. (\ref{ham70f}) and (\ref{ham71}), we get
\begin{eqnarray}
\delta{\hat\psi}(k,y,\sigma)=\gamma\int dy'\, G_{\rm bare}(k,y,y')
\frac{k\frac{\partial\omega'}{\partial y'}}{kU(y')-\sigma}G_{\rm bare}(k,y',y_0)
 e^{-i k x_0} \delta(kU(y_0)-\sigma).
\label{ham75}
\end{eqnarray}
The Fourier transform of the corresponding velocity field in the $y$-direction
$V_y=(\delta u)_y=-{\partial\delta\psi}/{\partial
x}$ is 
\begin{eqnarray}
{\hat V}_y(k,y,\sigma)&=&-ik\delta{\hat\psi}(k,y,\sigma) \nonumber\\
&=&-ik \gamma\int dy'\, G_{\rm bare}(k,y,y') \frac{k\frac{\partial\omega'}{\partial y'}}{kU(y')-\sigma}G_{\rm bare}(k,y',y_0)  e^{-i k x_0} \delta(kU(y_0)-\sigma).
\label{ham76}
\end{eqnarray}
Returning to physical space, we get 
\begin{eqnarray}
V_y(x,y,t)&=&-i\gamma\int dk\int\frac{d\sigma}{2\pi}\, e^{i(kx-\sigma t)}\int dy'\, k G_{\rm bare}(k,y,y') \frac{k\frac{\partial\omega'}{\partial y'}}{kU(y')-\sigma}G_{\rm bare}(k,y',y_0)  e^{-i k x_0} \delta(kU(y_0)-\sigma)\nonumber\\
&=&\pi\gamma\int dk\int\frac{d\sigma}{2\pi}\, e^{i(kx-\sigma t)}\int
dy'\, k G_{\rm bare}(k,y,y') k\frac{\partial\omega'}{\partial y'}
\delta(kU(y')-\sigma)G_{\rm bare}(k,y',y_0)  e^{-i k x_0}
\delta(kU(y_0)-\sigma)\nonumber\\
&=&\frac{\gamma}{2}\int dk\, e^{ik(x-x_0-U(y_0) t)}\int dy'\, k
G_{\rm bare}(k,y,y')
k\frac{\partial\omega'}{\partial y'} \delta(k(U(y')-U(y_0))) G_{\rm bare}(k,y',y_0)\nonumber\\
&=&\frac{\gamma}{2}\int dk\, e^{ik(x-x_0-U(y_0) t)}\int dy'\, |k|
G_{\rm bare}(k,y,y')
\frac{\partial\omega'}{\partial y'} \delta(U(y')-U(y_0)) G_{\rm bare}(k,y',y_0),
\label{ham77}
\end{eqnarray}
where we have used the Landau prescription $\sigma\rightarrow \sigma+i0^{+}$ 
and the Sokhotski-Plemelj \cite{sokhotski,plemelj} formula
(\ref{plemelj}) to get the second
line, and Eq.
(\ref{gei7b}) to get the last line.

Applying Eq. (\ref{ham77}) at the position of the test vortex at time $t$
($x=x_0+U(y_0)t$, $y=y_0$) and using the identity $G_{\rm bare}(k,y',y)=G_{\rm
bare}(k,y,y')$ (see Appendix \ref{sec_gfn}),
we obtain the drift by
polarization
\begin{eqnarray}
V_{\rm pol}=\frac{\gamma}{2}\int dk\int dy'\, |k| G_{\rm bare}(k,y,y')^2 \delta(U(y')-U(y))\frac{\partial\omega'}{\partial y'}.
\label{ham80}
\end{eqnarray}
This is the general expression of the drift
by polarization
when collective effects are neglected.  It can be directly obtained from
Eq.
(\ref{ham66}) by
replacing the dressed Green function $G(k,y,y',kU(y))$ by the bare Green
function $G_{\rm bare}(k,y,y')$ (see footnote 17). 
Introducing the function from Eq. (\ref{gei10b}), it can be written in the more
compact form
\begin{eqnarray}
V_{\rm pol}&=&\gamma \int dy'\,  \chi_{\rm bare}(y,y') \delta(U(y')-U(y))
\frac{\partial\omega'}{\partial y'}\nonumber\\
&=&\gamma \sum_r    \frac{\chi_{\rm
bare}(y,y_r)}{|U'(y_r)|}\frac{\partial\omega}{\partial y}(y_r).
\label{ham66qbb}
\end{eqnarray}
Explicit expressions of the function $\chi_{\rm bare}(y,y')$ are given in
Appendix \ref{sec_gfn}. In the
dominant approximation, we have $\chi_{\rm bare}(y,y')\simeq
(1/4)\ln\Lambda$.
If the velocity profile is monotonic, we obtain 
\begin{eqnarray}
V_{\rm pol}=\gamma \frac{\chi_{\rm bare}(y,y)}{|U'(y)|}\frac{\partial\omega}{\partial y}=\frac{1}{4}\gamma  \frac{\ln\Lambda}{|U'(y)|} \frac{\partial\omega}{\partial y},
\label{ham69b}
\end{eqnarray}
where we have used Eq. (\ref{hamg10}).

If the vorticity $\omega$ is due to a collection of field vortices, the drift by
polarization takes the form
\begin{eqnarray}
V_{\rm pol}=\frac{\gamma}{2}\sum_b \int dk \int dy'\,  |k| G_{\rm bare}(k,y,y')^2 \delta(U(y')-U(y))\frac{\partial\omega'_b}{\partial y'}.
\label{ham81}
\end{eqnarray}
The other equations remain valid with $\omega=\sum_b\omega_b$. If
the field vortices have the same circulation $\gamma_b$, the
velocity profile is monotonic (see footnote 15) and Eq. (\ref{ham81}) reduces
to 
\begin{eqnarray}
V_{\rm pol}=\frac{1}{4}\gamma  \frac{\ln\Lambda}{|U'(y)|} \frac{\partial\omega_b}{\partial y}.
\label{ham69bis}
\end{eqnarray}

\section{Einstein relation}
\label{sec_ein}

We consider here the evolution of a test vortex is a sea of field vortices and
establish the Einstein relation for a thermal bath and its generalization for
an out-of-equilibrium bath.

\subsection{Einstein relation for a thermal bath}

If the field vortices are at statistical equilibrium with the Boltzmann
distribution (\ref{ham52}), 
the drift by polarization from Eq. (\ref{ham66q}) can be rewritten as
\begin{eqnarray}
V_{\rm pol}(y)&=&-\beta\sum_b \gamma_b\frac{\gamma}{2}\int dk \int dy'\,  |k| |G(k,y,y',kU(y))|^2 \delta(U(y')-U(y))\frac{d\psi'}{d y'}\omega_b(y')\nonumber\\
&=&-\beta\sum_b\gamma_b\frac{\gamma}{2}\int dk \int dy'\,  |k| |G(k,y,y',kU(y))|^2 \delta(U(y')-U(y))U(y')\omega_b(y')\nonumber\\
&=&-\beta\sum_b\gamma_b U(y)\frac{\gamma}{2}\int dk \int dy'\,  |k| |G(k,y,y',kU(y))|^2 \delta(U(y')-U(y))\omega_b(y').
\label{ham87}
\end{eqnarray}
To get the second line, we have used Eq. (\ref{ham10}) and to get the 
third line, we have used the properties of the $\delta$-function. Recalling the
expression (\ref{gei7diff}) of the diffusion coefficient, we obtain
\begin{eqnarray}
V_{\rm pol}(y)=-D\beta\gamma  U(y)=- D\beta\gamma\frac{d\psi}{dy}.
\label{ham88}
\end{eqnarray}
We see that the drift by polarization ${\bf V}_{\rm
pol}=-D\beta\gamma\nabla\psi$ is 
proportional to the gradient of the stream function  (hence perpendicular  to
the mean field velocity ${\bf u}=-{\bf z}\times \nabla\psi$) and that the drift
coefficient is given by a
form of Einstein relation \cite{preR,pre} 
\begin{eqnarray}
\mu=D\beta\gamma,
\label{ham88ein}
\end{eqnarray}
like in the theory of Brownian motion \cite{chandrabrown}.\footnote{We note that
the temperature arising in this expression can be positive
or negative. The temperature is negative in situations of most physical
interest \cite{onsager}.} The Einstein relation
connecting the drift to the diffusion coefficient is a
manifestation of the fluctuation-dissipation theorem (see the Remark below). We
note that
the Einstein relation is
valid for the drift by
polarization $V_{\rm pol}$, not for the total drift which has a more
complicated expression due to the term $\partial D/\partial y$
[see Eqs. (\ref{fp9}) and (\ref{totdrift})]. We do
not
have this subtlety
for the usual Brownian motion where the diffusion coefficient is constant.
When $\beta<0$ the
drift by polarization is directed towards $y=0$  so the
point vortices tend to
accumulate at the center of the domain. When $\beta>0$ the drift by
polarization is directed away from $y=0$ so the point vortices tend to move to
infinity. This is consistent with the results obtained by Onsager
\cite{onsager} from very general
considerations. The drift by polarization therefore provides a mechanism for
the 
self-organization of point vortices at negative temperatures.
The existence of a systematic drift for 2D point vortices in a
background vorticity gradient, and the analogy with Brownian theory, were first
discussed by Chavanis \cite{preR,pre}. The drift
experienced by a point vortex is the counterpart of the Chandrasekhar
dynamical friction experienced by a star in a stellar system
\cite{chandra1,chandra2,nice}. They both arise from a polarization process
\cite{preR,kandrupfriction}. The
necessity  of the drift of point vortices and the Einstein relation were
discussed in
Sec. III of \cite{pre} by using very general arguments similar to those given by
Chandrasekhar for stellar systems  \cite{chandra1,chandra2,nice}.

{\it Remark:} These results can also be obtained by substituting the 
fluctuation-dissipation theorem (\ref{ham50}) valid at statistical equilibrium
into the expression (\ref{ham65}) of the drift by polarization. This yields
\begin{eqnarray}
V_{\rm pol}(y,t)= -\beta\frac{\gamma}{2}U(y)\int dk \, k^2   P(k,y,kU(y)).
\label{ham92}
\end{eqnarray}
Recalling the expression (\ref{gei4}) of the diffusion coefficient, we recover
Eq. (\ref{ham88}). 
In this sense, the Einstein relation is another formulation of the
fluctuation-dissipation theorem. On the other hand, combining Eqs.
(\ref{dcorr}), (\ref{ham88}) and (\ref{ham88ein}) we obtain the relation
\begin{eqnarray}
\mu=\beta\gamma\int_0^{+\infty} \langle
V_y(0)V_y(t)\rangle\, dt.
\label{kuborela}
\end{eqnarray}
This is a form of
Green-Kubo \cite{green,kubo0,kubo} relation expressing the
fluctuation-dissipation theorem. There is a
similar relation in Brownian theory, fluids, plasmas and stellar systems
relating
the friction coefficient to the force auto-correlation function
\cite{kirkwood,green,ross,gnr,kubo0,mori,rubin,zwanzig,resibois,kubo,
kandruprep,kandrup1,kandrupfriction,hb4,aa}.

\subsection{Generalized Einstein relation}

We assume here that the vorticity field $\omega$ is due to 
a single population of point vortices with circulation $\gamma_b$ that is not
necessarily at statistical equilibrium. In that case, the velocity field is
monotonic (see footnote 15) and the distribution of field vortices
does not change on a timescale of the order $N\, t_D$ (see Sec.
\ref{sec_mono}). It forms therefore an out-of-equilibrium bath. The diffusion
coefficient and the drift by polarization are given by   Eqs. (\ref{gei11}) and
(\ref{ger1}). Combining these relations, we obtain
\begin{eqnarray}
V_{\rm pol}=\frac{\gamma}{\gamma_b}D\frac{\partial\ln|\omega_b|}{\partial y}.
\label{ger1h}
\end{eqnarray}
This equation can be seen
as a generalized Einstein relation for an out-of-equilibrium bath. If the field
vortices are at statistical equilibrium with the Boltzmann distribution
from Eq. (\ref{ham52}), we recover the expression of the drift by polarization
from Eq. (\ref{ham88}).

{\it Remark:} Eq. (\ref{ger1h}) can also be obtained by substituting into Eq.
(\ref{ham65}) the out-of-equilibrium fluctuation-dissipation theorem
(\ref{out4}) derived in Appendix \ref{sec_outfd} and using Eq. (\ref{gei4}).

\section{Kinetic equation with an arbitrary velocity profile}
\label{sec_avp}

The kinetic equation of 2D point vortices can
be obtained by substituting the expressions of the diffusion
coefficient (\ref{gei7diff}) and drift by polarization (\ref{ham66q}) into the
Fokker-Planck equation (\ref{fp8}). This provides
an alternative  derivation of the Lenard-Balescu equation of 2D point
vortices
as compared to the one given in \cite{klim}  which is based on the
Klimontovich
formalism. In this section, we study the general properties of this
equation for an arbitrary velocity profile. 

\subsection{Multi-species systems}

Substituting Eqs. (\ref{gei7diff}) and (\ref{ham66q}) into the Fokker-Planck
equation (\ref{fp8}) and
introducing the vorticity $\omega_a=N_a\gamma_a P_1^{(a)}$ of each species of
point vortices, we obtain the integrodifferential
equation 
\begin{eqnarray}
\frac{\partial\omega_a}{\partial t}=\frac{1}{2}\frac{\partial}{\partial y} 
\sum_b \int dy'\int dk\, |k| |G(k,y,y',kU(y))|^2   \delta(U(y')-U(y))\left
(\gamma_b\omega'_b \frac{\partial \omega_a}{\partial
y}-\gamma_a\omega_a\frac{\partial \omega'_b}{\partial y'}\right ),
\label{ham105}
\end{eqnarray}
where $\omega_a$ stands for $\omega_a(y,t)$ and $U(y)$ stands for $U(y,t)$. The
mean velocity $U(y,t)$ is determined by the total vorticity
$\omega(y,t)=\sum_a\omega_a(y,t)$. Equation (\ref{ham105}) is the counterpart
of the multi-species Lenard-Balescu 
equation in plasma physics. When collective effects are neglected, i.e., when
$G(k,y,y',kU(y))$ is replaced by $G_{\rm bare}(k,y,y')$, it reduces to the
multi-species Landau equation. It can be shown
\cite{cl,prep1} that the
kinetic equation (\ref{ham105}) conserves the total
energy $E=\frac{1}{2}\int\omega\psi\, dy$, the total impulse $P=\int\omega y\,
dy$ and the vortex number $N_a$ (or the
circulation $\Gamma_a=N_a\gamma_a=\int\omega_a\, dy$) of each species of point
vortices, and that the Boltzmann entropy
$S=-\sum_a\int
\frac{\omega_a}{\gamma_a}\ln\frac{\omega_a}{\gamma_a}\, dy$ increases
monotonically:  $\dot S\ge 0$ (H-theorem). Furthermore, the multi-species
Boltzmann distribution\footnote{The Boltzmann distribution is the ``most
probable'' distribution of point vortices. It can be obtained by
maximizing the
Boltzmann entropy
$S$ at fixed energy $E$, impulse $P$ and vortex numbers $N_a$ by introducing
appropriate Lagrange multipliers
$\beta$ (inverse temperature), $V$ (translation velocity) and $\mu_a$ (chemical
potentials) \cite{cl,prep1}. In the following, for simplicity, we shall work in
a frame of reference where $V=0$. More generally, we have to
replace the stream function $\psi$ by the relative stream function $\psi_{\rm
eff}=\psi-Vy$ \cite{lastepjb}.}
\begin{eqnarray}
\omega_a^{\rm eq}(y)=A_a e^{-\beta\gamma_a\psi(y)},
\label{aham108}
\end{eqnarray}
where the inverse temperature $\beta$ is the same for all species of point
vortices, 
is always a steady state of the kinetic equation (\ref{ham105}). We note that
the Boltzmann distribution of the different species of point vortices satisfies
the relation
\begin{eqnarray}
\omega_a^{\rm eq}(y)=C_{ab} [\omega_b^{\rm eq}(y)]^{\gamma_a/\gamma_b},
\label{aham109}
\end{eqnarray}
where $C_{ab}$ is a constant (independent of $y$). 
As discussed in 
Sec. \ref{sec_mono}, the kinetic equation (\ref{ham105}) does not necessarily
relax towards the Boltzmann distribution because of
the phenomenon of
kinetic blocking \cite{cl}.  The Lenard-Balescu equation is valid at the order
$1/N$ so it describes the evolution of the system on a timescale of the order
$Nt_D$. It may be necessary to develop the kinetic theory at higher order to 
describe the relaxation of the system of point vortices towards the Boltzmann
distribution.

{\it Remark:} Substituting Eqs. (\ref{gei4}) and (\ref{ham65}) into the
Fokker-Planck equation (\ref{fp8}) we obtain the kinetic  equation
\begin{eqnarray}
\frac{\partial \omega_a}{\partial t}=\frac{1}{2}\frac{\partial}{\partial y}\int dk\, k^2\left \lbrack P(k,y,kU(y))\frac{\partial\omega_a}{\partial y}-\omega_a \frac{\gamma_a}{\pi k} {\rm Im}\,  G(k,y,y,kU(y))\right \rbrack.
\label{ham102}
\end{eqnarray}
Using Eqs. (\ref{ham33}) and
(\ref{hami4}), we can check that Eq. (\ref{ham102}) is equivalent to Eq.
(\ref{ham105}).

\subsection{Moment equations}

Introducing the notation from Eq. (\ref{gei10}) the kinetic 
equation (\ref{ham105}) can be written as 
\begin{eqnarray}
\frac{\partial\omega_a}{\partial t}=\frac{\partial}{\partial y} \sum_b \int
dy'\, \chi(y,y',U(y)) \delta[U(y')-U(y)]\left (\gamma_b\omega'_b \frac{\partial
\omega_a}{\partial y}-\gamma_a\omega_a\frac{\partial \omega'_b}{\partial
y'}\right ).
\label{aham105}
\end{eqnarray}
Using the identity from Eq. (\ref{gei7d}), we get
\begin{eqnarray}
\frac{\partial\omega_a}{\partial t}=\frac{\partial}{\partial y} \sum_b\sum_r
\frac{\chi(y,y_r,U(y))}{|U'(y_r)|}\left (\gamma_b\omega^r_b \frac{\partial
\omega_a}{\partial y}-\gamma_a\omega_a\frac{\partial \omega^r_b}{\partial
y_r}\right ).
\end{eqnarray}
Equation (\ref{aham105}) has the form of a Fokker-Planck equation
\begin{eqnarray}
\frac{\partial\omega_a}{\partial t}=\frac{\partial}{\partial y}\left (D \frac{\partial \omega_a}{\partial y}-\omega_a V_{\rm pol}^{(a)}\right )
\label{aham105b}
\end{eqnarray}
with a diffusion coefficient [see Eq. (\ref{gei7c})]
\begin{eqnarray}
D=\sum_b    \int dy'\, \chi(y,y',U(y))   \delta[U(y')-U(y)] \gamma_b\omega'_b=    \int dy'\, \chi(y,y',U(y))   \delta[U(y')-U(y)] \omega'_2
\label{aham94}
\end{eqnarray}
and a drift by polarization [see Eq. (\ref{ham66q})]
\begin{eqnarray}
V_{\rm pol}^{(a)}=\gamma_a\sum_b  \int dy'\,  \chi(y,y',U(y)) \delta[U(y')-U(y)]\frac{\partial\omega'_b}{\partial y'}=\gamma_a \int dy'\,  \chi(y,y',U(y)) \delta[U(y')-U(y)]\frac{\partial\omega'}{\partial y'}.
\label{aham95}
\end{eqnarray}
In the second equalities of Eqs. (\ref{aham94}) and
(\ref{aham95}) we have introduced  the total vorticity
$\omega=\sum_a\omega_a=\sum_a N_a\gamma_a P_1^{(a)}$  and the second moment
$\omega_2=\sum_a \gamma_a\omega_a=\sum_a N_a\gamma_a^2 P_1^{(a)}$ of the
vorticity distribution. As explained
previously, the diffusion coefficient $D$ of a test vortex is due to the
fluctuation
of all the field vortices so it depends on $\lbrace \gamma_b\rbrace$ through
$\omega_2$. By
contrast, the
drift by polarization of a test vortex of species $a$ is due to the
retroaction (response) of the perturbation that this point vortex caused on the
mean flow
$\omega$. As a
result, $V_{\rm pol}^{(a)}$ is proportional to $\gamma_a$.

The equation for the total vorticity is
\begin{eqnarray}
\frac{\partial\omega}{\partial t}=\frac{\partial}{\partial y} \int dy'\, \chi(y,y',U(y)) \delta[U(y')-U(y)]\left (\omega'_2\frac{\partial \omega}{\partial y}-\omega_2\frac{\partial \omega'}{\partial y'}\right ).
\label{aham106}
\end{eqnarray}
This equation depends on the second moment $\omega_2$. We
can write down a hierarchy of equations for the moments $\omega_n=\sum_a
\gamma_a^{n-1}\omega_a=\sum_a N_a\gamma_a^n P_1^{(a)}$. The generic term of this
hierarchy is
\begin{eqnarray}
\frac{\partial\omega_n}{\partial t}=\frac{\partial}{\partial y} \int dy'\, \chi(y,y',U(y)) \delta[U(y')-U(y)]\left (\omega'_{2} \frac{\partial \omega_n}{\partial y}-\omega_{n+1}\frac{\partial \omega'}{\partial y'}\right ).
\label{aham107}
\end{eqnarray}
This hierarchy is not closed since the equation for $\omega_n$ depends
on $\omega_{n+1}$.

{\it Remark:} If we neglect collective effects, the kinetic equation
(\ref{aham105}) reduces to
\begin{eqnarray}
\frac{\partial\omega_a}{\partial t}=\frac{\partial}{\partial y} \sum_b \int
dy'\, \chi_{\rm bare}(y,y') \delta[U(y')-U(y)]\left (\gamma_b\omega'_b
\frac{\partial \omega_a}{\partial y}-\gamma_a\omega_a\frac{\partial
\omega'_b}{\partial y'}\right ),
\label{aham105nc}
\end{eqnarray}
where $\chi_{\rm bare}(y,y')$ is given in Appendix \ref{sec_gfn}. In the
dominant approximation, we have $\chi_{\rm bare}(y,y')\simeq
(1/4)\ln\Lambda$.

\subsection{Thermal bath approximation}
\label{sec_tbaq}

We consider a test vortex\footnote{This
can be a single point vortex or an ensemble of noninteracting point vortices
of the same species. In the second case, we take into
account the collisions
between the test vortices $\gamma$ and the field vortices
$\lbrace\gamma_b\rbrace$,
but we ignore the collisions between the test vortices.} of
circulation
$\gamma$
in ``collision'' with field vortices  of circulations $\lbrace
\gamma_b\rbrace$. We assume that the vorticity $\omega_b$ of the field vortices 
is prescribed (the validity of this approximation is discussed below). The
velocity profile $U(y)$, which is determined
by $\omega=\sum_b \omega_b$, is also prescribed. This situation
corresponds to the bath approximation in its general form. Under these
conditions, the Lenard-Balescu equation (\ref{aham105}) reduces  to 
\begin{eqnarray}
\frac{\partial P}{\partial t}=\frac{\partial}{\partial y}\sum_b \int dy'\, \chi(y,y',U(y))  \delta[U(y')-U(y)]\left (\gamma_b\omega'_b \frac{\partial P}{\partial y}-\gamma P \frac{d\omega'_b}{d y'}\right ).
\label{tb1}
\end{eqnarray}
Equation (\ref{tb1}) governs the evolution of the probability
density $P(y,t)$ of finding the test vortex of circulation $\gamma$ in $y$ at
time $t$. It can be written under the form of a Fokker-Planck equation
\begin{eqnarray}
\frac{\partial P}{\partial t}=\frac{\partial}{\partial y}\left (D \frac{\partial P}{\partial y}-P V_{\rm pol}\right )
\label{tb2}
\end{eqnarray}
with 
\begin{eqnarray}
D=\sum_b \int dy'\, \chi(y,y',U(y))  \delta[U(y')-U(y)]\gamma_b\omega'_b=\int dy'\, \chi(y,y',U(y))  \delta[U(y')-U(y)]\omega'_2
\label{tb3}
\end{eqnarray}
and
\begin{eqnarray}
V_{\rm pol}=\gamma\sum_b \int dy'\, \chi(y,y',U(y))  \delta[U(y')-U(y)] \frac{d\omega'_b}{dy'}=\gamma \int dy'\, \chi(y,y',U(y))  \delta[U(y')-U(y)] \frac{d\omega'}{dy'}.
\label{tb4}
\end{eqnarray}
In this manner, we have transformed 
an integrodifferential equation of the Lenard-Balescu or Landau form [see Eq.
(\ref{aham105})] into a  differential equation  of the Fokker-Planck form [see
Eq. (\ref{tb1})]. 

This bath approach is self-consistent in the general case only if the 
field vortices are at statistical equilibrium, otherwise their distribution
$\omega_b$ evolves in time due to discreteness effects. If we assume that
the field vortices are at statistical equilibrium with the Boltzmann
distribution (\ref{ham52}), corresponding to the
thermal bath
approximation, the Fokker-Planck
equation (\ref{tb1}) becomes
\begin{eqnarray}
\frac{\partial P}{\partial t}&=&\frac{\partial}{\partial y} \sum_b \int dy'\,   \chi(y,y',U(y))\delta[U(y')-U(y)]\left (\gamma_b\omega'_b \frac{\partial P}{\partial y}+\beta\gamma_b\gamma P U(y') \omega'_b\right )\nonumber\\
&=&\frac{\partial}{\partial y} \sum_b \int dy'\,   \chi(y,y',U(y))\delta[U(y')-U(y)]\gamma_b\omega'_b\left (\frac{\partial P}{\partial y}+\beta\gamma P U(y)\right ),
\label{tb5}
\end{eqnarray}
where we have used Eqs. (\ref{ham10}) and (\ref{ham52})  to get the first
equality, and the properties of the $\delta$-function to get the second
equality.
Using Eq. (\ref{ham10}) again, we can write the Fokker-Planck 
equation under the form
\begin{eqnarray}
\frac{\partial P}{\partial t}=\frac{\partial}{\partial y}\left \lbrack D(y)\left (\frac{\partial P}{\partial y}+\beta \gamma P\frac{d\psi}{dy} \right )\right\rbrack,
\label{tb6}
\end{eqnarray}
where $D$
is given by Eq. (\ref{tb3}) with the Boltzmann distribution (\ref{ham52}).
Equation (\ref{tb6}) can also be directly obtained from Eq. (\ref{tb2}) by using
the expression (\ref{ham88}) of the drift by polarization valid for a thermal
bath (Einstein relation). The Fokker-Planck equation (\ref{tb6})
conserves the normalization condition $\int P\, dy=1$ and decreases the free
energy $F=E-TS$ with $E=\int \gamma P\psi\, dy$ and $S=-\int P\ln P\, dy$
monotonically: $\dot F\le 0$ (canonical H-theorem). It relaxes towards the
Boltzmann distribution
\begin{eqnarray}
P_{\rm eq}(y)=A\, e^{-\beta\gamma\psi(y)}.
\label{tb7}
\end{eqnarray} 
Since the
Fokker-Planck equation (\ref{tb6})  is valid at the
order $1/N$, the
relaxation time of a test vortex in a thermal bath scales as
\begin{equation}
t_{\rm R}^{\rm bath}\sim {N}t_D,
\label{rbath1}
\end{equation}
where $t_D$ is the dynamical time. The
Fokker-Planck equation (\ref{tb6}) for 
a test vortex is similar to the Smoluchowski \cite{smoluchowski} equation
describing the
evolution
of an overdamped Brownian particle submitted to  an external potential $\psi$. 
The Smoluchowski equation for a test vortex evolving in a sea of field vortices
is the
counterpart of the Klein-Kramers-Chandrasekhar
equation
\cite{klein,kramers,chandrabrown,chandrany,chandra1,chandra2,chandra3,nice} for
a test star evolving in a stellar system. The
relaxation
of a
system of point
vortices towards statistical equilibrium is due to a competition between
diffusion and
drift \cite{preR,pre}. Similarly, the relaxation of a stellar system is  due to
a competition
between diffusion and friction \cite{chandra1,chandra2,nice}. We note that 2D
point vortices do not have
inertia. This is why they resemble  overdamped Brownian particles described by
the Smoluchowski equation in configuration space rather than inertial Brownian
particles
described by the Kramers equation in phase space.

{\it Remark:} In the thermal bath approximation, substituting the
fluctuation-dissipation theorem (\ref{ham50}) into the kinetic equation
(\ref{ham102}) and using Eq. (\ref{ham10}), we obtain
\begin{eqnarray}
\frac{\partial P}{\partial t}=\frac{1}{2}\frac{\partial}{\partial
y}\int dk\, k^2 P(k,y,kU(y))\left ( \frac{\partial P}{\partial
y}+\beta \gamma P \frac{d\psi}{dy}\right ).
\end{eqnarray}
This equation  is equivalent to the Fokker-Planck equation (\ref{tb6}). The
diffusion coefficient $D$
is given by Eqs. (\ref{ham33}) and (\ref{gei4}) which return Eq.
(\ref{tb3}).

\subsection{Pure diffusion}
\label{sec_purediff}

When $\gamma\ll \gamma_b$, the drift by polarization can be neglected ($V_{\rm
pol}=0$)  and the test vortex  has a purely diffusive evolution. In that case,
the Fokker-Planck equation (\ref{tb2}) reduces to 
\begin{eqnarray}
\frac{\partial P}{\partial t}=\frac{\partial}{\partial y}\left ( D\frac{\partial
P}{\partial y}\right ),
\label{pd1}
\end{eqnarray}
where $D$ is given by Eq. (\ref{tb3}). We note that the diffusion coefficient
$D$ depends on the circulations $\lbrace\gamma_b\rbrace$ of the field vortices
through the second moment  $\omega_2$. This reflects the discrete nature of the
field vortices.  On the other hand, the diffusion coefficient does not depend on
the circulation $\gamma$ of the test vortex.

{\it Remark:} Since the diffusion coefficient depends on $y$ and 
since it is placed between the two spatial derivatives $\partial/\partial y$,
Eq.
(\ref{pd1}) is not exactly a diffusion equation. It can be rewritten as Eq.
(\ref{fp5}) showing that the test vortex experiences a drift [see Eq.
(\ref{fp9})]
\begin{eqnarray}
V_{\rm tot}=\frac{\partial D}{\partial y}.
\label{pd2}
\end{eqnarray}

\subsection{Pure drift}

When $\gamma\gg \gamma_b$, the diffusion can be neglected ($D=0$) and the test
vortex has a purely deterministic evolution. In that case, the Fokker-Planck
equation (\ref{tb2})  reduces to 
\begin{eqnarray}
\frac{\partial P}{\partial t}=\frac{\partial}{\partial y}\left (-P V_{\rm pol}\right ),
\label{pr1}
\end{eqnarray}
where $V_{\rm pol}$ is given by Eq. (\ref{tb4}).
We note that the drift by polarization  $V_{\rm pol}$ is proportional to the
circulation  $\gamma$
of the test vortex. On the other hand, it depends only on the mean
vorticity field $\omega$, not on the individual
circulations $\lbrace\gamma_b\rbrace$  of the
field vortices reflecting their discrete nature. As a
result, Eq. (\ref{pr1}) remains valid when a test vortex of circulation
$\gamma$ evolves in a continuous
vorticity field $\omega$ which is not necessarily due to a collection of point
vortices.

{\it Remark:} The deterministic motion of the test
vortex induced by the
drift term can be written as
\begin{eqnarray}
\frac{dy}{dt}=V_{\rm pol}(y,t),
\label{ham96}
\end{eqnarray}
and we have $V_{\rm tot}=V_{\rm pol}$ [see Eq.
(\ref{fp9})].  This equation describes the drift of a point vortex in a mean
flow. This is the counterpart of the sinking satellite problem in
stellar dynamics \cite{bt}. The change of
energy of the test vortex is
$\dot \epsilon=\gamma UV_{\rm pol}$. This relation can be obtained by taking the
time variation of $E=\gamma\int P\psi \, dy$, using Eq. (\ref{pr1}), and
integrating by parts. Using the general expression  (\ref{ham66i}) of the drift
term, we obtain
\begin{eqnarray}
\dot\epsilon=\frac{\gamma^2}{2}\int dk \int dy'\,  k \lbrack kU(y')\rbrack
|G(k,y,y',kU(y))|^2 \delta(kU(y')-kU(y))\frac{d\omega'}{d y'}.
\label{ham66ik}
\end{eqnarray}
For a stationary flow of the form $\omega=\omega(\psi)$, we find that
\begin{eqnarray}
\dot\epsilon=\frac{\gamma^2}{2}\int dk \int dy'\,  \lbrack kU(y')\rbrack^2
|G(k,y,y',kU(y))|^2 \delta(kU(y')-kU(y))\frac{d\omega'}{d\psi'}.
\label{ham66ik2}
\end{eqnarray}
Therefore $\dot\epsilon$ is negative if $d\omega/d\psi<0$ and positive
if $d\omega/d\psi>0$. In the first case, the test vortex loses energy to the
flow and in the second case it gains energy from the flow. For the Boltzmann
distribution (\ref{ham52}), we have
$d\omega/d\psi=-\beta\omega_2$ \cite{lastepjb}. Therefore, $d\omega/d\psi<0$
corresponds to $\beta>0$ (positive temperature) and $d\omega/d\psi>0$
corresponds to $\beta<0$ (negative temperature).

\section{Kinetic equation with a monotonic velocity profile}
\label{sec_mono}

In this section, we study the general properties of the kinetic equation of
2D point vortices when the  velocity profile is monotonic. 

\subsection{Multi-species systems}
\label{sec_mscq}

If we velocity profile is monotonic, using identity (\ref{gei8}), the
kinetic equation (\ref{aham105}) becomes\footnote{It is possible that,
initially, the velocity profile is
non-monotonic 
but that it becomes monotonic during the evolution. In that case, the evolution
of the vorticity is first described by Eq. (\ref{aham105}), then by Eq.
(\ref{aham105bg}).}
\begin{eqnarray}
\frac{\partial\omega_a}{\partial t}=\frac{\partial}{\partial y} \sum_b  \frac{\chi(y,y,U(y))}{|U'(y)|}\left (\gamma_b\omega_b \frac{\partial \omega_a}{\partial y}-\gamma_a\omega_a\frac{\partial \omega_b}{\partial y}\right ).
\label{aham105bg}
\end{eqnarray}
It can be written as a Fokker-Planck equation of the form of Eq.
(\ref{aham105b}) with a
diffusion coefficient [see Eq. (\ref{gei9})]
\begin{eqnarray}
D=\sum_b  \frac{\chi(y,y,U(y))}{|U'(y)|}\gamma_b \omega_b= \frac{\chi(y,y,U(y))}{|U'(y)|} \omega_2
\label{aham94b}
\end{eqnarray}
and a drift by polarization [see Eq.  (\ref{ham69})]
\begin{eqnarray}
V_{\rm pol}^{(a)}=\gamma_a\sum_b   \frac{\chi(y,y,U(y))}{|U'(y)|} \frac{\partial\omega_b}{\partial y}=\gamma_a   \frac{\chi(y,y,U(y))}{|U'(y)|} \frac{\partial\omega}{\partial y}.
\label{aham95b}
\end{eqnarray}
Equation
(\ref{aham105bg}) conserves the energy, the impulse and the vortex number of
each species, and
satisfies an $H$-theorem for the Boltzmann entropy. At equilibrium, the
currents $J_a$ defined by $\partial\omega_a/\partial t=-\partial
J_a/\partial y$
vanish (for each species), and  we have the relation
\begin{eqnarray}
\omega_a^{\rm eq}(y)=C_{ab} [\omega_b^{\rm eq}(y)]^{\gamma_a/\gamma_b},
\label{aham109b}
\end{eqnarray}
where $C_{ab}$ is a constant (independent of $y$).  This relation is similar to
Eq. (\ref{aham109})
which was obtained in the case where the equilibrium vorticity is given by the
Boltzmann distribution (\ref{aham108}). However, it is important to mention
that
Eq.
(\ref{aham105bg})  does {\it not} relax towards the Boltzmann distribution (see
below). Therefore, the equilibrium vorticity distribution $\omega_a^{\rm eq}(y)$
in Eq. (\ref{aham109b}) is generally not given by Eq. (\ref{aham108}).

Equation (\ref{aham106}) for the total vorticity  reduces to
\begin{eqnarray}
\frac{\partial\omega}{\partial t}=0.
\label{aham106b}
\end{eqnarray}
This equation is closed and shows that the average vorticity profile does not
change (the total current vanishes). This kinetic blocking is discussed in more
detail in the following section. We note, by contrast, that the vorticity
$\omega_b$ of the different species evolves in time. 
Because of the drift term (see Sec. \ref{sec_ifpol}) the vortices with large
positive (resp.
negative) circulation tend to concentrate around maxima (resp. minima) of
vorticity. The hierarchy of equations
for the moments of the vorticity, Eq. (\ref{aham107}),  reduces to
\begin{eqnarray}
\frac{\partial\omega_n}{\partial t}=\frac{\partial}{\partial y}\left\lbrack 
\frac{\chi(y,y,U(y))}{|U'(y)|}\left (\omega_{2} \frac{\partial
\omega_n}{\partial y}-\omega_{n+1}\frac{\partial \omega}{\partial y}\right
)\right\rbrack.
\label{aham107b}
\end{eqnarray}
This hierarchy of moments   is not closed (for $n\ge 2$) since the equation for
$\omega_{n}$ depends on $\omega_{n+1}$.

{\it Remark:} If we neglect collective effects, the function $\chi(y,y,U(y))$
can be replaced
by $\chi_{\rm bare}(y,y)=\frac{1}{4}\ln\Lambda$ (see Appendix \ref{sec_gfn})
and the kinetic equation (\ref{aham105bg}) reduces to
\begin{eqnarray}
\frac{\partial\omega_a}{\partial
t}=\frac{1}{4}\ln\Lambda\frac{\partial}{\partial y} \sum_b
\frac{1}{|U'(y)|}\left (\gamma_b\omega_b \frac{\partial
\omega_a}{\partial y}-\gamma_a\omega_a\frac{\partial \omega_b}{\partial y}\right
).
\end{eqnarray}
The subsequent equations (\ref{aham94b})-(\ref{aham107b}) can be simplified
similarly.

\subsection{Single-species systems}
\label{sec_ssc}

For a single species gas of point vortices, the Lenard-Balescu equation 
(\ref{aham105})  reduces to
\begin{eqnarray}
\frac{\partial\omega}{\partial t}=\gamma\frac{\partial}{\partial y} \int dy'\,  \chi(y,y',kU(y))   \delta(U(y')-U(y))\left (\omega' \frac{\partial \omega}{\partial y}-\omega\frac{\partial \omega'}{\partial y'}\right ).
\label{ham106}
\end{eqnarray}
Since the velocity profile of a single species gas of point vortices is
monotonic (see footnote 15) we find, using identity
(\ref{gei8}), that
 \begin{eqnarray}
\frac{\partial\omega}{\partial t}=\gamma\frac{\partial}{\partial y} \int dy'\, \chi(y,y',kU(y))   \frac{1}{|U'(y)|}\delta(y'-y)\left (\omega' \frac{\partial \omega}{\partial y}-\omega\frac{\partial \omega'}{\partial y'}\right )=0.
\label{ham107}
\end{eqnarray}
We recall that the Lenard-Balescu equation (\ref{ham106})  is 
valid at the order $1/N$ so it describes the evolution of the average vorticity
on a timescale $N t_D$ under the effect of two-body correlations. Equation
(\ref{ham107}) shows that the vorticity does not change on this timescale (the
current vanishes at the order $1/N$). This is a situation of kinetic
blocking due to the absence of resonance at the order $1/N$.\footnote{For an
axisymmetric distribution of point vortices, the
angular velocity  profile is not necessarily monotonic, even in the
single-species case. As long as the angular
velocity profile is nonmonotonic there are resonances leading to a
nonzero current ($J\neq
0$). However, the relaxation stops ($J=0$) when the profile of angular velocity
becomes monotonic even if the system has not reached the Boltzmann distribution
of statistical equilibrium. This ``kinetic blocking'' for axisymmetric flows is
illustrated in \cite{cl}. } The vorticity may evolve on a longer timescale due
to higher order correlations  between the point vortices. For example,
three-body correlations are of order $1/N^2$ and induce an evolution of the
vorticity on a timescale $N^2\, t_D$.

{\it Remark:} The same situation of 
kinetic blocking at the order $1/N$ occurs for 1D homogeneous systems of
material particles with long-range interactions such as 1D plasmas and the HMF
model above the critical energy \cite{epjp,kindetail}. In that case, an explicit
kinetic equation has been derived at the order
$1/N^2$ \cite{fbcn2,fcpn2}. This equation does not display kinetic blocking
and relaxes towards the Boltzmann distribution. In that case, the relaxation
time scales as $N^2 t_D$. The same results are expected to hold for
unidirectional or axisymmetric distributions of 2D point vortices.
By contrast, for more general flows that are neither unidirectional
nor axisymmetric, the kinetic equation of 2D point vortices  is given by Eq.
(128) or Eq. (137) of \cite{pre} (see also Eq. (54) in \cite{bbgky} and Eq.
(16) in \cite{cl}) and the collision
term does not necessarily vanish.\footnote{This is
because there are potentially more resonances at the order $1/N$ for
complicated flows than for unidirectional and axisymmetric flows. Similarly,
resonances appear for inhomogeneous 1D systems with long-range interactions
that are not present for homogeneous 1D systems \cite{angleactionini}.}  In that
case, the system
may relax towards the
Boltzmann distribution on a timescale $Nt_D$. The same situation holds
for 1D inhomogeneous systems of material particles with long-range interactions
such as 1D self-gravitating systems and for homogeneous or inhomogeneous systems
of material particles with long-range interactions in $d\ge 2$ which  relax
towards the Boltzmann distribution on a timescale $Nt_D$ \cite{epjp,kindetail}.

\subsection{Out-of-equilibrium bath}
\label{sec_ooeb}

The kinetic equation governing the evolution of a test vortex of circulation
$\gamma$ in a 
bath of field vortices with circulations $\lbrace\gamma_b\rbrace$ and
vorticities $\lbrace\omega_b\rbrace$ is given by
Eq. (\ref{tb1}). If the velocity profile is monotonic, using identity
(\ref{gei8}), this Fokker-Planck equation becomes
\begin{eqnarray}
\frac{\partial P}{\partial t}=\frac{\partial}{\partial y}\sum_b \frac{\chi(y,y,U(y))}{|U'(y)|}\left (\gamma_b\omega_b \frac{\partial P}{\partial y}-\gamma P \frac{d\omega_b}{d y}\right ).
\label{tb1b}
\end{eqnarray}
It can be written as Eq. (\ref{tb2}) with
\begin{eqnarray}
D=\sum_b \frac{\chi(y,y,U(y))}{|U'(y)|}\gamma_b\omega_b=\frac{\chi(y,y,U(y))}{|U'(y)|}\omega_2,
\label{tb3b}
\end{eqnarray}
and
\begin{eqnarray}
V_{\rm pol}=\gamma\sum_b \frac{\chi(y,y,U(y))}{|U'(y)|} \frac{d\omega_b}{dy}=\gamma \frac{\chi(y,y,U(y))}{|U'(y)|} \frac{d\omega}{dy}.
\label{tb4b}
\end{eqnarray}
If we neglect collective effects, the function $\chi(y,y,U(y))$ can be replaced
by $\chi_{\rm bare}(y,y)=\frac{1}{4}\ln\Lambda$ (see Appendix
\ref{sec_gfn}). As explained in Sec. \ref{sec_tbaq}, this approach is
self-consistent in the
multispecies case only if the field vortices are at statistical equilibrium with
the Boltzmann distribution from Eq. (\ref{ham52}) otherwise their distribution
evolves under
the effect of collisions (see Sec.
\ref{sec_mscq}). However, if the
field vortices have the same circulation, their distribution does not change on
a timescale $N\, t_D$ (see Sec. \ref{sec_ssc}). They are in a blocked state. In
that case, the bath approximation is justified (on this timescale) for an
arbitrary vorticity field $\omega_b$, not only for the Boltzmann
distribution. The Fokker-Planck equation (\ref{tb1b}) can then be rewritten as
 \begin{eqnarray}
\frac{\partial P}{\partial t}=\frac{\partial}{\partial y}\left\lbrack D(y)\left (\frac{\partial P}{\partial y}-\frac{\gamma}{\gamma_b}P\frac{d\ln|\omega_b|}{dy}\right )\right\rbrack
\label{ham98}
\end{eqnarray}
with
\begin{eqnarray}
D=\frac{\chi(y,y,U(y))}{|U'(y)|}\gamma_b\omega_b=\chi(y,y,U(y))|\gamma_b|.
\label{tb3c}
\end{eqnarray}
This equation can also be directly obtained from the Fokker-Planck equation
(\ref{tb2}) by using the expression of the drift by polarization given by Eq.
(\ref{ger1h}) and the diffusion coefficient from Eq. (\ref{gei11}). The
distribution of the test vortex relaxes towards the equilibrium distribution
\begin{eqnarray}
P_{\rm eq}(y)=A|\omega_b|^{\gamma/\gamma_b}
\end{eqnarray}
on a relaxation time $t_R^{\rm bath}\sim Nt_D$.

{\it Remark:} In the bath approximation,
substituting the out-of-equilibrium
fluctuation-dissipation theorem (\ref{out4}) into
the kinetic
equation
(\ref{ham102}), we obtain
\begin{eqnarray}
\frac{\partial P}{\partial t}=\frac{1}{2}\frac{\partial}{\partial
y}\int dk\, k^2 P(k,y,kU(y))\left ( \frac{\partial P}{\partial
y}-\frac{\gamma}{\gamma_b} P \frac{d\ln|\omega_b|}{dy}\right ).
\end{eqnarray}
This equation  is equivalent to the Fokker-Planck equation (\ref{ham98}). The
diffusion coefficient  $D$
is given by Eqs. (\ref{gei4}) and (\ref{out3})  which return Eq.
(\ref{tb3c}).

\subsection{Thermal bath}
\label{sec_tb}

In the thermal bath approximation, using Eq. (\ref{ham52}), the Fokker-Planck
equation (\ref{ham98}) reduces to 
\begin{eqnarray}
\frac{\partial P}{\partial t}=\frac{\partial}{\partial y}\left \lbrack D\left
(\frac{\partial P}{\partial y}
+\beta \gamma P  \frac{d\psi}{dy} \right )\right\rbrack,
\label{ham103}
\end{eqnarray}
in agreement with Eq. (\ref{tb6}). When the field vortices have the same
circulation $\gamma_b$, the equilibrium stream function can be
calculated analytically (see Appendix \ref{sec_sol}) yielding
\begin{eqnarray}
 \psi=-\frac{2}{|\beta|\gamma_b}\ln\left\lbrace \cosh\left
(\frac{|\beta|\gamma_b\Gamma_b y}{4}\right ) \right\rbrace\qquad (\beta<0).
\label{ham103b}
\end{eqnarray}
The equilibrium
distribution of the test vortex is given by [see Eq. (\ref{tb7}) with Eq.
(\ref{ham103b})]
\begin{eqnarray}
P_{\rm eq}(y)=\frac{A}{\cosh^{\frac{2\gamma}{\gamma_b}}\left
(\frac{|\beta|\gamma_b\Gamma_b y}{4}\right )},
\end{eqnarray}
where $A$ is a constant determined by the normalization condition
$\int_{-\infty}^{+\infty} P_{\rm eq}(y)\, dy=1$. On the other hand, if we
neglect collective
effects, the diffusion coefficient (\ref{tb3c}) has a  constant value given by
\begin{eqnarray}
D=\chi_{\rm bare}(y,y)|\gamma_b|=\frac{1}{4}|\gamma_b|\ln\Lambda,
\end{eqnarray}
where we have used Eq. (\ref{hamg10}). When the diffusion  coefficient it
constant, it is 
possible to transform the Smoluchowski equation (\ref{ham103}) into a
Schr\"odinger equation with imaginary time \cite{risken,prep1}. Indeed, making
the change
of variables
\begin{eqnarray}
P(y,t)=\Phi(y,t)e^{-\frac{1}{2}\beta\gamma\psi(y)},
\label{q1}
\end{eqnarray}
we obtain after simplification the Schr\"odinger-like equation
\begin{eqnarray}
\frac{\partial\Phi}{\partial t}=D\frac{\partial^2\Phi}{\partial y^2}-V(y)\Phi
\label{q2}
\end{eqnarray}
with the effective potential
\begin{eqnarray}
V(y)=-\frac{1}{2}D\beta\gamma\frac{d^2\psi}{dy^2}+\frac{1}{4}D\beta^2\gamma^2\left (\frac{d\psi}{dy}\right )^2.
\label{q3}
\end{eqnarray}
When the stream function is given by Eq. (\ref{ham103b}) we explicitly obtain
\begin{eqnarray}
V(y)=\frac{1}{16}D\beta^2\gamma^2\Gamma_b^2\left \lbrack 1-\frac{1+\frac{\gamma_b}{\gamma}}{\cosh^2\left (\frac{|\beta|\gamma_b\Gamma_b y}{4}\right )}\right \rbrack.
\label{q4}
\end{eqnarray}
This is a Rosen-Morse \cite{rosenmorse} (or
P\"oschl-Teller
\cite{poschlteller}) potential. Interestingly, the
Smoluchowski equation (\ref{ham103}) with the potential from Eq. (\ref{ham103b})
can be solved analytically \cite{prep1}.

\section{2D Brownian point vortices}
\label{sec_tdbv}

In the previous sections,  we have considered an
isolated system of 2D point vortices 
described by $N$-body Hamiltonian equations (see Appendix \ref{sec_pvmod}).
This is the Kirchhoff \cite{kirchhoff} model. It is associated with the
microcanonical ensemble where the energy $E$ of the system is conserved.  When
$N\rightarrow +\infty$ with $\gamma\sim 1/N$, the collisions between vortices
are negligible and the
evolution of the mean vorticity is described by the 2D Euler-Poisson equations
(\ref{ham2bb}) and (\ref{ham7ep}). These equations generically experience a
process of
violent relaxation towards a quasistationary state (QSS) on a few dynamical
times
$t_D$.\footnote{This metaequilibrium state, which is a stable
steady  state of the 2D Euler-Poisson equations, may be an
unidirectional or axisymmetric flow (as considered in this paper) or have a
more complicated structure.} On a
longer timescale $\sim N t_D$, the evolution of
the mean vorticity is governed by the Lenard-Balescu equation (\ref{ham105}),
which is valid at the order $1/N$. This equation takes into account
the collisions between the vortices. For general flows that are not
unidirectional
or axisymmetric, the kinetic equation of point vortices [see Eq.
(128) or Eq. (137) of \cite{pre}] is expected to relax towards the  Boltzmann
distribution with energy $E$ on a timescale $t_R\sim N\, t_D$. For
unidirectional flows and for axisymmetric flows with a monotonic
profile of angular velocity, the Lenard-Balescu equation (for a single
species system of point
vortices) reduces to
$\partial\omega/\partial
t=0$. This leads to a situation of kinetic blocking \cite{cl}. In that case,
the relaxation towards the
Boltzmann distribution
(\ref{ham52}) with energy $E$ is described  by a kinetic equation valid at the
order $1/N^2$ but the collision term of this equation is not explicitly known
(see, however, Refs. \cite{fbcn2,fcpn2} in the case of material particles with
long-range interactions). It is expected to relax towards the Boltzmann
distribution on a timescale $t_R\sim N^2 t_D$.

In Ref. \cite{bv} 
we have formally introduced a system of 2D Brownian
point vortices described by $N$-body stochastic Langevin equations\footnote{We
have similarly introduced a system of self-gravitating
Brownian particles in Ref. \cite{crs} and a system of Brownian particles with a
cosine interaction, called the Brownian mean field (BMF) model, in Ref.
\cite{bmf}.} 
\begin{equation}
\frac{d{\bf r}_i}{dt}={\bf z}\times \sum_j \frac{\gamma_j}{2\pi} \frac{{\bf
r}_i-{\bf
r}_j}{|{\bf r}_i-{\bf r}_j|^2}+\beta D\gamma_i
\sum_j \frac{\gamma_j}{2\pi} \frac{{\bf
r}_i-{\bf
r}_j}{|{\bf r}_i-{\bf r}_j|^2} +\sqrt{2D}{\bf R}_i(t),
\end{equation}
where $i=1,...,N$ label the point vortices and ${\bf R}_i(t)$ is a Gaussian
white noise satisfying $\langle {\bf R}_i(t)\rangle ={\bf 0}$ and $\langle
{R}^{\alpha}_i(t){R}^{\beta}_j(t')\rangle=\delta_{ij}\delta_{\alpha\beta}
\delta(t-t')$. As compared to the Kirchhoff model (first term),
this Brownian model includes a drift velocity (second term) and a random
velocity (third term). This corresponds to the canonical ensemble where the
temperature $T$ is
fixed. This Brownian model could describe the motion of
quantized 2D point vortices in Bose-Einstein condensates and superfluids,
where the fluctuations and the dissipation (drift) are caused by
impurities.
When
$N\rightarrow +\infty$  with $\gamma\sim 1/N$,  the collisions between
the vortices
are negligible and the
evolution of the mean
vorticity is described by a mean field Fokker-Planck
equation which has the form of a system of 2D Euler-Smoluchowski-Poisson
equations
\begin{eqnarray}
\frac{\partial\omega}{\partial t}+{\bf u}\cdot \nabla\omega=\nabla\cdot
\left\lbrack D(\nabla\omega+\beta\gamma\omega\nabla\psi)\right\rbrack,
\label{bv1}
\end{eqnarray}
\begin{eqnarray}
\omega=-\Delta\psi.
\label{bv2}
\end{eqnarray}
These equations relax towards the Boltzmann distribution (\ref{ham52}) with
temperature $T$ on a diffusive (Brownian) timescale $t_{B}\sim L^2/D$, where
$L$ is the system size. When $D\rightarrow 0$, these equations first
experience  a process of violent relaxation towards a QSS  on a few dynamical
times
$t_D$ before slowly relaxing towards
the  Boltzmann distribution (\ref{ham52}), as discussed in \cite{bco} in the
context of the BMF model.\footnote{When $N$ is finite and  $N^2\ll 1/D$, the
system  may first achieve a QSS
on a timescale $t_D$, followed by a microcanonical equilibrium state at
energy $E$ on a timescale $t_R\sim N\, t_D$ or $t_R\sim N^2\, t_D$ (depending
on the structure of the flow), itself followed by a canonical
equilibrium state at temperature
$T$ on a timescale $t_B\sim L^2/D$ (see Ref. \cite{bco}). When the fluctuations
are
taken into account, the phenomenology is even richer as discussed at the end of
this section and in Sec. \ref{sec_spv}.}

The 2D Euler-Smoluchowski-Poisson equations (\ref{bv1}) and (\ref{bv2})
associated
with the canonical ensemble are structurally
very different from the Lenard-Balescu equation (\ref{ham105}) associated
with
the microcanonical ensemble. They are also
very
different from the Smoluchowski equation (\ref{tb6}) describing the
evolution 
of the mean vorticity of a system of noninteracting test vortices in a
thermal bath of field vortices at statistical equilibrium.  In that case,
the stream function $\psi(y)$ is determined by the equilibrium distribution of
the field vortices whereas in Eqs. (\ref{bv1}) and (\ref{bv2}) the stream
function $\psi({\bf r},t)$  is self-consistently produced by the distribution
of the
point vortices itself. Furthermore, the 2D Euler-Smoluchowski-Poisson
equations (\ref{bv1}) and (\ref{bv2}) are valid when $N\rightarrow
+\infty$ while the Lenard-Balescu equation (\ref{ham105}) and the
Smoluchowski
equation (\ref{tb6}) are valid at the
order $1/N$.

The
2D Boltzmann-Poisson equations
may admit several equilibrium states.\footnote{This is the case in a shear layer
where the statistical equilibrium state may be either unidirectional (jet) or
have the form of a large scale vortex \cite{staquet}. This is also the case
when the point vortices have positive and negative circulations. In that case,
the statistical equilibrium states have the form of monopoles, dipoles or even
tripoles \cite{jfm1,jfm2}. In the present section, we
write the equations for a single species gas of point vortices, but these
equations can be
straightforwardly generalized to a multi-species gas (see e.g.
\cite{bv,lastepjb} and Appendix \ref{sec_sim}).} The 2D
Euler-Smoluchowski-Poisson equations
describe
the evolution of the mean vorticity towards one of these equilibrium
states which is stable in the canonical ensemble. Following
\cite{paper5,bv}, if we
consider  the evolution of 2D Brownian point vortices on a mesoscopic scale, we
have to add a stochastic term in the kinetic equation. This noise term arises
from finite $N$
effects and takes fluctuations into account (note that we are still neglecting
the collisions between the vortices). This leads to the stochastic 2D
Euler-Smoluchowski-Poisson
equations \cite{bv}\footnote{The evolution of the
discrete (exact) vorticity field
$\omega_d({\bf r},t)=\sum_{i=1}^{N}\gamma_i \delta({\bf r}-{\bf r}_i(t))$ is
also determined  by stochastic 2D
Euler-Smoluchowski-Poisson
equations of the form of Eqs. (\ref{bv3}) and (\ref{bv4}) with $\omega$ and
$\psi$  replaced by $\omega_d$ and $\psi_d$ \cite{paper5,bv}. Taking the
ensemble average of
these equations and making a mean field approximation ($N\rightarrow +\infty$
with $\gamma\sim 1/N$),
we obtain Eqs. (\ref{bv1}) and (\ref{bv2}). Keeping track of the fluctuations at
a
mesoscopic scale, we obtain  Eqs. (\ref{bv3}) and (\ref{bv4}).}
\begin{eqnarray}
\frac{\partial\omega}{\partial t}+{\bf u}\cdot \nabla\omega=\nabla\cdot
\left\lbrack
D(\nabla\omega+\beta\gamma\omega\nabla\psi)\right\rbrack+\nabla\cdot
\left \lbrack \sqrt{2D\gamma\omega}\, {\bf R}({\bf r},t)\right \rbrack,
\label{bv3}
\end{eqnarray}
\begin{eqnarray}
\omega=-\Delta\psi,
\label{bv4}
\end{eqnarray}
where ${\bf R}({\bf r},t)$ is a Gaussian white noise such that $\langle {\bf
R}({\bf r},t)\rangle={\bf 0}$ and  $\langle R_i({\bf r},t)R_j({\bf
r}',t)\rangle=\delta_{ij}\delta({\bf r}-{\bf r}')\delta(t-t')$. When the 2D 
 Boltzmann-Poisson equations admit different
equilibrium states,  Eqs. (\ref{bv3}) and
(\ref{bv4}) can be used to study random
transitions from one state to the other (see Ref. \cite{random} in a similar
context). The probability of transition is given
by the Kramers formula, which can be established from the 
instanton theory associated with the Onsager-Machlup functional
\cite{random,gsse}.

The equation for the velocity field corresponding to Eqs.
(\ref{bv3}) and  (\ref{bv4}) reads \cite{bv}
\begin{eqnarray}
\frac{\partial {\bf u}}{\partial t}+({\bf u}\cdot \nabla){\bf
u}=-\frac{1}{\rho}\nabla P+D\Delta{\bf u}-\mu\omega{\bf
u}+\sqrt{2D\gamma\omega}\, {\bf z}\times {\bf R}({\bf r},t),
\end{eqnarray}
where $\mu=D\beta\gamma$ is the drift coefficient. Since $\beta<0$ in the most
relevant situations, the
term $-\mu\omega {\bf u}$ represents a nonlinear anti-friction (forcing)
proportional to $\omega{\bf u}$ that opposes itself to the diffusion term
$D\Delta {\bf u}$.

For unidirectional flows (possibly resulting from a process of violent
relaxation), the stochastic Smoluchowski-Poisson equations
(\ref{bv3}) and (\ref{bv4}) become
\begin{eqnarray}
\frac{\partial\omega}{\partial t}=\frac{\partial}{\partial y}\left
(D\frac{\partial \omega}{\partial y}+\mu \omega
\frac{\partial\psi}{\partial y}\right )+\frac{\partial}{\partial y}\left \lbrack
\sqrt{2D\gamma\omega}\, {R}(y,t)\right \rbrack,
\label{burgers1}
\end{eqnarray}
\begin{eqnarray}
\omega=-\frac{\partial^2\psi}{\partial y^2}.
\label{burgers2}
\end{eqnarray}
Using Eq.
(\ref{ham10}), they can be combined into a single equation for the velocity
field
\begin{eqnarray}
\frac{\partial U}{\partial t}=D\frac{\partial^2 U}{\partial y^2}+\mu
U\frac{\partial U}{\partial y}-\sqrt{-2D\gamma U'(y)} R(y,t).
\label{burgers3b}
\end{eqnarray}
If we set $v=-\mu U$, the foregoing equation may be rewritten as
\begin{eqnarray}
\frac{\partial v}{\partial t}+v\frac{\partial v}{\partial y}=D\frac{\partial^2
v}{\partial y^2}+\sqrt{2D\mu\gamma v'(y,t)} R(y,t).
\label{burgers4}
\end{eqnarray}
It can be viewed as a stochastic (noisy) viscous Burgers equation for
a pseudo ``velocity'' field $v(y,t)$ with a viscosity $D$. If we neglect the
noise,
it
reduces to the  viscous Burgers equation 
\begin{eqnarray}
\frac{\partial v}{\partial t}+v\frac{\partial v}{\partial y}=D\frac{\partial^2
v}{\partial y^2}.
\label{burgers5}
\end{eqnarray}
This analogy allows us to solve the 1D Smoluchowski-Poisson
equation analytically by using the Cole-Hopf transformation
\cite{prep1}. We note that the stationary solution of Eq. (\ref{burgers5}) is
[see Eqs. (\ref{bpe6b})]
\begin{eqnarray}
v=\frac{1}{2}\mu\Gamma \tanh\left (\frac{|\beta|\gamma\Gamma y}{4}\right )
\label{burgers5stat}
\end{eqnarray}

{\it Remark:} Equations (\ref{bv3}) and  (\ref{bv4}) or Eqs. (\ref{burgers1})
and 
(\ref{burgers2}) take into account finite $N$ effects ($\gamma\sim 1/N$) which
are responsible for the noise term (fluctuations) but they ignore the collisions
between the
point
vortices that would lead to the Lenard-Balescu collision term, as well as the
correlations
arising from the noise term. The correlations arising from the noise term induce
an additional nonlinear diffusion of the mean vorticity (on a timescale $Nt_D$)
which is discussed in Sec. \ref{sec_spv}.

\section{Secular dressed diffusion equation}
\label{sec_sdd}

In this section, we consider 
the case where a continuous incompressible 2D
flow\footnote{This could be an intrinsically continuous flow
 (see the Remark at the end of Sec.
\ref{sec_inhos}) or a gas of point
vortices in the mean field limit $N\rightarrow +\infty$ with $\gamma\sim 1/N$
where the collisions between the point vortices are negligible.} is
submitted to an external
stochastic velocity field ${\bf u}_{e}$ [see Eqs. (\ref{ham1})-(\ref{ham2b})].
We consider a
rather general situation where the external forcing is not
necessarily induced by point vortices. We
show that, under the effect of the
external forcing, the evolution of the mean flow satisfies a SDD
equation. We derive the SDD equation from the Klimontovich equation
and from the Fokker-Planck equation and analyze its main properties.

\subsection{From the Klimontovich equation}
\label{sec_sddk}

Under the
assumptions of Sec. \ref{sec_inhos},  the basic equations governing the
evolution
of the mean vorticity $\omega(y,t)$ of a unidirectional
flow\footnote{This configuration, which is a steady
state of
the 2D Euler equation, may be imposed initially or result from a process
of violent relaxation.} forced
by an external 
perturbation are given by Eqs. (\ref{ham12}) and
(\ref{ham13}). Introducing the Fourier transforms of the fluctuations of
vorticity and stream
function, these equations can be rewritten as  
\begin{eqnarray}
\frac{\partial\omega}{\partial t}=\frac{\partial}{\partial y}\int dk\int\frac{d\sigma}{2\pi}\int dk'\int\frac{d\sigma'}{2\pi}(ik')e^{i(kx-\sigma t)}e^{i(k'x-\sigma' t)}\langle \delta{\hat \omega}(k,y,\sigma)\delta{\hat \psi}_{\rm tot}(k',y,\sigma')
\rangle,
\label{ham16}
\end{eqnarray}
\begin{eqnarray}
\delta{\hat\omega}(k,y,\sigma)=\frac{k\frac{\partial\omega}{\partial
y}}{kU(y)-\sigma}
\delta{\hat\psi}_{\rm tot}(k,y,\sigma),
\label{ham17a}
\end{eqnarray}
and they can be combined into
\begin{eqnarray}
\frac{\partial\omega}{\partial t}=\frac{\partial}{\partial y}\int dk\int\frac{d\sigma}{2\pi}\int dk'\int\frac{d\sigma'}{2\pi}(ik')e^{i(kx-\sigma t)}e^{i(k'x-\sigma' t)} \frac{k\frac{\partial\omega}{\partial y}}{kU(y)-\sigma} \langle \delta{\hat\psi}_{\rm tot}(k,y,\sigma)\delta{\hat \psi}_{\rm tot}(k',y,\sigma')
\rangle.
\label{ham110}
\end{eqnarray}
Introducing the
power spectrum of the fluctuations from Eq. (\ref{ham25}) we obtain
\begin{eqnarray}
\frac{\partial\omega}{\partial t}=-i\frac{\partial}{\partial y}\int dk\int\frac{d\sigma}{2\pi}k\frac{k\frac{\partial\omega}{\partial y}}{kU(y)-\sigma} P(k,y,\sigma).
\label{ham112}
\end{eqnarray}
Recalling the Landau prescription $\sigma\rightarrow
\sigma+i0^+$ and using the Sokhotski-Plemelj
formula (\ref{plemelj}), we can replace  $1/(kU(y)-\sigma-i0^+)$
by $+i\pi\delta(kU(y)-\sigma)$.  Accordingly,
\begin{eqnarray}
\frac{\partial\omega}{\partial t}=\pi \frac{\partial}{\partial y}\int dk\int\frac{d\sigma}{2\pi}k^2\delta(kU(y)-\sigma) P(k,y,\sigma)\frac{\partial\omega}{\partial y}.
\label{ham114}
\end{eqnarray}
Integrating over the $\delta$-function (resonance), we get
\begin{eqnarray}
\frac{\partial\omega}{\partial t}
=\frac{1}{2}\frac{\partial}{\partial y}\int dk\, k^2 P(k,y,k U(y))\frac{\partial\omega}{\partial y}.
\label{ham114b}
\end{eqnarray}
Therefore, the secular evolution of
the mean vorticity $\omega(y,t)$ sourced by the
external stochastic perturbation is governed by a nonlinear diffusion equation
of the form
\begin{eqnarray}
\frac{\partial\omega}{\partial t}=\frac{\partial}{\partial y}\left
(D[y,\omega]\frac{\partial \omega}{\partial y}\right )
\label{ham116}
\end{eqnarray}
with a diffusion coefficient
\begin{eqnarray}
D[y,\omega]=\frac{1}{2} \int dk \, k^2 P(k,y,kU(y)).
\label{ham115}
\end{eqnarray}
Using Eq.
(\ref{ham28}), we can express the diffusion coefficient in terms of the
correlation function of the external vorticity field as
\begin{eqnarray}
D[y,\omega]=\frac{1}{2} \int dy'\int dk \, k^2 |G(k,y,y',kU(y))|^2
{\hat C}(k,y',kU(y)).
 \label{ham118}
\end{eqnarray}
The diffusion coefficient depends on the correlation function of the
external perturbation ${\hat C}(k,y',\sigma)$ and on the response
function $G(k,y,y',\sigma)$ of the flow  evaluated at
the resonance
frequencies $\sigma=kU(y)$.  As a
result, the
diffusion coefficient $D[y,\omega]$ depends on the position $y$
and on the vorticity $\omega(y,t)$ itself through the response function
$G(k,y,y',kU(y))$ defined by Eq. (\ref{ham21}). It also depends
implicitly on $\omega(y,t)$ through $U(y,t)$. Equation (\ref{ham116}) with the
diffusion coefficient from Eq. (\ref{ham118}) is
therefore
a complicated integrodifferential equation called the SDD
equation. This is the counterpart of
the SDD equation derived in Ref. \cite{epjp} for homogeneous systems of
material particles with
long-range
interactions forced by an external stochastic perturbation.\footnote{A similar,
but different, equation is derived in Refs. \cite{nardini,nardini2}. See
Ref. \cite{epjp2} for a
comparison between these two approaches.}
When collective
effects
are neglected, i.e., when we replace $G(k,y,y',kU(y))$ by
$G_{\rm bare}(k,y,y')$ in Eqs. (\ref{ham118}), we obtain
\begin{eqnarray}
D_{\rm bare}[y,\omega]=\frac{1}{2} \int dy'\int dk \, k^2 G_{\rm bare}(k,y,y')^2
{\hat C}(k,y',kU(y)).
 \label{ham118bh}
\end{eqnarray}
In that case, Eq. (\ref{ham116}) is called the SBD
equation.

{\it Remark:} If our system consists in a gas of point
vortices with individual circulation $\gamma\sim 1/N$ and $N\gg 1$ (large but
finite), the SDD
equation describes the diffusive evolution of this near-equilibrium system 
caused by the fluctuations of the stream function
induced by the external
perturbation. The evolution timescale is intermediate between the violent
collisionless relaxation time $\sim t_D$ and the collisional relaxation time
$\sim N t_D$ or $\sim N^2 t_D$.

\subsection{From the Fokker-Planck equation}
\label{sec_sddfp}

The SDD equation (\ref{ham116}) can also be derived from the Fokker-Planck
equation (\ref{fp8}). If our system is a continuous vorticity field or an
ensemble of point vortices with circulation $\gamma\sim 1/N$ in
the collisionless
limit $N\rightarrow +\infty$, the drift by
polarization, which is proportional to $\gamma$ (see Sec. \ref{sec_ifpol}),
vanishes
\begin{equation}
\label{ham119}
{V}_{\rm pol}={0}.
\end{equation}
Indeed, the perturbation on the system caused by the
test particle is
negligible. In that case, the Fokker-Planck equation (\ref{fp8}) reduces to
\begin{equation}
\label{ham121}
\frac{\partial \omega}{\partial t}=\frac{\partial}{\partial y} \left
(D\frac{\partial \omega}{\partial y}\right ).
\end{equation}
The diffusion coefficient can be calculated as in Sec. \ref{sec_gei} 
returning the expression from Eqs. (\ref{ham115})-(\ref{ham118bh}). Therefore,
the Klimontovich approach  and the Fokker-Planck approach coincide.

{\it Remark:} We note that, in Eq. (\ref{ham121}), the diffusion coefficient is ``sandwiched'' between
the two spatial derivatives $\partial/\partial y$ in agreement with Eq.
(\ref{ham116}).
As explained 
in Sec. \ref{sec_fp}, this is not the usual form of the Fokker-Planck equation
which is given by Eq. (\ref{fp5}). Therefore, the
test vortex experiences a drift [see
Eq. (\ref{fp9})] 
\begin{equation}
\label{ham120}
V_{\rm tot}=\frac{\partial D}{\partial y},
\end{equation}
arising from the spatial inhomogeneity of the diffusion
coefficient.\footnote{This
formula is established by a direct calculation in Sec. 4.4 of \cite{klim}.}
Using Eqs.
(\ref{fp6}) and (\ref{fp7}), the relation (\ref{ham120}) can be written as
\begin{equation}
\label{ham120b}
\frac{\langle\Delta y\rangle}{\Delta t}=\frac{1}{2}\frac{\partial}{\partial y}\frac{\langle (\Delta y)^2\rangle}{\Delta t}.
\end{equation}

\subsection{Properties of the SDD equation}

Some general properties of the SDD equation (\ref{ham116}) can be
given. First
of all, the circulation of the system $\Gamma=\int \omega\, d{y}$ is conserved
since
the right hand side of
Eq. (\ref{ham116})
is the divergence of a current. By contrast, the energy and the impulse of the
system are not
conserved, contrary to the case of the Lenard-Balescu equation \cite{cl},
since the system
is forced by an external medium. Taking the time
derivative
of the energy
\begin{equation}
E=\frac{1}{2}\int \omega\psi\, dy,
\label{n30}
\end{equation}
using Eq. (\ref{ham116}),  and integrating by parts, we get
\begin{equation}
\dot E=-\int D[y,\omega]\frac{\partial \omega}{\partial y} \frac{\partial
\psi}{\partial y} \, dy.
\label{n31}
\end{equation}
In general, $\dot E$ has not a definite sign. However, for
$\omega=\omega(\psi)$ we find that
$\dot E=-\int \omega'(\psi) D (\partial\psi/\partial y)^2 \,
dy\ge 0$. When $\omega'(\psi)\le 0$, the energy is injected in the
system and when $\omega'(\psi)\ge 0$ the energy is absorbed from the system. For
the impulse $P=\int \omega y\, dy$, we find that $\dot P=-\int D
\frac{\partial\omega}{\partial y}\, dy$. Finally, introducing
the
$H$-functions
\begin{equation}
S=-\int C(\omega) \, dy,
\label{n32}
\end{equation}
where $C(\omega)$ is any convex function, we get
\begin{equation}
\dot S=\int C''(f) D[y,\omega] \left (\frac{\partial \omega}{\partial y}\right
)^2 \, d{y}.
\label{n33}
\end{equation}
Because of
the convexity condition $C''\ge 0$ and the fact that $D$ is positive (see
Sec.
\ref{sec_sddk}),
we find that $\dot S\ge 0$. Therefore, all the $H$-functions increase
monotonically with time. This is
different from the case of the
Lenard-Balescu equation where only the Boltzmann entropy
increases
monotonically \cite{cl}.

\subsection{Connection between the SDD equation and the multi-species
Lenard-Balescu equation}
\label{sec_conn}

Let us discuss the connection between the SDD
equation (\ref{ham116}) with Eq. (\ref{ham118})   and the
multi-species Lenard-Balescu equation (\ref{ham105}). The Lenard-Balescu 
equation governs the evolution of the vorticity $\omega_a(y,t)$ of
point vortices of species $a$
under the effects of ``collisions'' with point vortices of all species ``$b$''
(including the point vortices of
species $a$) with vorticity $\omega_b({y}',t)$. The dressed Green function is
given by Eq. (\ref{ham21}) where $\omega(y,t)$ denotes the
total vorticity $\sum_b \omega_b(y,t)$ and $U(y,t)$ is the corresponding
velocity field. The set of equations
(\ref{ham105}), in which all
the vorticities $\omega_a({y},t)$ evolve in a self-consistent manner, is
closed.

We now make the following approximations to simplify these
equations. The point vortices of circulation $\gamma_a$ with vorticity
$\omega_a(y,t)$ form our
system. They will
be called the test vortices. The vortices of circulation $\lbrace
\gamma_b\rbrace_{b\neq a}$ with vorticities
$\lbrace \omega_b({y},t)\rbrace_{b\neq a}$ form 
the external -- background -- medium. They will be called the field vortices. We
take into account the
collisions induced by the field vortices on the test vortices but we neglect
the
collisions induced by the test vortices on the field vortices and on
themselves. This approximation is valid for very light test vortices 
$\gamma_a\ll \gamma_b$ or, more precisely, in the limit $N_a\rightarrow +\infty$
with $\gamma_a\sim
1/N_a$ (see Sec. \ref{sec_purediff}). Finally, the vorticities
$\lbrace
\omega_b({y},t)\rbrace_{b\neq a}$ of the field vortices are
either assumed to be fixed
(bath) or evolve according to their own dynamics (i.e. following equations that
we do not write explicitly). Under these
conditions, the
kinetic equation (\ref{ham105}) reduces to
\begin{eqnarray}
\frac{\partial\omega_a}{\partial t}=\frac{1}{2}\frac{\partial}{\partial y} 
\sum_{b\neq a} \int dy'\int dk\, |k| |G(k,y,y',kU(y))|^2  
\delta(U(y')-U(y))\gamma_b\omega'_b \frac{\partial \omega_a}{\partial
y}.
\label{ham105qw}
\end{eqnarray}
This equation can be
interpreted as a nonlinear diffusion equation. The diffusion arises from the
discrete
distribution of the field vortices which creates a fluctuating velocity field
(Poisson shot noise) acting on the test vortices.
For that reason, the diffusion coefficient of the test vortices is proportional
to the
circulations $\lbrace \gamma_b\rbrace$ of the field vortices. The
diffusion coefficient has no contribution from vortices
of species $a$. The
condition $\gamma_a\ll
\gamma_b$ justifies neglecting the fluctuations induced  by the test vortices on
themselves.  On the
other hand, the drift by polarization vanishes (${V}_{\rm
pol}={0}$). Indeed, since the circulation $\gamma_a$ of the  test vortices is
small,
the test vortices do not significantly perturb the vorticity of the
medium, so there
is no drift by polarization (no retroaction).  As a
result, the circulation $\gamma_a\rightarrow 0$ of the vortices  of species $a$
does not
appear in the kinetic equation (\ref{ham105qw}).

Equation (\ref{ham105qw}) can be written as a SDD equation
\begin{equation}
\frac{\partial \omega}{\partial t}=\frac{\partial}{\partial y}\left
(D[y,\omega]\frac{\partial \omega}{\partial y}\right
)
\label{conn3}
\end{equation}
with a
diffusion tensor 
\begin{eqnarray}
D[y,\omega]=\frac{1}{2} 
\sum_{b} \int dy'\int dk\, |k| |G(k,y,y',kU(y))|^2  
\delta(U(y')-U(y))\gamma_b\omega'_b,
\label{ham105ju}
\end{eqnarray}
where we have droped the subscript $a$ for clarity. The expression
(\ref{ham105ju}) of the diffusion coefficient is consistent with Eq.
(\ref{ham118}) where ${\hat C}({k},y,\sigma)$ 
is the
bare correlation function of the vorticity field created by a discret
collection of
field vortices of circulations
$\lbrace \gamma_b\rbrace$ given by Eq. (\ref{ham32}). To exactly recover the SDD
equation (\ref{ham116}) with Eq. (\ref{ham118}), we
have to replace
$\omega({y},t)+\sum_{b} \omega_b({y})$ by  $\omega({y},t)$ in the dressed Green
function. This
assumes that the external medium -- field vortices -- is non polarizable (i.e.
collective effects
can be neglected) while our system -- test vortices -- is
polarizable (i.e.
collective effects must be taken into account).\footnote{In the present case,
Eq.
(\ref{conn3}) with Eq. (\ref{ham105ju})
is more accurate than the SDD equation (\ref{ham116}) with Eq.
(\ref{ham118})
because it takes into account the
polarizability of the external medium.} As we
have already mentioned,
the SDD equation (\ref{conn3}) is a nonlinear
diffusion equation involving a
diffusion coefficient which depends on the distribution function of the system
$\omega({y},t)$ itself. It is
therefore a complicated integrodifferential equation.

\subsection{SDD equation with damping}
\label{sec_sddwd}

Let us add a small linear damping term $-\alpha \omega$ on
the right hand side of the SDD equation (\ref{ham121}) in order to account for a
possible dissipation. This yields
\begin{eqnarray}
\frac{\partial\omega}{\partial t}=\frac{\partial}{\partial y}\left
(D[y,\omega]\frac{\partial \omega}{\partial y}\right )-\alpha\omega.
\label{fham116}
\end{eqnarray}
The general behavior of this nonlinear equation is difficult to
predict because it depends on the correlation function of the external
potential. Furthermore, since the diffusion tensor is a
functional of $\omega$, the SDD equation presents a rich and complex behavior.
It may relax towards a non-Boltzmannian steady state determined by
\begin{eqnarray}
\frac{\partial}{\partial y}\left
(D[y,\omega]\frac{\partial \omega}{\partial y}\right )-\alpha\omega=0,
\label{fham116eq}
\end{eqnarray}
or exhibit a
complicated (e.g. periodic) dynamics. Using Eq.
(\ref{ham10}), we can rewrite Eq. (\ref{fham116}) in terms of the
velocity field $U(y,t)$ as
\begin{eqnarray}
\frac{\partial U}{\partial t}=D[y,U]\frac{\partial^2 U}{\partial
y^2}-\alpha U.
\label{ham122}
\end{eqnarray}
The stationary solution of this equation is a jet profile
$U(y)$ determined by
\begin{eqnarray}
\frac{d^2 U}{dy^2}=\alpha \frac{U}{D[y,U]}.
\label{ham123}
\end{eqnarray}
Since $D$ generically depends on $U$ (the diffusion coefficient
is a functional of $U$), this equation is a very nonlinear equation that has
nontrivial solutions. If we
assume that the external noise is due to $N$ point
vortices and if the velocity field is monotonic we have (see
Sec. \ref{sec_geib}) 
\begin{eqnarray}
D=\sum_b \gamma_b  \frac{\chi(y,y,U(y)) }{|U'(y)|} \omega_b(y),
\label{gei11sdd}
\end{eqnarray}
where we have assumed that $U(y,t)$ is produced by $\omega(y,t)$ only. If
we neglect collective effects and assume that $\omega_b(y)$ is uniform, we find
that $D\sim 1/|U'(y)|$. On the other hand, when $D$ is constant,\footnote{This
is the case, for example, when the external perturbation is created by a random
distribution of point
vortices with equal circulation $\gamma_b$, when collective effects are
neglected, and when $U(y,t)$ in the diffusion coefficient is produced by
$\omega_b(y,t)$ only (see Sec. \ref{sec_dco}).} Eqs. (\ref{ham122}) and
(\ref{ham123}) reduce to
\begin{eqnarray}
\frac{\partial U}{\partial t}=D\frac{\partial^2 U}{\partial y^2}-\alpha U
\label{ham124b}
\end{eqnarray}
and
\begin{eqnarray}
D\frac{\partial^2 U}{\partial y^2}-\alpha U=0.
\label{ham124}
\end{eqnarray}
In that case, Eq. (\ref{ham124}) leads to an exponential
jet:
$U(y)=e^{-|y|}$. 

{\it Remark:} These results are similar to the
results obtained for zonal jets 
in the
context of forced 2D turbulence
\cite{ckl,bnt,lbfkl,kl1,kl2,kl3,flf,wb1,frishman,wb2,kk,kl4}. This connection is
not
unexpected since these approaches are based on the quasilinear theory
of the 2D Euler equation (see an early work in
Ref. \cite{qlprl} in the
context of the theory of violent relaxation). The fact that the
diffusion coefficient is generically inversely proportional to the local
shear, $D\sim 1/|\Sigma|$, with
$\Sigma=-U'(y)$ for unidirectional flows and  $\Sigma=r(d/dr)(U_{\theta}/r)$ 
for axisymmetric flows, was pointed out in \cite{preR,pre}. Therefore, the
diffusion is generically reduced as the shear increases.
However, the forcing that we consider in this section is
different from the forcing considered in
Refs. \cite{ckl,bnt,lbfkl,kl1,kl2,kl3,flf,wb1,frishman,wb2,kk,kl4}. Therefore,
the
approaches and the results are substantially distinct and complementary to each
other.

\subsection{SDD equation with drift}

By analogy with the Smoluchowski equation (\ref{tb6}) or (\ref{bv1}), we can
heuristically add
a drift term on
the right hand side of the SDD equation
(\ref{ham121}).
This leads to an equation of the form
\begin{eqnarray}
\frac{\partial\omega}{\partial t}=\frac{\partial}{\partial y}\left
(D[y,\omega]\frac{\partial \omega}{\partial y}+\mu \omega
\frac{\partial\psi}{\partial y}\right ),
\label{sddxi1}
\end{eqnarray}
where $\psi$ is either a given external stream function or
the self-consistent stream function produced by the vorticity field
$\omega$.  In this latter
case, using Eq. (\ref{ham10}), we can rewrite Eq. (\ref{sddxi1}) in terms of the
velocity field $U(y,t)$ alone as
\begin{eqnarray}
\frac{\partial U}{\partial t}=D[y,U]\frac{\partial^2 U}{\partial y^2}+\mu
U\frac{\partial U}{\partial y}.
\label{sddxi2}
\end{eqnarray}
If we set $v=-\mu U$, we get
\begin{eqnarray}
\frac{\partial v}{\partial t}+v\frac{\partial v}{\partial
y}=D[y,v]\frac{\partial^2
v}{\partial y^2}.
\label{sddxi3}
\end{eqnarray}
This equation is similar to a viscous Burgers equation with a viscosity $D[y,v]$
which is a functional of
the pseudo ``velocity'' $v(y,t)$.\footnote{When $\psi(y)$ is a given external
stream
function, $v\partial_yv$ is replaced by $v_g\partial_yv$ with $v_g=-\mu
d\psi/dy$.} The stationary
solutions of Eqs. (\ref{sddxi1}) and (\ref{sddxi2}) satisfy
\begin{eqnarray}
\frac{\partial}{\partial y}\left
(D[y,\omega]\frac{\partial \omega}{\partial y}+\mu \omega
\frac{\partial\psi}{\partial y}\right )=0,\qquad D[y,U]\frac{\partial^2
U}{\partial y^2}+\mu
U\frac{\partial U}{\partial y}=0.
\label{sddxi1st}
\end{eqnarray}

\subsection{Stochastic SDD equation}

The SDD equation (\ref{sddxi1}), which is a deterministic partial
differential
equation, describes the evolution of the mean vorticity $\omega(y,t)$. If
we take fluctuations into account, by
analogy with the results presented in \cite{bv}, we expect that the
mesoscopic vorticity $\overline{\omega}(y,t)$ will
satisfy a stochastic partial differential equation of the
form
\begin{eqnarray}
\frac{\partial\overline{\omega}}{\partial t}=\frac{\partial}{\partial y}\left
(D[y,\overline{\omega}]\frac{\partial \overline{\omega}}{\partial y}+\mu
\overline{\omega}
\frac{\partial\overline{\psi}}{\partial y}\right )+\frac{\partial
\zeta}{\partial y}(y,t),
\label{sddxi1s}
\end{eqnarray}
where $\zeta(y,t)$ is a noise term with zero mean that
generally depends
on
$\overline{\omega}(y,t)$.  When $D$ is
constant and when the
fluctuation-dissipation theorem is fulfilled so that $\mu=D\beta
\gamma$, as in the case of 2D Brownian vortices (see Sec. \ref{sec_tdbv}), the
noise term is given by \cite{bv} 
\begin{equation}
\zeta(y,t)=\sqrt{2D\gamma \overline{\omega}}\,
{Q}(y,t),
\end{equation}
where ${Q}(y,t)$ is a Gaussian white noise satisfying $\langle
{Q}(y,t)\rangle =0$ and $\langle
{Q}(y,t){Q}(y',t')\rangle=\delta({y}-{y}')\delta(t-t')$.
This
expression can be obtained from an adaptation of the theory of
fluctuating hydrodynamics \cite{paper5}. When $D[\overline{\omega}]$ is a
functional of
$\overline{\omega}$, the noise term may be more complicated.\footnote{A general
approach to obtain the noise term and the corresponding action is to use
the theory of large deviations \cite{bouchetld}.}  When the deterministic
equation (\ref{sddxi1}) admits several equilibrium states, the noise
term in Eq.
(\ref{sddxi1s}) can trigger random transitions from one state to the other (see,
e.g.,
\cite{random,gsse,nardini2,bouchetsimmonet,rbs,brs} in
different contexts).

{\it Remark:} Similarly, we can introduce a stochastic term (noise) in Eq.
(\ref{fham116}) to describe the evolution of the system on a mesoscopic scale.
This leads to the stochastic SDD equation
\begin{eqnarray}
\frac{\partial\overline{\omega}}{\partial t}=\frac{\partial}{\partial y}\left
(D[y,\overline{\omega}]\frac{\partial \overline{\omega}}{\partial y}\right
)-\alpha\overline{\omega}+\frac{\partial
\zeta}{\partial y}(y,t).
\label{fham116a}
\end{eqnarray}
Using Eq. (\ref{ham10}), it can be written in terms of the velocity field as 
\begin{eqnarray}
\frac{\partial \overline{U}}{\partial
t}=D[y,\overline{U}]\frac{\partial^2\overline{U}}{\partial
y^2}-\alpha \overline{U}-\zeta(y,t).
\label{fham116b}
\end{eqnarray}
The comments made previously also apply to these stochastic partial
differential equations.

\section{Stochastic damped 2D Euler equation}
\label{sec_fd}

In this section, we consider the stochastic damped
2D Euler equation
\begin{equation}
\frac{\partial\omega_c}{\partial t}+{\bf u}_c\cdot
\nabla\omega_c=-\alpha\omega_c+\sqrt{2\gamma\alpha \omega_c}\, Q({\bf r},t),
\label{fh1}
\end{equation}
where $\omega_c({\bf r},t)$ is a continuous vorticity field, ${Q}({\bf r},t)$
is a Gaussian white noise
satisfying $\langle {Q}({\bf
r},t)\rangle =0$ and $\langle {Q}({\bf r},t){Q}({\bf r}',t')\rangle=\delta({\bf
r}-{\bf r}')\delta(t-t')$, $\alpha$ is a small damping coefficient, and
$\gamma$ has the dimension of a circulation (we will
ultimately take the limit $\alpha\rightarrow 0$ and $\gamma\rightarrow
0$). We introduce this equation in
an {\it ad hoc} manner but we will show below that the stochastic term generates
a power spectrum that coincides with the power spectrum produced by an isolated
distribution of point vortices of circulation $\gamma\sim 1/N\ll 1$. This
approach therefore provides another manner to determine
the power
spectrum of a gas of point vortices. This is an additional
motivation
to study Eq. (\ref{fh1}) in detail. The calculations of this section are
inspired
by similar
calculations on
fluctuating hydrodynamics performed in \cite{paper5}.

When $\alpha\rightarrow 0$, the mean vorticity $\omega({\bf
r},t)=\langle \omega_c({\bf r},t)\rangle$ rapidly reaches a QSS
which is a steady state of the 2D Euler-Poisson equations. This process of
violent relaxation takes place on a few dynamical times. If the evolution is
ergodic, the QSS can be determined
by the MRS statistical 
theory \cite{miller,rs}. The kinetic theory of violent relaxation
is discussed in \cite{qlprl,qlnew}. On a longer timescale, the mean vorticity
evolves through
a sequence of QSSs sourced by the noise.  Adapting the procedure of
Sec. \ref{sec_inhos} to the present context, and assuming that the mean flow
(QSS) is unidirectional,
we obtain the quasilinear equations
\begin{equation}
\frac{\partial\omega}{\partial t}=-\alpha\omega+\frac{\partial}{\partial
y}\left\langle \delta\omega \frac{\partial\delta\psi}{\partial x}\right\rangle,
\label{fh2}
\end{equation}
 \begin{equation}
\frac{\partial \delta\omega}{\partial t}+U\frac{\partial \delta\omega}{\partial
x}-\frac{\partial\delta\psi}{\partial x} \frac{\partial \omega}{\partial
y}=-\alpha \delta\omega+\sqrt{2\gamma\alpha \omega}\, {Q}(x,y,t),
\label{fh3}
\end{equation} 
where ${Q}(x,y,t)$ is a Gaussian white noise satisfying $\langle
{Q}(x,y,t)\rangle =0$ and $\langle
{Q}(x,y,t){Q}(x',y',t')\rangle=\delta(x-x')\delta(y-y')\delta(t-t')$.
Introducing the Fourier transforms of the
fluctuations of vorticity and stream function, we can rewrite these
equations as
\begin{eqnarray}
\frac{\partial\omega}{\partial t}=-\alpha\omega+\frac{\partial}{\partial y}\int
dk\int\frac{d\sigma}{2\pi}\int dk'\int\frac{d\sigma'}{2\pi}(ik')e^{i(kx-\sigma
t)}e^{i(k'x-\sigma' t)}\langle \delta{\hat \omega}(k,y,\sigma)\delta{\hat
\psi}(k',y,\sigma')
\rangle,
\label{fh4}
\end{eqnarray}
\begin{equation}
\delta{\hat\omega}(k,y,\sigma)=\frac{k\frac{\partial\omega}{\partial
y}}{kU(y)-\sigma-i\alpha}\delta{\hat\psi}(k,y,\sigma)-\frac{i\sqrt{
2\gamma\alpha\omega(y)}}{kU(y)-\sigma-i\alpha}{\hat Q}(k,y,\sigma),
\label{fh4b}
\end{equation}
where ${\hat Q}(k,y,\sigma)$ is a Gaussian white noise satisfying $\langle {\hat
Q}(k,y,\sigma)\rangle=0$ and $\langle {\hat Q}(k,y,\sigma){\hat
Q}({k}',{y}',\sigma')\rangle=\delta({k}+{k}')\delta({y}-{y}
')\delta(\sigma+\sigma')$. Substituting Eq.
(\ref{fh4b}) into Eq. (\ref{fh4}), we get
\begin{eqnarray}
\frac{\partial\omega}{\partial t}=-\alpha\omega+\frac{\partial}{\partial y}\int
dk\int\frac{d\sigma}{2\pi}\int dk'\int\frac{d\sigma'}{2\pi}(ik')e^{i(kx-\sigma
t)}e^{i(k'x-\sigma' t)} \frac{k\frac{\partial\omega}{\partial
y}}{kU(y)-\sigma-i\alpha} \langle \delta{\hat\psi}(k,y,\sigma)\delta{\hat
\psi}(k',y,\sigma')
\rangle.
\label{ham110w}
\end{eqnarray}
In writing Eq. (\ref{ham110w}) we have only considered the contribution of the
term $\langle \delta{\hat \psi}(k,y,\sigma)\delta{\hat
\psi}(k',y,\sigma') \rangle$ which leads to a diffusive evolution. The
other terms will be investigated elsewhere \cite{prep1}.

Let us study Eq. (\ref{fh4b}) for the fluctuations and determine the
power
spectrum $P(k,y,\sigma)$ of the fluctuating stream function defined by
\begin{eqnarray}
\langle \delta{\hat \psi} (k,y,\sigma)\delta{\hat \psi}
(k',y,\sigma') \rangle=2\pi\delta(k+k')\delta(\sigma+\sigma')P(k,y,\sigma).
\label{ham25w}
\end{eqnarray}
The fluctuations of vorticity and stream function are related to each other by
the Poisson equation
\begin{equation}
\Delta\delta\psi=-\delta\omega.
\label{fh5}
\end{equation}
Writing this equation in Fourier space, we get
\begin{equation}
\left (\frac{d^2}{dy^2}-k^2\right )\delta{\hat \psi}=-\delta{\hat \omega}.
\label{fh6}
\end{equation}
Substituting Eq. (\ref{fh4b}) into Eq. (\ref{fh6}), we obtain
\begin{equation}
\left \lbrack \frac{d^2}{dy^2}-k^2+\frac{k\frac{\partial\omega}{\partial
y}}{kU(y)-\sigma-i\alpha}\right\rbrack
\delta{\hat\psi}(k,y,\sigma)=\frac{i\sqrt{2\gamma\alpha\omega(y)}}{
kU(y)-\sigma-i\alpha}{\hat Q}(k,y,\sigma).
\label{fh7}
\end{equation}
The formal solution of this differential equation is
\begin{equation}
\delta{\hat\psi}(k,y,\sigma)=-\int dy'\,
G(k,y,y',\sigma)\frac{i\sqrt{2\gamma\alpha\omega(y')}}{kU(y')-\sigma-i\alpha}{
\hat Q}(k,y',\sigma),
\label{fh8}
\end{equation}
where the Green function $G(k,y,y',\sigma)$ is defined in Eq. (\ref{ham21}).
The correlation function of the fluctuations of the stream function is
therefore
\begin{eqnarray}
\langle
\delta{\hat\psi}({k},y,\sigma)\delta{\hat\psi}({k}',y,\sigma')\rangle=-\int
dy'dy''\, G(k,y,y',\sigma)G(k',y,y'',\sigma')\nonumber\\
\times\frac{2\gamma\alpha\sqrt{\omega(y')\omega(y'')}}{
(kU(y')-\sigma-i\alpha)(k'U(y'')-\sigma'-i\alpha)}\langle {\hat
Q}(k,y',\sigma){\hat Q}(k',y'',\sigma')\rangle.
\label{fh9}
\end{eqnarray}
For a Gaussian white noise, we get
\begin{eqnarray}
\langle
\delta{\hat\psi}({k},y,\sigma)\delta{\hat\psi}({k}',y,\sigma')\rangle=\int dy'\,
|G(k,y,y',\sigma)|^2
\frac{2\gamma\alpha\omega(y')}{(kU(y')-\sigma)^2+\alpha^2}\delta({k}+{k}
')\delta(\sigma+\sigma'),\label{fh10}
\end{eqnarray}
where we have used Eq. (\ref{obv}) to simplify the expression. Taking the limit
$\alpha\rightarrow 0$ and using the identity
\begin{equation}
\lim_{\epsilon\rightarrow 0}\frac{\epsilon}{x^2+\epsilon^2}=\pi\, \delta(x),
\label{fh11}
\end{equation}
we finally obtain
\begin{eqnarray}
\langle
\delta{\hat\psi}({k},y,\sigma)\delta{\hat\psi}({k}',y,\sigma')\rangle=\pi\int
dy'\, |G(k,y,y',\sigma)|^2
2\gamma\omega(y')\delta(kU(y')-\sigma)\delta({k}+{k}
')\delta(\sigma+\sigma'),
\label{fh12}
\end{eqnarray}
leading to
\begin{eqnarray}
P(k,y,\sigma)=\gamma \int dy'\, |G(k,y,y',\sigma)|^2
\delta(kU(y')-\sigma)\omega(y').
\label{fh13}
\end{eqnarray}
This returns the power spectrum (\ref{ham33}) produced by a
random distribution of field vortices.

Substituting Eqs. (\ref{ham25w}) and (\ref{fh13}) into Eq. (\ref{ham110w}) and
repeating the calculations of Sec. \ref{sec_sdd}, we obtain the nonlinear
diffusion equation
\begin{eqnarray}
\frac{\partial\omega}{\partial t}=-\alpha\omega+\frac{\partial}{\partial y}\left
(D[y,\omega]\frac{\partial\omega}{\partial y}\right )
\label{fh14}
\end{eqnarray}
with a diffusion coefficient
\begin{eqnarray}
D[y,\omega]=\frac{1}{2}\int dk\, k^2 P(k,y,kU(y))=\frac{1}{2}\gamma\int dy'\int
dk\,
|k| |G(k,y,y',kU(y))|^2\delta(U(y')-U(y))\omega(y'),
\label{fh15}
\end{eqnarray}
which coincides with the diffusion coefficient of a gas of point vortices (see
Sec. \ref{sec_diffco}). Using Eq. (\ref{ham10}), we can write Eq. (\ref{fh14})
in terms of $U(y,t)$ as in Eq. (\ref{ham122}).  If $\omega(y)$ is
of constant sign, $U(y)$
is monotonic (see footnote 15) and we get
\begin{eqnarray}
D[y,\omega]=\frac{1}{2}\gamma\int dk\,
|k| |G(k,y,y,kU(y))|^2 \frac{\omega(y)}{|U'(y)|}=\frac{1}{2}|\gamma|\int dk\,
|k| |G(k,y,y,kU(y))|^2.
\label{fh16}
\end{eqnarray}
If we neglect collective effects, we find that the diffusion coefficient is
constant (see Appendix \ref{sec_gfn}):
\begin{eqnarray}
D=\frac{1}{2}|\gamma|\int dk\,
|k| G_{\rm bare}(k,y,y)^2=\frac{1}{4}|\gamma|\ln\Lambda.
\label{fh17}
\end{eqnarray}
In that case, the equation for $U(y,t)$ coincides with Eqs. (\ref{ham124b}) and
(\ref{ham124}). The term $-\alpha\omega$ describes the damping
of the system on a timescale $1/\alpha$ and
the   diffusion coefficient $D[y,\omega]\sim 1/N$ describes its
evolution
on a timescale $N t_D$.

On a mesoscopic scale, we can keep track of the fluctuations in the
evolution of the vorticity and write 
\begin{eqnarray}
\frac{\partial\overline{\omega}}{\partial
t}=-\alpha\overline{\omega}+\frac{\partial}{\partial y}\left
(D[y,\overline{\omega}]\frac{\partial\overline{\omega}}{\partial y}\right
)+\sqrt{2\alpha\gamma\overline{\omega}}\,
{Q}(y,t),
\label{fh17b}
\end{eqnarray}
where ${Q}(y,t)$ is a Gaussian white noise satisfying $\langle
{Q}(y,t)\rangle =0$ and $\langle
{Q}(y,t){Q}(y',t')\rangle=\delta(y-y')\delta(t-t')$. Equation (\ref{fh17b})
without the noise term may have several equilibrium
states. The noise term 
allows the system to switch from one equilibrium state to another one through
random transitions (see, e.g., 
\cite{random,gsse,nardini2,bouchetsimmonet,rbs,brs} in
different contexts).

{\it Remark:} The power spectrum of a gas of point vortices can
be obtained in
different manners. It can be obtained from the linearized
Klimontovich equation by solving
an initial value problem 
(see Eq. (34) of Ref. \cite{klim}). It can also be obtained by considering the
dressing of the bare correlation function of a random distribution of
point vortices viewed as an external perturbation (see Eq.
(\ref{ham33})
of Sec. \ref{sec_inhoscf}). Finally, in this section,  we have determined the
power spectrum [see Eq. (\ref{fh13})] directly from  the stochastic damped
2D Euler equation (\ref{fh1}) in the spirit of fluctuating
hydrodynamics \cite{paper5}.

\section{Stochastically forced 2D point vortices}
\label{sec_spv}

In this section, we consider a stochastic model of 2D point vortices described
by the $N$
coupled Langevin equations 
\begin{equation}
\frac{d{\bf r}_i}{dt}=\frac{1}{2\pi}{\bf z}\times \sum_j \gamma_j \frac{{\bf
r}_i-{\bf
r}_j}{|{\bf r}_i-{\bf r}_j|^2}+\sqrt{2\nu}{\bf R}_i(t),
\label{spv1}
\end{equation}
where $i=1,...,N$ label the point vortices and ${\bf R}_i(t)$ is a Gaussian
white noise satisfying $\langle {\bf R}_i(t)\rangle ={\bf 0}$ and $\langle
{R}^{\alpha}_i(t){R}^{\beta}_j(t')\rangle=\delta_{ij}\delta_{\alpha\beta}
\delta(t-t')$. The variable $\nu$ can be interpreted as a diffusion
coefficient or as a small viscosity (we will ultimately take the limit
$\nu\rightarrow 0$). This stochastic model of 2D point vortices was introduced
by Marchioro and Pulvirenti \cite{mp}. It also corresponds to the model of 2D
Brownian point vortices introduced in \cite{bv} with $\beta=0$ (see Sec.
\ref{sec_tdbv}). 

The exact equation satisfied by the discrete vorticity field
$\omega_d({\bf r},t)=\sum_{i=1}^{N}\gamma_i \delta({\bf r}-{\bf r}_i(t))$ is
\cite{bv}
\begin{eqnarray}
\frac{\partial\omega_d}{\partial t}+{\bf u}_d\cdot
\nabla\omega_d=\nu\Delta\omega_d+\nabla\cdot
\left \lbrack \sqrt{2\nu\gamma\omega}\, {\bf R}({\bf r},t)\right \rbrack,
\label{spv2}
\end{eqnarray}
where ${\bf R}({\bf r},t)$ is a Gaussian white noise such that $\langle
{\bf
R}({\bf r},t)\rangle={\bf 0}$ and  $\langle R_{\alpha}({\bf r},t)R_{\beta}({\bf
r}',t)\rangle=\delta_{\alpha\beta}\delta({\bf r}-{\bf r}')\delta(t-t')$. For
simplicity, we consider a single species gas of point
vortices but the generalization to multiple species of point vortices is
straightforward.

In this section, we neglect the collisions (finite $N$ effects) between the
point vortices that would lead to a Lenard-Balescu collision term (see
Secs. \ref{sec_avp} and \ref{sec_mono}) and focus on the effect of the noise. In
that case, the mean
vorticity $\omega({\bf r},t)=\langle
\omega_{d}({\bf r},t)\rangle$ satisfies the equation
\begin{eqnarray}
\frac{\partial\omega}{\partial t}+{\bf u}\cdot
\nabla\omega=\nu\Delta\omega
\label{spv2b}
\end{eqnarray}
and the  mesoscopic
vorticity satisfies the equation
\begin{eqnarray}
\frac{\partial\omega}{\partial t}+{\bf u}\cdot
\nabla\omega=\nu\Delta\omega+\nabla\cdot
\left \lbrack \sqrt{2\nu\gamma\omega}\, {\bf R}({\bf r},t)\right \rbrack.
\label{spv2bmeso}
\end{eqnarray}

When $\nu\rightarrow 0$ the mean vorticity $\omega({\bf
r},t)$ rapidly reaches a QSS
which is a steady state of the 2D Euler-Poisson equations. On a longer
timescale, the mean vorticity evolves through
a sequence of QSSs sourced by the noise. This is similar to the problem
discussed in Sec.  \ref{sec_fd}. Adapting the procedure of
Sec. \ref{sec_inhos} to the present context, and assuming that the mean flow
(QSS) is unidirectional,
we obtain the quasilinear equations
\begin{equation}
\frac{\partial\omega}{\partial
t}=\nu\frac{\partial^2\omega}{\partial y^2}+\frac{\partial}{\partial
y}\left\langle \delta\omega \frac{\partial\delta\psi}{\partial x}\right\rangle,
\label{spv4}
\end{equation}
 \begin{equation}
\frac{\partial \delta\omega}{\partial t}+U\frac{\partial \delta\omega}{\partial
x}-\frac{\partial\delta\psi}{\partial x} \frac{\partial \omega}{\partial
y}=\nu\frac{\partial^2\delta\omega}{\partial
x^2}+\frac{\partial}{\partial x}\left\lbrack \sqrt{2\nu\gamma\omega}\,
{Q}(x,y,t)\right\rbrack,
\label{spv5}
\end{equation} 
where ${Q}(x,y,t)$ is a Gaussian white noise satisfying $\langle
{Q}(x,y,t)\rangle =0$ and $\langle
{Q}(x,y,t){Q}(x',y',t')\rangle=\delta(x-x')\delta(y-y')\delta(t-t')$. Repeating
the calculations of Sec. \ref{sec_sdd} with only minor modifications, we obtain
the nonlinear
diffusion equation
\begin{eqnarray}
\frac{\partial\omega}{\partial t}=\nu\frac{\partial^2\omega}{\partial
y^2}+\frac{\partial}{\partial y}\left
(D[y,\omega]\frac{\partial\omega}{\partial y}\right )
\label{spv6}
\end{eqnarray}
with the diffusion coefficient $D[y,\omega]$ from Eq. (\ref{fh15}) which
coincides with the
diffusion
coefficient of a gas of point vortices (see
Sec. \ref{sec_diffco}). The diffusion coefficient $\nu$ describes the evolution
of the system on a diffusive timescale $L^2/\nu$ (where $L$ is the size of the
system) and the   diffusion coefficient $D[y,\omega]\sim 1/N$ describes its
evolution
on a timescale $N t_D$. We should also take into account
the collisions between the point vortices, leading to the Lenard-Balescu
current from Eq. (\ref{ham105}),
which develop on the same timescale.

At a mesoscopic level,\footnote{We stress that there are several levels
of description. The mesoscopic description leading to Eq. (\ref{spv7}) takes
place at a higher scale than the mesoscopic description leading to Eq.
(\ref{spv2bmeso}).} we can keep track of the
forcing in the evolution of the vorticity and write 
\begin{eqnarray}
\frac{\partial\omega}{\partial t}=\nu\frac{\partial^2\omega}{\partial
y^2}+\frac{\partial}{\partial y}\left
(D[y,\omega]\frac{\partial\omega}{\partial y}\right )+\frac{\partial}{\partial
y}\left\lbrack \sqrt{2\nu\gamma\omega}\,
{Q}(y,t)\right\rbrack,
\label{spv7}
\end{eqnarray}
where ${Q}(y,t)$ is a Gaussian white noise satisfying $\langle
{Q}(y,t)\rangle =0$ and $\langle
{Q}(y,t){Q}(y',t')\rangle=\delta(y-y')\delta(t-t')$. 
Using Eq.
(\ref{ham10}), we can rewrite Eq. (\ref{spv7}) in terms of  the
velocity
field $U(y,t)$ as
\begin{eqnarray}
\frac{\partial U}{\partial t}=(\nu+D[y,U])\frac{\partial^2 U}{\partial
y^2}-\sqrt{-2D\gamma U'(y)} R(y,t).
\label{burgers3}
\end{eqnarray}

\section{Summary of the different kinetic equations}
\label{sec_diff}

In this section, we recapitulate the different kinetic equations derived in
our paper.

\subsection{Lenard-Balescu equation}

The Lenard-Balescu equation (\ref{ham105}) governs the
mean evolution of an isolated  system of 2D point vortices due to discreteness
effects (``collisions'').

It can be derived from the Klimontovich formalism by
taking ${\bf u}_e=0$ in Eq. (\ref{ham1}) and considering an initial
value problem as explained in Sec. 3 of \cite{klim}. In that case, Eq.
(\ref{ham13}) with $\psi_e=0$ has to
be solved by using a Fourier transform in space and a Laplace transform in
time. This introduces in Eq. (\ref{ham17}) a term related to the initial
condition [see Eq. (23) in \cite{klim}] instead of the term related to $\psi_e$.
We can then compute the
collision term in Eq. (\ref{ham12}) with $\psi_e=0$ as in Sec. 3 of \cite{klim}
and obtain
the
Lenard-Balescu equation (\ref{ham105}).

Another approach is to start from the Fokker-Planck equation (\ref{fp5}) or
(\ref{fp8}) and compute the
diffusion and drift coefficients individually. 

(i) To compute the diffusion coefficient from Eq. (\ref{fp6}) leading to Eq.
(\ref{gei4}), we have to
evaluate
the power spectrum of the stream function fluctuations created by a
random distribution  of
field vortices. This can be done
in two manners. 

(a) The first possibility is to take  $\psi_e=0$ in Eq. (\ref{ham1}) and
solve Eq. (\ref{ham13}) with
$\psi_e=0$ by using a Fourier transform in space and a Laplace
transform in time as mentioned above. This leads to the expression (23) of
\cite{klim} for the fluctuations, which involves the initial
condition. The
power spectrum is then given by Eq. (34) of \cite{klim} and the diffusion
coefficient by Eq. (68) of \cite{klim}.

(b) Another possibility is to introduce a stochastic perturbation $\psi_e$
in Eq. (\ref{ham1}) and 
solve Eq. (\ref{ham13}) by introducing Fourier transforms in space and time.
This leads to Eq. (\ref{ham17}) for the fluctuations, which involves the
external perturbation. The power
spectrum is then given by Eq. (\ref{ham28}). If we assume that the external
perturbation is due to a collection of field vortices, we can use  Eq.
(\ref{ham32}) to obtain the expression (\ref{ham33}) of the power spectrum. The
diffusion
coefficient is then given by Eq. (\ref{gei7diff}).

(ii) We can compute  the drift in two manners.

(a) The first possibility is to
compute the total drift (\ref{fp7}) arising in Eq. (\ref{fp5}) by proceeding
like in Secs. 4.3 and 4.4 of \cite{klim}. This leads to Eq. (102) of
\cite{klim}.
We then find that the total drift splits in two terms: a term 
interpreted as a ``drift by polarization'' (see  Sec. 4.3 of
\cite{klim}) and another term related to the
gradient of the diffusion coefficient (see  Sec. 4.4 of
\cite{klim}). Substituting the diffusion coefficient (Eq. (23) of \cite{klim})
and the total drift (Eq. (102) of \cite{klim}) in the ordinary expression
(\ref{fp5}) of the Fokker-Planck
equation, and using an integration by parts, we obtain the Lenard-Balescu
equation (\ref{ham105}). This is the
approach followed in Sec. 4.5 of \cite{klim}.

(b) Alternatively, we can compute
the drift by polarization arising in Eq. (\ref{fp8}) by considering the response
of the mean flow to the
perturbation created by the test vortex (see Sec. \ref{sec_ifpol}). This
leads to Eq. (\ref{ham66}). This
calculation convincingly shows that $V_{\rm pol}$ can be interpreted as a drift
by polarization. Substituting the diffusion coefficient (\ref{gei7diff}) and the
drift by
polarization (\ref{ham66}) in the expression (\ref{fp8}) of the Fokker-Planck
equation, we obtain the
Lenard-Balescu equation (\ref{ham105}). This derivation is simpler (less
technical) and
more physical that the one given in \cite{klim}.

\subsection{SDD equation}

The SDD equation (\ref{ham116}) with Eqs. (\ref{ham115}) and (\ref{ham118})
governs the mean evolution of a gas of point
vortices submitted to an external
stochastic perturbation in the limit where the collisions between the
point
vortices are negligible, i.e., in the limit $N\rightarrow +\infty$ with
$\gamma\sim 1/N\rightarrow 0$. In that case, we can solve Eq. (\ref{ham13})
by using Fourier transforms in space and time. This leads to the
expression (\ref{ham28}) of the power spectrum. The SDD equation can be
derived from the Klimontovich formalism (see Sec. \ref{sec_sddk}) or from
the Fokker-Planck formalism (see Sec. \ref{sec_sddfp}).  Since there is
no drift by polarization ($\gamma\rightarrow 0$), the Fokker-Planck equation
(\ref{fp8}) reduces to Eq. (\ref{ham121}). Using Eq. (\ref{ham28}), the
diffusion coefficient
(\ref{gei4}) can be written as in Eq. (\ref{gei5}). Substituting Eqs.
(\ref{gei4}) and
(\ref{gei5}) into Eq. (\ref{ham121}), we obtain the SDD equation (\ref{ham116})
with Eqs. (\ref{ham115}) and (\ref{ham118}). The SDD equation also describes
the mean evolution of 
a continuous vorticity field submitted to an external stochastic perturbation
(see the
Remark at
the end of Sec. \ref{sec_inhos}).

\subsection{General kinetic equation}

We now present a kinetic equation that generalizes the Lenard-Balescu equation 
(\ref{ham105}) and the SDD equation (\ref{ham116})-(\ref{ham118}). We consider a
collection of point
vortices of circulation $\gamma$
submitted to a stochastic perturbation that can be internal or external to the
system (or both). The test vortices experience a diffusion
due to the
stochastic
perturbation and a drift by polarization due to retroaction (response) of the
mean flow
to the deterministic perturbation that they induce. The evolution of
their density (mean vorticity)
is
thus governed by
a general Fokker-Planck equation of the form of Eq. (\ref{fp8}) where $D$ is
given by Eq. (\ref{gei4})
and ${V}_{\rm pol}$ is given by Eq. (\ref{ham65}). Explicitly, 
\begin{eqnarray}
\frac{\partial \omega}{\partial t}=\frac{1}{2}\frac{\partial}{\partial y}\int
dk\, k^2\left \lbrack P(k,y,kU(y))\frac{\partial \omega}{\partial y}-\omega
\frac{\gamma}{\pi k} {\rm Im}\,  G(k,y,y,kU(y))\right \rbrack.
\label{hybrid1}
\end{eqnarray}
Alternative expressions of this kinetic equation can be obtained by using Eqs.
(\ref{gei5}) and (\ref{ham66}) instead of Eqs.  (\ref{gei4}) and 
(\ref{ham65}). This kinetic equation is more general than the Lenard-Balescu
equation (\ref{ham105}) because the noise is not necessarily due to a discrete
collection of point vortices. It is
also more general than the SDD equation (\ref{ham116})-(\ref{ham118})
because it takes into account the
drift by polarization of the test vortices. If we neglect the drift
by polarization (i.e. $\gamma\rightarrow 0$) we recover the SDD equation. If we
assume that the external perturbation is only due to
field vortices and use Eqs. (\ref{gei7diff}) and (\ref{ham66q}), we recover the
Lenard-Balescu equation.  If we assume that a part of the stochastic
perturbation is due to
point vortices and another part is due to an external noise, we get an hybrid 
(mixed) kinetic equation with a Lenard-Balescu term and a SDD term (see Sec.
\ref{sec_hybrid}).

\subsection{Hybrid kinetic equation and its subcases}
\label{sec_hybrid}

In order to be as general as possible, we consider a system of test vortices
of circulation $\gamma$ in ``collision'' with field vortices of
circulations $\lbrace\gamma_b\rbrace$ and submitted in addition to an
external noise. The
evolution of the mean vorticity of the test vortices is governed by a mixed
kinetic equation of the
form\footnote{For simplicity, we assume that the test vortices of
circulation $\gamma$ evolve in a bath of field vortices of circulations
$\lbrace\gamma_b\rbrace$. We can then treat the test vortices as representing a
particular species ``a'' and write self-consistent kinetic equations for all the
species.}
\begin{eqnarray}
\frac{\partial \omega}{\partial t}=C_{\rm LB}+C_{\rm SDD},
\label{hybrid2}
\end{eqnarray}
with a Lenard-Balescu collision term $C_{\rm LB}$ due to
the collisions between the point vortices (finite $N$ effects) and a collision
term $C_{\rm SDD}$ due to the external noise. This corresponds to Eq.
(\ref{hybrid1}) with a power spectrum $P=P_{\rm LB}+P_{\rm SDD}$.

Let us consider particular cases of this equation:

(i) When $\gamma\rightarrow 0$, we can neglect the drift by polarization
and we get a diffusion equation with two terms of diffusion $D_{\rm LB}$ and
$D_{\rm SDD}$.

(i-a)  In the absence of external noise, we recover the diffusion equation
(\ref{pd1}) with Eq. (\ref{tb3}).

(i-b) When $\gamma_b\rightarrow 0$ we can neglect the diffusion induced by the
field vortices and we recover the SDD equation (\ref{ham116}) with Eqs.
(\ref{ham115}) and (\ref{ham118}).

(ii) When $\gamma_b\rightarrow 0$, we can neglect the diffusion induced by
the field vortices and we get a Fokker-Planck equation of the form of Eq.
(\ref{hybrid1}) with a diffusion term
$D_{\rm SDD}$ due to the external noise and a drift by polarization.

(ii-a) In the absence of external noise, we recover the deterministic
equation (\ref{pr1}) with Eq. (\ref{tb4}).

(ii-b) When $\gamma\rightarrow 0$, we
can neglect the drift by polarization and 
we recover the SDD equation (\ref{ham116}) with Eqs.
(\ref{ham115}) and (\ref{ham118}).

(ii-c) In the case where the field vortices are at
statistical equilibrium, we can simplify the drift by polarization and we get an
equation of the form of Eq. (\ref{sddxi1}).

(iii) In the absence of external noise, we recover the
Lenard-Balescu equation (\ref{ham105}).

(iii-a) When $\gamma\rightarrow 0$, we can neglect the drift by polarization
and  we recover the diffusion equation (\ref{pd1}) with Eq. (\ref{tb3}).

(iii-b) When $\gamma_b\rightarrow 0$, we can neglect the diffusion induced by 
the field vortices and we recover the deterministic
equation (\ref{pr1}) with Eq. (\ref{tb4}).

(iii-c) In the case where the field vortices are at
statistical equilibrium we can simplify the drift by polarization and we recover
the Smoluchowski equation (\ref{tb6}).

\section{Conclusion}
\label{sec_conc}

In this paper, we have completed the kinetic theory of 2D point vortices
initiated in previous works. We have proposed a new and more physical
derivation of the kinetic equation for a 
multispecies system of point vortices. 

We started from the Fokker-Planck
equation written in
the form of Eq. (\ref{fp8}) and we computed the diffusion coefficient $D$ and
the drift by polarization $V_{\rm pol}$ of a test vortex.

In order to take
collective effects into account, we 
considered the response of the system to a small perturbation of arbitrary
origin. We showed that the response function of the flow is determined by
the
dressed Green function $G(k,y,y',\sigma)$ defined by Eq. (\ref{ham21}). 

To derive the diffusion coefficient $D$, we  assumed
that the perturbation is a stationary stochastic process characterized by a 
bare correlation function ${\hat C}(k,y,\sigma)$ and we determined  the dressed
power
spectrum $P(k,y,\sigma)$ of the total fluctuating stream function  experienced
by
the test
vortex [see Eq. (\ref{ham28})]. The diffusion coefficient is then given by Eqs.
(\ref{gei4}) and (\ref{gei5}). We considered the case of an arbitrary
external perturbation and the case of a perturbation produced by a collection
of $N$ point vortices. In that latter case, we explicitly
determined the bare
correlation function [see Eq. (\ref{ham32})], the dressed power spectrum [see
Eq. (\ref{ham33})], and the diffusion coefficient [see Eq.
(\ref{gei7diff})].

To derive the drift by polarization $V_{\rm pol}$, we used the fact that the
test vortex induces a small perturbation on the flow and we determined the
response of the flow to that perturbation. The drift velocity of the test
vortex then corresponds to the velocity produced by the perturbation that it has
caused [see Eqs. (\ref{ham65}) and (\ref{ham66})].

It is interesting to contrast the origin of the diffusion and drift by
polarization of the test vortex. The diffusion of the test vortex is due to the
stochastic perturbation caused by the field vortices or, more generally, by an
external random potential. The drift by polarization of the test vortex is due
to
the retroaction (response) of the perturbation that it causes on the mean flow.
This
is a purely
deterministic process. This is why  the drift by polarization is
proportional to the circulation $\gamma$ of the test vortex while the diffusion
coefficient is proportional
to the circulations $\lbrace\gamma_b\rbrace$ of the field vortices.

Substituting the expressions of the diffusion coefficient and drift by
polarization into the Fokker-Planck equation (\ref{fp8}) we obtained the
general kinetic equation (\ref{hybrid1}). When the collisions between the point
vortices are neglected ($N\rightarrow +\infty$), and when the system is
subjected to an external perturbation, it reduces to the SDD equation
(\ref{ham116})
with Eqs.
(\ref{ham115}) and (\ref{ham118}). When the system is isolated and the noise is
due to the system of point vortices itself (finite $N$ effects), it reduces to
the
Lenard-Balescu equation (\ref{ham105}). In that latter case, we
discussed the phenomenon of kinetic
blocking that occurs when the velocity
profile is monotonic. In the present paper, we
considered unidirectional flows
but similar
results can be obtained for axisymmetric
flows \cite{cl,klim,onsagerkin}. On the other hand, when collective effects are
neglected, a general kinetic equation 
can be derived for an arbitrary distribution of
point vortices [see Eq. (128) or Eq. (137) of \cite{pre}].

The previous results are valid for an isolated Hamiltonian system of point
vortices. We also considered the case of 2D Brownian (stochastically forced)
point
vortices in the canonical ensemble and we established the mean field
Fokker-Planck equation (\ref{bv1})
and the stochastic Fokker-Planck  equation (\ref{bv3}). This last equation can
describe random transitions between different equilibrium states caused by
finite-$N$ fluctuations. We
showed furthermore that the fluctuations induce a nonlinear diffusion of the
point vortices [see Eqs.
(\ref{spv6})
and (\ref{spv7})].

Another goal  of the paper was to emphasize the fluctuation-dissipation theorem
for 2D point vortices. 
The velocity of a  test vortex moving in a sea of field
vortices can be
decomposed in two components. There is a mean field velocity due to the
average distribution of point vortices and a ``microscopic'' velocity due to the
discrete interaction between vortices (collisions). In
turn, this microscopic
velocity can be decomposed in two parts. There is a random part giving rise to a
diffusion and a deterministic part giving rise to a systematic drift. The
drift velocity and the random velocity must be related at statistical
equilibrium because they
both come from the same origin (finite $N$ effects). This internal
relationship between the
systematic drift and the random part of the microscopic velocity is of a
very general nature which is manifested in the so-called
fluctuation-dissipation theorem \cite{kubo}. A similar relationship between the
friction and
the random part of the microscopic force arises in the theory of
Brownian motion and in the kinetic theory of systems with long-range
interactions (self-gravitating
systems, plasmas, HMF model...). The fluctuation-dissipation theorem states a
general
relationship between the
response of a given system to an external perturbation and the internal
fluctuations of the system in the absence of the perturbation. Specifically, it
provides a relation between the  response function of the system,
the correlation function of the fluctuations, and the temperature. In the case
of 2D
point vortices at statistical equilibrium it takes the form of Eq. (\ref{ham50})
between the power spectrum
and the
imaginary part of the Green function.  This implies a relation between the drift
velocity and the diffusion
coefficient given by the Einstein relation (\ref{ham88ein}) or by the
Kubo formula (\ref{kuborela}). These equations involve the temperature of the
point vortex gas which may be positive or negative.

In a companion paper \cite{prep1}, we shall study in more detail
the kinetic equations derived in the present contribution and complement further
the kinetic theory of 2D point vortices. Our results can also be exported to
other systems with long-range interactions such as self-gravitating systems.

\appendix

\section{The point vortex model}
\label{sec_pvmod}

We consider an incompressible and inviscid flow described by the Euler equations
\begin{eqnarray}
\label{pv1}
 \nabla\cdot {\bf u}=0,\qquad \frac{\partial {\bf u}}{\partial t}+({\bf u}\cdot
\nabla){\bf u}=-\frac{1}{\rho}\nabla P.
\end{eqnarray}
For a 2D flow, the incompressibility condition becomes  $\partial_x
u_x+\partial_y u_y=0$. In that case, 
we can introduce a stream function $\psi({\bf r},t)$ such that
$u_x=\partial_y\psi$ and $u_y=-\partial_x\psi$. The velocity field can be
written as 
\begin{eqnarray}
\label{pv2}
{\bf u}=-{\bf z}\times\nabla\psi,
\end{eqnarray}
where ${\bf z}$ is a unit vector normal to the plane of the flow. The vorticity 
is defined by $\nabla\times {\bf u}$. For a 2D incompressible flow, the
vorticity is parallel to ${\bf z}$ and related to the stream function through
the Poisson equation
\begin{eqnarray}
\label{pv3}
\omega=-\Delta\psi.
\end{eqnarray}
Using the identity $({\bf u}\cdot \nabla){\bf u}=\nabla ({\bf u}^2/2)-{\bf
u}\times (\nabla\times {\bf u})$ and taking the curl of Eq. (\ref{pv1}), we
obtain 
\begin{eqnarray}
\label{pv4}
\frac{\partial \omega}{\partial t}+{\bf u}\cdot \nabla\omega=0.
\end{eqnarray}
This equation expresses the advection of the vorticity by the flow. It can be
written as $D\omega/Dt=0$, where $D=\partial/\partial t+{\bf u}\cdot \nabla$ is
the
material derivative (Stokes operator). Equations (\ref{pv3}) and (\ref{pv4})
define the 2D
Euler-Poisson system. In an infinite domain, the Poisson equation (\ref{pv3})
can be integrated into
\begin{eqnarray}
\label{pv5}
\psi({\bf r},t)=-\frac{1}{2\pi}\int\ln|{\bf r}-{\bf r}'|\omega({\bf r}',t)\,
d{\bf r}',
\end{eqnarray}
leading to the velocity field
\begin{eqnarray}
{\bf u}({\bf r},t)=-\frac{1}{2\pi}{\bf z}\times \int \frac{{\bf r}'-{\bf
r}}{|{\bf r}'-{\bf r}|^2}\omega({\bf r}',t)\, d{\bf r}'.
\label{ham127}
\end{eqnarray}

For a system of $N$ point vortices with circulation $\gamma_i$ the vorticity
field can be written as
\begin{eqnarray}
\label{pv6}
\omega({\bf r},t)=\sum_i \gamma_i \delta({\bf r}-{\bf r}_i(t)).
\end{eqnarray}
The discrete vorticity is a sum of Dirac $\delta$-functions.
Substituting this expression into Eq. (\ref{pv4}) we find after straightforward
manipulations (see below)  that the velocity of a test vortex is given by
\begin{eqnarray}
\label{pv7}
{\bf V}_{i}(t)=\frac{d{\bf r}_i}{dt}={\bf u}({\bf r}_i(t),t).
\end{eqnarray}
From Eqs. (\ref{pv5})-(\ref{pv6}), we get
\begin{eqnarray}
\label{pv8}
\psi({\bf r},t)=-\frac{1}{2\pi}\sum_i \gamma_i \ln|{\bf r}-{\bf r}_i|,\qquad
{\bf u}({\bf r},t)=\frac{1}{2\pi}{\bf z}\times \sum_i \gamma_i \frac{{\bf
r}-{\bf r}_i}{|{\bf r}-{\bf r}_i|^2},
\end{eqnarray}
and
\begin{eqnarray}
\label{pv9}
{\bf V}_i=\frac{d{\bf r}_i}{dt}=-{\bf
z}\times \nabla \psi({\bf r}_i)=\frac{1}{2\pi}{\bf z}\times
\sum_j \gamma_j \frac{{\bf r}_i-{\bf r}_j}{|{\bf r}_i-{\bf r}_j|^2}.
\end{eqnarray}
The velocity of a point vortex is induced by the other point
vortices. This is different from the case of
material particles where the interaction between the particles produces an
acceleration (or a force), not a velocity. In a sense, a point vortex does not
have inertia.  The equations of motion of the point vortices can be written in
Hamiltonian form
as
\begin{eqnarray}
\label{pv10}
\gamma_i\frac{dx_i}{dt}=\frac{\partial H}{\partial y_i},\qquad
\gamma_i\frac{dy_i}{dt}=-\frac{\partial H}{\partial x_i},\end{eqnarray}
with the Hamiltonian
\begin{eqnarray}
\label{pv11}
H=-\frac{1}{2\pi}\sum_{i<j} \gamma_i\gamma_j \ln|{\bf r}_i-{\bf r}_j|.
\end{eqnarray}
These are the so-called Kirchhoff equations \cite{kirchhoff}. We note
that the coordinates $(x,y)$ of the point vortices are canonically conjugate.
We can also write the equations of motion of the point
vortices under the form
\begin{eqnarray}
\label{pv11vec}
\gamma_i \frac{d{\bf r}_i}{dt}=-{\bf
z}\times \nabla H.
\end{eqnarray}

{\it Proof of Eq. (\ref{pv7}):} From Eq. (\ref{pv6}) we have
\begin{eqnarray}
\label{pv12}
\frac{\partial\omega}{\partial t}=-\sum_i \gamma_i \nabla\delta({\bf r}-{\bf
r}_i(t))\cdot \frac{d{\bf r}_i}{dt},
\end{eqnarray}
and
\begin{eqnarray}
\label{pv13}
{\bf u}\cdot \nabla\omega&=&{\bf u}({\bf r},t)\cdot\sum_i \gamma_i 
\nabla\delta({\bf r}-{\bf r}_i(t))\nonumber\\
&=&\sum_i \gamma_i  \nabla\left (\delta({\bf r}-{\bf r}_i(t)){\bf u}({\bf
r},t)\right )\nonumber\\
&=&\sum_i \gamma_i  \nabla\left (\delta({\bf r}-{\bf r}_i(t)){\bf u}({\bf
r}_i(t),t)\right )\nonumber\\
&=&\sum_i \gamma_i {\bf u}({\bf r}_i(t),t)\cdot \nabla \delta({\bf r}-{\bf
r}_i(t)),
\end{eqnarray}
where we have used the incompressibility of the flow to get the second line of
Eq. (\ref{pv13}). Substituting these expressions into Eq. (\ref{pv4}), we obtain
Eq. (\ref{pv7}) and the Kirchhoff equations (\ref{pv10}) and (\ref{pv11}).
Inversely, starting from Eq. (\ref{pv7}) or from the Kirchhoff equations
(\ref{pv10}) and (\ref{pv11}), we find that the discrete vorticity field defined
by Eq. (\ref{pv6}) satisfies Eq. (\ref{pv4}). The discrete 2D Euler equation
(\ref{pv4}) expressed in terms of $\delta$-functions is the counterpart of the
Klimontovich equation in plasma physics.

\section{Green function without collective effects}
\label{sec_gfn}

The stream function $\psi$ produced by the vorticity field $\omega$ is
determined by the Poisson equation (\ref{pv3}). Introducing a system of
cartesian coordinates and taking its Fourier transform in the $x$-direction, we
obtain 
\begin{eqnarray}
\frac{d^2{\hat \psi}}{dy^2}-k^2 {\hat\psi}=-{\hat\omega}.
\label{hamg2}
\end{eqnarray}
The general solution of this equation is given by
\begin{eqnarray}
{\hat\psi}(k,y,\sigma)=\int G_{\rm bare}(k,y,y') {\hat\omega}(k,y',\sigma)\,
dy',
\label{hamg3}
\end{eqnarray}
where  $G_{\rm bare}(k,y,y')$ is the bare Green function determined by the
differential equation
\begin{eqnarray}
\frac{d^2G_{\rm bare}}{dy^2}-k^2G_{\rm bare}=-\delta(y-y').
\label{hamg4}
\end{eqnarray}
In an unbounded domain, this equation can be solved analytically as follows. For
$y\neq y'$, we have
\begin{eqnarray}
\frac{d^2G_{\rm bare}}{dy^2}-k^2G_{\rm bare}=0.
\label{hamg5}
\end{eqnarray}
This equation can be integrated into
\begin{eqnarray}
G_{\rm bare}(k,y,y')=Ae^{-|k||y-y'|},
\label{hamg5b}
\end{eqnarray}
where we have selected the solution that decays to zero at infinity. To
determine the constant  $A$ we integrate Eq. (\ref{hamg4}) between $-\infty$ and
$+\infty$, giving
\begin{eqnarray}
\int_{-\infty}^{+\infty} \frac{d^2G_{\rm bare}}{dy^2}\,
dy-k^2\int_{-\infty}^{+\infty} G_{\rm bare}\, dy=-1.
\label{hamg6}
\end{eqnarray}
Since $G_{\rm bare}(k,y,y')$ and its derivatives vanish at infinity, the
foregoing equation reduces to
\begin{eqnarray}
2k^2A\int_{0}^{+\infty} e^{-|k|y}\, dy=1,
\label{hamg7}
\end{eqnarray}
yielding
\begin{eqnarray}
A=\frac{1}{2|k|}.
\label{hamg7b}
\end{eqnarray}
Therefore, the bare Green function in an infinite domain is given by
\begin{eqnarray}
G_{\rm bare}(k,y,y')=\frac{1}{2|k|}e^{-|k||y-y'|}.
\label{hamg8}
\end{eqnarray}

Introducing the function 
\begin{eqnarray}
\chi_{\rm bare}(y,y')=\frac{1}{2}\int |k| G_{\rm bare}(k,y,y')^2 \, dk,
\label{hamg9}
\end{eqnarray}
and using Eq. (\ref{hamg8}), we get
\begin{eqnarray}
\chi_{\rm bare}(y,y')=\int_0^{+\infty} \frac{1}{4k}e^{-2k|y-y'|} \, dk.
\label{hamg11}
\end{eqnarray}
When $y'=y$, this integral reduces to
\begin{eqnarray}
\chi_{\rm bare}(y,y)=\frac{1}{4}\int_0^{+\infty}
\frac{dk}{k}=\frac{1}{4}\ln\Lambda,
\label{hamg10}
\end{eqnarray}
where $\ln\Lambda=\int_0^{+\infty} {dk}/{k}=\ln(\lambda_{\rm max}/\lambda_{\rm
min})$. We note that $\chi_{\rm bare}(y,y)$ involves an integral that diverges
logarithmically at small and large scales. It can be regularized by introducing
appropriate cut-offs (see Refs. \cite{preR,pre,dubin2} for more details) leading
to a
logarithmic factor $\ln\Lambda$ similar to the Coulomb
logarithm in plasma physics. 
When $y'\neq y$, the integral from Eq. (\ref{hamg11}) is convergent at small
scales ($k\rightarrow +\infty$) but  
divergent at large scales ($k\rightarrow 0$). In the dominant approximation, we
can write\footnote{Collective effects are usually
negligible when $y'\rightarrow y$. In that case, $\chi(y,y,U(y))$ can be
replaced by  $\chi_{\rm bare}(y,y)=
(1/4)\ln\Lambda$. More generally, in the dominant
approximation,  $\chi(y,y',U(y))$ may be
replaced by  $(1/4)\ln\Lambda$ with good accuracy. }
\begin{eqnarray}
\chi_{\rm bare}(y,y')\simeq \frac{1}{4}\ln\Lambda.
\label{hamg10m}
\end{eqnarray}

In order to regularize the large-scale divergence in Eq. (\ref{hamg11}) we can 
replace the Poisson equation (\ref{pv3}) by the screened Poisson
equation
\begin{eqnarray}
\Delta\psi-k_R^2\psi=-\omega,
\label{hamg12}
\end{eqnarray}
and ultimately take the limit $k_R\rightarrow 0$ \cite{pre}. Equation
(\ref{hamg12}) can be introduced in an {\it ad hoc} manner but it is interesting
to note that it also corresponds to the quasigeostrophic (QG) model describing
geophysical flows \cite{pedlosky}. In that context, $k_R^{-1}$ is the so-called
Rossby radius. The bare Green function corresponding to Eq. (\ref{hamg12}) is
obtained from Eq. (\ref{hamg8}) by making the substitution
$k^2\rightarrow k^2+k_R^2$. This yields
\begin{eqnarray}
G_{\rm bare}(k,y,y')=\frac{1}{2\sqrt{k^2+k_R^2}}e^{-\sqrt{k^2+k_R^2}\, |y-y'|}.
\label{hamg13}
\end{eqnarray}
The function defined by Eq. (\ref{hamg9}) takes the form
\begin{eqnarray}
\chi_{\rm bare}(y,y')=\frac{1}{4}\int_0^{+\infty} \frac{k}{k^2+k_R^2}
e^{-2\sqrt{k^2+k_R^2}\, |y-y'|}\, dk.
\label{hamg14}
\end{eqnarray}
It can be written as
\begin{eqnarray}
\chi_{\rm bare}(y,y')=\frac{1}{4}E_1(2k_R|y-y'|),
\label{hamg15}
\end{eqnarray}
where
\begin{eqnarray}
E_1(x)=\int_x^{+\infty}\frac{e^{-t}}{t}\, dt
\label{hamg16}
\end{eqnarray}
is the exponential integral. For $x\rightarrow 0$, we have the expansion
$E_1(x)=-\gamma_E-\ln x+...$, where $\gamma_E=0.57721$ is Euler's constant.
Therefore, for $k_R\rightarrow 0$, we get
\begin{eqnarray}
\chi_{\rm bare}(y,y')\simeq \frac{1}{4}\left \lbrack
-\gamma_E-\ln(2k_R|y-y'|)\right \rbrack,
\label{hamg17}
\end{eqnarray}
which is perfectly well-defined for $y'\neq y$.  When
$y'=y$, the integral (\ref{hamg14}) converges at large scales ($k\rightarrow
0$) but diverges logarithmically at small scales ($k\rightarrow +\infty$).

{\it Remark:} The previous results can be generalized to an arbitrary  potential
of interaction of the form  $u(|{\bf r}-{\bf r}'|)$ such that $\psi({\bf
r})=\int u(|{\bf r}-{\bf r}'|)\omega({\bf r}')\, d{\bf r}'$. For the 2D Euler
equation in an infinite domain we have $u(|{\bf r}-{\bf
r}'|)=-\frac{1}{2\pi}\ln|{\bf r}-{\bf r}'|$ and for the QG equations in an
infinite domain we have $u(|{\bf r}-{\bf r}'|)=\frac{1}{2\pi}K_0(k_R|{\bf
r}-{\bf r}'|)$, where $K_0(x)$ is the modified Bessel function of zeroth order. 
We note that $G_{\rm bare}(k,|y-y'|)={\hat u}(k,|y-y'|)$ is the Fourier
transform of the potential of interaction $u(|{\bf r}-{\bf r}'|)$ with respect
to the variable $x$. Using Eqs. (\ref{hamg8}) and (\ref{hamg13}), we find that
\begin{eqnarray}
\ln|{\bf r}-{\bf r}'|=-\pi\int e^{ik(x-x')}\frac{1}{|k|}e^{-|k||y-y'|}\, dk,
\label{hamg18}
\end{eqnarray}
\begin{eqnarray}
K_0(k_R|{\bf r}-{\bf r}'|)=\pi\int e^{ik(x-x')} 
\frac{1}{\sqrt{k^2+k_R^2}}e^{-\sqrt{k^2+k_R^2}\, |y-y'|}\, dk.
\label{hamg18b}
\end{eqnarray}

\section{An important identity}
\label{sec_id}

The Green function $G(k,y,y',\sigma)$ introduced in Sec. \ref{sec_inhosrf} is
determined by the equation 
\begin{eqnarray}
\frac{d^2G}{dy^2}-k^2G+\frac{k\frac{\partial\omega}{\partial
y}}{kU(y)-\sigma}G=-\delta(y-y')
\label{hami1}
\end{eqnarray}
with the Landau prescription $\sigma\rightarrow \sigma+i0^+$. Multiplying Eq.
(\ref{hami1}) by
$G(k,y,y',\sigma)^*$ and integrating over $y$ between $-\infty$ and $+\infty$,
we get
\begin{eqnarray}
-\int_{-\infty}^{+\infty}  \left |\frac{dG}{dy}\right |^2\,
dy-k^2\int_{-\infty}^{+\infty} |G|^2\,
dy+\int_{-\infty}^{+\infty}\frac{k\frac{\partial\omega}{\partial
y}}{kU(y)-\sigma}|G|^2\, dy=-G(k,y',y',\sigma)^*,
\label{hami2}
\end{eqnarray}
where we have integrated the first term by parts. Taking the imaginary part of
this equation, we find that
\begin{eqnarray}
{\rm Im}\,  G(k,y',y',\sigma)={\rm Im}\, \int_{-\infty}^{+\infty}
\frac{k\frac{\partial\omega}{\partial y}}{kU(y)-\sigma}|G|^2(k,y,y',\sigma)\,
dy.
\label{hami3}
\end{eqnarray}
Using the Sokhotski-Plemelj formula
\begin{eqnarray}
\frac{1}{x\pm i0^+}={\cal P}\left (\frac{1}{x}\right )\mp i\pi\delta(x),
\label{plemelj}
\end{eqnarray}
we obtain the important identity
\begin{eqnarray}
{\rm Im}\,  G(k,y,y,\sigma)=\pi\int_{-\infty}^{+\infty}
k\frac{\partial\omega'}{\partial y'} \delta(kU(y')-\sigma)
|G(k,y',y,\sigma)|^2\, dy'.
\label{hami4}
\end{eqnarray}
We also mention the identity
\begin{eqnarray}
G(-k,y,y',-\sigma)=G(k,y,y',\sigma)^*,
\label{obv}
\end{eqnarray}
which can be derived from Eq. (\ref{hami1}) by using the Landau prescription.

\section{Alternative derivations of the diffusion coefficient}

\subsection{General expression of the diffusion coefficient using Fourier
transforms in position and time}
\label{sec_geia}

The change in position (in the $y$-direction) of a test vortex due to the total
fluctuating stream
function is
\begin{eqnarray}
\frac{dy}{dt}=V_y=-\frac{\partial \delta\psi_{\rm tot}}{\partial x}(x,y,t).
\label{hama1}
\end{eqnarray}
Integrating this equation between $0$ and $t$, we obtain
\begin{eqnarray}
\Delta y&=&-\int_0^t \frac{\partial \delta\psi_{\rm tot}}{\partial
x}({x}(t'),{y}(t'),t')\, dt'\nonumber\\
&=&-\int_0^t \frac{\partial \delta\psi_{\rm tot}}{\partial
x}(x+U(y)t',y,t')\, dt',
\label{hama2}
\end{eqnarray}
where we have used the unperturbed equation of motion (\ref{lin}) in the second
equation (this accounts for the fact that the
point vortex follows the mean field trajectory at leading order).
Decomposing the stream function in Fourier modes, we get
\begin{eqnarray}
\Delta y&=&-\int_0^t dt'\, \frac{\partial}{\partial x}\int dk\int
\frac{d\sigma}{2\pi}e^{i k(x+U(y)t')}e^{-i\sigma
t'}\delta{\hat \psi}_{\rm tot}({k},{y},\sigma)\nonumber\\
&=&-\int dk\int \frac{d\sigma}{2\pi} i {k}  e^{i{k}x}\delta{\hat \psi}_{\rm
tot}({k},{y},\sigma)\int_0^t e^{i({k}U(y)-\sigma)t'}\,
dt'\nonumber\\
&=&-\int dk\int \frac{d\sigma}{2\pi} i {k}  e^{i{k}x}\delta{\hat
\psi}_{\rm
tot}({k},{y},\sigma)\frac{e^{i({k}U(y)-\sigma)t}-1}{i({k}U(y)-\sigma)}.
\label{hama3}
\end{eqnarray}
The diffusion coefficient is defined by [see Eq. (\ref{fp6})]
\begin{eqnarray}
D=\lim_{t\rightarrow +\infty}\frac{\langle (\Delta y)^2 \rangle}{2t}.
\label{hama5}
\end{eqnarray}
Substituting Eq. (\ref{hama3}) into Eq. (\ref{hama5}), we obtain
\begin{equation}
D=-\lim_{t\rightarrow +\infty}\frac{1}{2t}\int dk\int dk'\int
\frac{d\sigma}{2\pi} \int \frac{d\sigma'}{2\pi}  k k'  e^{i(k+k')x}\langle
\delta{\hat \psi}_{\rm tot}({k},{y},\sigma)\delta{\hat \psi}_{\rm
tot}({k}',{y},\sigma')\rangle
\frac{e^{i({k}U(y)-\sigma)t}-1}{i({k}U(y)-\sigma)}
\frac{e^{i({k}'U(y)-\sigma')t}-1}{i({k}'U(y)-\sigma')}.
\label{hama6}
\end{equation}
Introducing the power spectrum from Eq. (\ref{ham25}), the foregoing equations
can be
rewritten as 
\begin{equation}
D=\lim_{t\rightarrow +\infty}\frac{1}{2t}\int dk\int
\frac{d\sigma}{2\pi}  k^2 P({k},{y},\sigma)
\frac{|e^{i(kU(y)-\sigma)t}-1|^2}{({k}U(y)-\sigma)^2}.
\label{hama8}
\end{equation}
Using the identity 
\begin{equation}
\lim_{t\rightarrow +\infty}\frac{|e^{ixt}-1|^2}{x^2t}=2\pi\delta(x),
\label{heyid}
\end{equation}
we find that
\begin{equation}
D=\pi\int dk\int \frac{d\sigma}{2\pi}  k^2 
P({k},{y},\sigma)\delta(kU(y)-\sigma).
\label{hama9}
\end{equation}
Integrating over the $\delta$-function (resonance), we get
\begin{equation}
D=\frac{1}{2}\int dk \, k^2  P({k},{y},kU(y)),
\label{hama10}
\end{equation}
which returns Eq. (\ref{gei4}). Then, using Eq. (\ref{ham28}) we obtain Eq.
(\ref{gei5}).

{\it Remark:} If we do not take the limit $t\rightarrow
+\infty$ in Eq. (\ref{hama8}), we obtain a
time-dependent diffusion coefficient of the form
\begin{equation}
D(t)=\pi\int dk\int
\frac{d\sigma}{2\pi}  k^2  P({k},{y},\sigma)
\Delta({k}U(y)-\sigma,t)
\label{hama8del}
\end{equation}
with the regularized function
\begin{equation}
\Delta(x,t)=\frac{1}{2\pi t}\frac{|e^{ixt}-1|^2}{x^2}=\frac{1-\cos(xt)}{\pi t
x^2}.
\label{hama8def}
\end{equation}
When $t\rightarrow +\infty$, we can make the replacement
$\Delta(x,t)\rightarrow \delta(x)$ corresponding to the diffusive regime. When
$t\rightarrow 0$, we have $\Delta(x,t)\sim t/2\pi$ corresponding to the
ballistic regime.

\subsection{General expression of the diffusion coefficient using
a Fourier transform in position}
\label{sec_geiab}

We can make the calculations of the previous section in a slightly different
manner. In Eq. (\ref{hama2}) we decompose the total fluctuating stream function
in
Fourier modes in position but not in time. In that case, we get
\begin{eqnarray}
\Delta y&=&-\int_0^t dt'\, \frac{\partial}{\partial x}\int dk\,
e^{i{k} (x+U(y) t')}\delta{\hat \psi}_{\rm tot}({k},{y},t')\nonumber\\
&=&-\int_0^t dt'\int dk\, i{k} e^{i{k} ({x}+{U(y)}t')}\delta{\hat
\psi}_{\rm tot}({k},{y},t').
\label{hamq}
\end{eqnarray}
Substituting Eq. (\ref{hamq}) into Eq. (\ref{hama5}), we obtain
\begin{eqnarray}
D=-\lim_{t\rightarrow +\infty}  \frac{1}{2t}\int_0^t dt'\int_0^t dt'' \int
dk\int dk'\, kk'  e^{i({k}+{k}')x}e^{i{k}U(y)t'} e^{i{k}'U(y)t''} \left\langle
\delta{\hat \psi}_{\rm tot}({k},{y},t')\delta{\hat \psi}_{\rm
tot}({k}',y,t'')\right\rangle. 
\label{hamq1}
\end{eqnarray}
Introducing the inverse Fourier transform in time of the power spectrum from Eq.
(\ref{sta1}), we can rewrite the foregoing equation as
\begin{eqnarray}
D=\lim_{t\rightarrow +\infty} \frac{1}{2t}\int_0^t dt'\int_0^t dt'' \int
dk\, k^2 e^{i{k}U(y)(t'-t'')} {\cal P}({k},y,t'-t'').
\label{hamq3}
\end{eqnarray}
Using the identity (\ref{diff4}), we get
\begin{eqnarray}
D=\lim_{t\rightarrow +\infty} \frac{1}{t}\int_0^t ds\, (t-s) \int dk\, k^2 
e^{i k U(y) s}  {\cal P}(k,y,s).
\label{hamq4}
\end{eqnarray}
Assuming that ${\cal P}({k},{y},s)$ decreases
more rapidly than $s^{-1}$, we obtain
\begin{eqnarray}
D=\int_0^{+\infty} ds\, \int dk\, k^2  e^{i {k} U(y) s} 
{\cal P}({k},{y},s).
\label{hamq5}
\end{eqnarray}
Making the change of
variables $s\rightarrow -s$ and ${k}\rightarrow -{k}$, and using the
fact that ${\cal P}(-{k},y,-s)={\cal P}({k},y,s)$, we see
that we can replace
$\int_{0}^{+\infty}ds$ by $(1/2)\int_{-\infty}^{+\infty}ds$. Therefore,
\begin{eqnarray}
D=\frac{1}{2}\int_{-\infty}^{+\infty} ds\, \int dk\, k^2  e^{i {k} U(y) s} 
{\cal P}({k},{y},s).
\label{hamq6}
\end{eqnarray}
Finally, taking the inverse Fourier transform in time of ${\cal P}({k},{y},s)$
we find that
\begin{eqnarray}
D=\frac{1}{2} \int dk\, k^2  {P}({k},y,kU(y)),
\label{hamq7}
\end{eqnarray}
which returns Eq. (\ref{gei4}). Then, using Eq. (\ref{ham28}) we obtain Eq.
(\ref{gei5}). We note that
${\cal P}({k},{y},s)$ is complex while $P({k},{y},\sigma)$ is real. They
satisfy the identities ${\cal P}({-k},{y},s)={\cal P}({k},{y},s)^*={\cal
P}({k},{y},-s)$ and
$P({k},{y},\sigma)=P({k},{y},\sigma)^*=P(-{k},{y},-\sigma)$
which can be directly obtained from the definition of ${\cal
P}({k},{y},s)$ and 
${P}({k},{y},\sigma)$ in Sec. \ref{sec_cf}.

{\it Remark:} If we introduce the temporal Fourier
transform of  ${\cal P}({k},{y},t)$ in Eq. (\ref{hamq3}) we get
\begin{eqnarray}
D=\lim_{t\rightarrow +\infty} \frac{1}{2t}\int_0^t dt'\int_0^t dt'' \,
\int dk\int \frac{d\sigma}{2\pi}\, k^2  e^{i{k}U(y)(t'-t'')}
e^{-i\sigma(t'-t'')} {P}({k},{y},\sigma),
\label{an8}
\end{eqnarray}
which is equivalent to Eq. (\ref{diff1}) with Eq.
(\ref{diff3}). If we
integrate over $t'$ and $t''$, we recover Eq. (\ref{hama8}).

\subsection{Diffusion coefficient created by $N$ point vortices}
\label{sec_dn}

According to Eqs.  (\ref{ham20}) and (\ref{ham30}), the total fluctuating stream
function created by a collection of $N$
point vortices is
\begin{eqnarray}
\delta{\hat \psi}_{\rm tot}({k},{y},\sigma)
=\sum_i\gamma_i G(k,y,y_i,\sigma) e^{-i{k}x_i}\delta(\sigma-{k}U(y_i)).
\label{dn7}
\end{eqnarray}
Substituting this expression into Eq. (\ref{hama3}) and integrating over
$\sigma$, we obtain
\begin{eqnarray}
\Delta y
= -\frac{1}{2\pi} \sum_i \gamma_i\int dk\, i {k}  e^{i{k}x}
\frac{e^{ik(U(y)-U(y_i))t}-1}{i{k}(U(y)-U(y_i))} G(k,y,y_i,kU(y_i)) 
e^{-i{k}x_i}.
\label{dn8}
\end{eqnarray}
The diffusion coefficient from Eq. (\ref{hama5}) is then given by
\begin{eqnarray}
D=-\lim_{t\rightarrow +\infty}\frac{1}{2t}\Biggl\langle
\frac{1}{4\pi^2}\sum_{ij}\gamma_i\gamma_j \int dk\int dk'\,
kk'e^{i({k}+k')x}
\frac{e^{i{k}(U(y)-U(y_i))t}-1}{ik(U(y)-U(y_i))}
\frac{e^{i{k'}(U(y)-U(y_j))t}-1}{ik'(U(y)-U(y_j))}\nonumber\\
\times G(k,y,y_i,kU(y_i)) G(k',y,y_j,k'U(y_j))
e^{-i{k}x_i}e^{-i{k'}x_j}
\Biggr\rangle.
\label{dn10}
\end{eqnarray}
Since the point vortices are initially uncorrelated, and since the point
vortices of the
same
species are identical, we get (see the similar steps detailed after Eq.
(\ref{ham31}) in Sec. \ref{sec_inhoscf})
\begin{eqnarray}
D=-\lim_{t\rightarrow +\infty}\frac{1}{2t}\sum_b \frac{1}{4\pi^2}\int dx'\int
dy'\int dk\int dk'\, kk' e^{i(k+k')x} 
\frac{e^{i{k}(U(y)-U(y'))t}-1}{ik(U(y)-U(y'))}
\frac{e^{i{k'}(U(y)-U(y'))t}-1}{ik'(U(y)-U(y'))}\nonumber\\
\times G(k,y,y',kU(y')) G(k',y,y',k'U(y'))
e^{-i(k+k')x'} \gamma_b \omega_b(y').
\label{dn11}
\end{eqnarray}
Integrating over $x'$, then over $k'$, and using the identity from Eq.
(\ref{obv}), we get
\begin{eqnarray}
D=\lim_{t\rightarrow +\infty}\frac{1}{2t}\sum_b \frac{1}{2\pi} \int
dy'\int dk\, k^2
\frac{|e^{i{k}(U(y)-U(y'))t}-1|^2}{k^2(U(y)-U(y'))^2}|G(k,y,y',kU(y'))|^2 
\gamma_b \omega_b(y').
\label{dn15}
\end{eqnarray}
Finally, using Eq. (\ref{heyid}), we obtain
\begin{eqnarray}
D=\frac{1}{2}\sum_b \int dy' \int dk\, k^2 \delta\left\lbrack
{k}(U(y)-U(y'))\right\rbrack |G(k,y,y',kU(y'))|^2 
\gamma_b \omega_b(y'),
\label{dn17}
\end{eqnarray}
which returns Eq. (\ref{gei7}). Integrating
over the
$\delta$-function (resonance) with the identity from Eq.
(\ref{gei7b}), we recover Eq. (\ref{gei7diff}).

{\it Remark:} If we do not take the limit $t\rightarrow +\infty$ in the
foregoing equations, we obtain a
time-dependent diffusion coefficient
\begin{eqnarray}
D(t)=\frac{1}{2}\sum_b \int dy' \int dk\, k^2
\Delta\left\lbrack
{k}(U(y)-U(y')),t\right\rbrack |G(k,y,y',kU(y'))|^2 
\gamma_b \omega_b(y'),
\label{dn17w}
\end{eqnarray}
where the regularized function $\Delta(x,t)$ is defined in Eq. (\ref{hama8def}).

\section{Velocity auto-correlation function and
diffusion coefficient with collective effects}
\label{sec_geic}

The $y$-component of the velocity of a test vortex is 
\begin{eqnarray}
V_y=-\frac{\partial\delta\psi_{\rm tot}}{\partial x},
\label{ham178}
\end{eqnarray}
where $\delta\psi_{\rm tot}(x,y,t)$ is the total fluctuating stream function
acting on the test vortex. Introducing the Fourier transform of the total stream
function, the velocity
auto-correlation function of the test vortex accounting for collective effects
can be written as
\begin{equation}
\left\langle V_y(x,y,0)V_y(x+U(y)t,y,t)\right \rangle=-\int dk
\int\frac{d\sigma}{2\pi}\int dk' \int\frac{d\sigma'}{2\pi}
 k k' e^{i k x} e^{ik' (x+U(y)t)} e^{-i\sigma' t}\left\langle
\delta\hat\psi_{\rm tot}(k,y,\sigma)  \delta\hat\psi_{\rm
tot}(k',y,\sigma')\right\rangle.
 \label{ham179}
\end{equation}
To define the correlation function, we have used a Lagrangian point of view
and we have used the fact that the test vortex follows the mean field trajectory
from Eq. (\ref{lin})  at leading order. Using the expression (\ref{ham25})
of the correlation function of the
total fluctuating stream function (power spectrum), we get
\begin{eqnarray}
\langle V_y(x,y,0)V_y(x+U(y)t,y,t)\rangle=\int dk \int\frac{d\sigma}{2\pi}
 k^2  e^{i(\sigma-kU(y)) t}P(k,y,\sigma).
 \label{ham180}
\end{eqnarray}
Recalling the relation (\ref{ham28}) between the dressed correlation function of
the
total fluctuating stream function and the bare correlation function of the
external
vorticity field, we obtain
\begin{eqnarray}
\left\langle V_y(x,y,0)V_y(x+U(y)t,y,t)\right \rangle=\int dy'\int dk
\int\frac{d\sigma}{2\pi}\,
 k^2  e^{i(\sigma-kU(y)) t} |G(k,y,y',\sigma)|^2 {\hat C}(k,y',\sigma).
 \label{ham181}
\end{eqnarray}
If the external vorticity field is created by $N$ point vortices then, using Eq.
(\ref{ham32}), the foregoing equation becomes 
\begin{eqnarray}
\left\langle V_y(x,y,0)V_y(x+U(y)t,y,t)\right \rangle&=&\sum_b \gamma_b\int
dy'\int dk \int\frac{d\sigma}{2\pi}\,
 k^2  e^{i(\sigma-kU(y)) t} |G(k,y,y',\sigma)|^2  \delta(kU(y')-\sigma)
\omega_b(y') \nonumber\\
 &=&\sum_b \frac{\gamma_b}{2\pi}\int dy'\int dk\,
 k^2  e^{ik(U(y')-U(y)) t} |G(k,y,y',kU(y'))|^2   \omega_b(y'). \label{ham182}
\end{eqnarray}
Explicit expressions of the velocity auto-correlation function of a point
vortex are given in \cite{pre,prep1}. Using
Eq. (\ref{gei1}), we
find that the diffusion coefficient of the test vortex is
\begin{eqnarray}
D&=&\frac{1}{2}\sum_b  \gamma_b\int dy' \int dk \, k^2 |G(k,y,y',kU(y))|^2   
\delta(kU(y')-kU(y))\omega_b(y')\nonumber\\
&=&\frac{1}{2} \sum_b \gamma_b \int dy'\int dk  \, |k| |G(k,y,y',kU(y))|^2  
\delta(U(y')-U(y)) \omega_b(y').
\label{ham183}
\end{eqnarray}
This returns the result from Eq. (\ref{gei7diff}).

\section{Polarization cloud}
\label{sec_pc}

In this Appendix we determine the polarization cloud created by a test vortex
moving in the flow.

\subsection{With collective effects}

If we take into account collective effects, the change of vorticity of the
flow due to an external perturbation $\omega_e$ is given in Fourier space by
Eqs.
(\ref{ham17}) and (\ref{ham20}). We assume here that the perturbation is caused
by a point vortex of circulation $\gamma$. The
Fourier transform of the 
vorticity created by the point vortex is [see Eq. (\ref{ham30})] 
\begin{eqnarray}
{\hat\omega}_e(k,y,\sigma)=\gamma  e^{-i k x_0} \delta(kU(y)-\sigma)
\delta(y-y_0),
\label{ham36}
\end{eqnarray}
where $(x_0,y_0)$ denotes the initial position of the test vortex. From Eqs.
(\ref{ham17}), (\ref{ham20}) and (\ref{ham36}), we get
\begin{eqnarray}
\delta{\hat\psi}_{\rm tot}(k,y,\sigma)= \gamma \, G(k,y,y_0,\sigma)  e^{-i k
x_0} \delta(kU(y_0)-\sigma)
\label{ham37}
\end{eqnarray}
and
\begin{eqnarray}
\delta{\hat\omega}(k,y,\sigma)= \gamma \frac{k\frac{\partial\omega}{\partial
y}}{kU(y)-\sigma} G(k,y,y_0,\sigma)  e^{-i k x_0} \delta(kU(y_0)-\sigma).
\label{ham38}
\end{eqnarray}
Returning to physical space, we obtain
\begin{eqnarray}
\delta{\omega}(x,y,t)&=&\gamma\int dk\int\frac{d\sigma}{2\pi}\, e^{i(kx-\sigma
t)} \frac{k\frac{\partial\omega}{\partial y}}{kU(y)-\sigma} G(k,y,y_0,\sigma) 
e^{-i k x_0} \delta(kU(y_0)-\sigma)\nonumber\\
&=&\frac{\gamma}{2\pi}\frac{\frac{\partial\omega}{\partial y}}{U(y)-U(y_0)}
\int dk \, e^{ik(x-x_0-U(y_0) t)} G(k,y,y_0,kU(y_0)).
\label{ham39}
\end{eqnarray}
If we measure the position $x$ with respect to the position of the test vortex
at the instant $t$, writing $X=x-x_0-U(y_0) t$, we get 
\begin{eqnarray}
\delta{\omega}(X,y)=\frac{\gamma}{2\pi}\frac{\frac{\partial\omega}{\partial
y}}{U(y)-U(y_0)} \int dk \, e^{ikX}G(k,y,y_0,kU(y_0)).
\label{ham40}
\end{eqnarray}

If we neglect collective effects, we just have to replace the dressed Green
function by the bare Green function. This yields 
\begin{eqnarray}
\delta{\omega}(X,y)=\frac{\gamma}{2\pi}\frac{\frac{\partial\omega}{\partial
y}}{U(y)-U(y_0)} \int dk \, e^{ikX} G_{\rm bare}(k,y,y_0).
\label{ham41}
\end{eqnarray}
Using the expression (\ref{hamg8})  of the bare Green function in an infinite
domain, we find that
\begin{eqnarray}
\delta{\omega}(X,y)=\frac{\gamma}{2\pi}\frac{\frac{\partial\omega}{\partial
y}}{U(y)-U(y_0)} \int dk \, e^{ikX} \frac{1}{2|k|}e^{-|k||y-y_0|}.
\label{ham43}
\end{eqnarray}
The integral displays a logarithmic divergence when $k\rightarrow 0$. In the
dominant approximation, we can write
\begin{eqnarray}
\delta{\omega}(X,y)=\frac{\gamma}{2\pi}\frac{\frac{\partial\omega}{\partial
y}}{U(y)-U(y_0)} \ln\Lambda.
\end{eqnarray}
For $y\rightarrow y_0$, we obtain the equivalent
\begin{eqnarray}
\delta{\omega}(X,y)\sim \frac{\gamma}{2\pi}\frac{\omega'(y_0)}{U'(y_0)(y-y_0)}
\ln\Lambda,
\label{ham44}
\end{eqnarray}
provided that $U'(y_0)\neq 0$.

\subsection{Without collective effects}

If we neglect collective effects from the start, the change of vorticity due to
the external field is given by
\begin{eqnarray}
\delta{\hat\omega}(k,y,\sigma)=\frac{k\frac{\partial\omega}{\partial
y}}{kU(y)-\sigma}{\hat\psi}_{e}(k,y,\sigma)
\label{ham45}
\end{eqnarray}
with Eq. (\ref{ham20bare}). If the external vorticity is created  by a point
vortex,
using Eqs. (\ref{ham20bare}), (\ref{ham36}) and (\ref{ham45}), we obtain
\begin{eqnarray}
{\hat\psi}_{e}(k,y,\sigma)= \gamma \, G_{\rm bare}(k,y,y_0)  e^{-i k x_0}
\delta(kU(y_0)-\sigma)
\label{ham48}
\end{eqnarray}
and
\begin{eqnarray}
\delta{\hat\omega}(k,y,\sigma)= \gamma \frac{k\frac{\partial\omega}{\partial
y}}{kU(y)-\sigma} G_{\rm bare}(k,y,y_0)  e^{-i k x_0} \delta(kU(y_0)-\sigma).
\label{ham49}
\end{eqnarray}
Equations (\ref{ham48}) and (\ref{ham49}) correspond to Eqs. (\ref{ham37})
and (\ref{ham38}) with the
dressed Green function replaced by the bare Green function. They finally lead
to Eqs. (\ref{ham41})-(\ref{ham44}).

\section{Solution of the Boltzmann-Poisson equation}
\label{sec_sol}

We consider a distribution of $N$ point vortices with equal circulation $\gamma$
at statistical equilibrium in an infinite domain. We assume that the mean flow
is unidirectional. The equilibrium vorticity is given by the Boltzmann
distribution 
(\ref{ham52}) coupled to the Poisson equation (\ref{pv3}). This leads to the
Boltzmann-Poisson equation
\begin{eqnarray}
\label{bpe1}
-\frac{d^2\psi}{dy^2}=\omega=A e^{-\beta\gamma\psi}.
\end{eqnarray}
The constant $A$ is determined by the circulation (or vortex number)
$\Gamma=N\gamma$ and 
the inverse temperature $\beta$ is determined by the energy of the flow $E$ 
(see below). The vorticity can be written as
\begin{eqnarray}
\label{bpe2}
\omega=\omega_0 e^{-\phi}\quad {\rm with}\quad \phi=\beta\gamma(\psi-\psi_0),
\end{eqnarray}
where $\omega_0$ and $\psi_0$ are the vorticity and the stream function at the
origin $y=0$. Equilibrium states exist in an infinite domain
only for $\beta<0$. Introducing the rescaled distance
$\xi=(|\beta|\gamma\omega_0)^{1/2}y$, we can write the Boltzmann-Poisson
equation (\ref{bpe1}) as
\begin{eqnarray}
\label{bpe3}
\frac{d^2\phi}{d\xi^2}=e^{-\phi}
\end{eqnarray}
with boundary conditions $\phi(0)=\phi'(0)=0$. This equation is similar to the 
Emden equation in astrophysics \cite{chandrass,sirechav}. It is also similar to
the
equation of motion of a particle of unit mass moving in a potential
$V(\phi)=e^{-\phi}$, where $\phi$ plays the role of the position and $\xi$ the
role of time. It has the analytical solution (see, e.g., \cite{sirechav})
\begin{eqnarray}
\label{bpe4}
e^{-\phi}=\frac{1}{\cosh^2\left (\frac{\xi}{\sqrt{2}}\right )}.
\end{eqnarray}
Computing the total circulation $\Gamma=\int_{-\infty}^{+\infty}\omega(y)\,
dy=-2\psi'(+\infty)$, we find that the central vorticity is given by
$\omega_0=|\beta|\gamma\Gamma^2/8$. We can then write the equilibrium vorticity
profile as
\begin{eqnarray}
\label{bpe5}
\omega=\frac{|\beta|\gamma\Gamma^2}{8}\frac{1}{\cosh^2\left
(\frac{|\beta|\gamma\Gamma y}{4}\right )}.
\end{eqnarray}
Taking $\psi_0=0$ by convention, we obtain the equilibrium steam function
\begin{eqnarray}
\label{bpe6}
\psi=-\frac{2}{|\beta|\gamma}\ln\left\lbrace \cosh\left
(\frac{|\beta|\gamma\Gamma y}{4}\right )\right\rbrace.
\end{eqnarray}
The corresponding velocity field is
\begin{eqnarray}
\label{bpe6b}
U(y)=-\frac{1}{2}\Gamma  \tanh\left (\frac{|\beta|\gamma\Gamma y}{4}\right ).
\end{eqnarray}
Finally, the inverse temperature $\beta$ is related to the energy by
\begin{eqnarray}
\label{bpe7}
E=\frac{1}{2}\int_{-\infty}^{+\infty} \omega(y)\psi(y)\,
dy=-\frac{\Gamma}{|\beta|\gamma}\int_0^{+\infty}\frac{\ln\lbrack\cosh(x)\rbrack}
{\cosh^2(x)}\, dx=-\frac{\Gamma}{|\beta|\gamma}(1-\ln 2).
\end{eqnarray}

\section{Out-of-equilibrium fluctuation-dissipation theorem}
\label{sec_outfd}

In this Appendix, we consider an arbitrary distribution of point
vortices with a monotonic velocity profile (see Sec. \ref{sec_mono}) and we
establish a form of out-of-equilibrium fluctuation-dissipation theorem.

If the velocity field is monotonic, we have the identity [see Eq. (\ref{gei7d})]
\begin{eqnarray}
\delta(kU(y')-\sigma)=\frac{1}{|k U'(y_*)|}\delta(y'-y_*),
\label{out1}
\end{eqnarray}
where $y_*=U^{-1}(\sigma/k)$ is the (unique) root of the equation
$kU(y_*)=\sigma$.
Substituting Eq. (\ref{out1}) into Eq. (\ref{hami4}), we obtain
\begin{eqnarray}
{\rm Im}\,  G(k,y,y,\sigma)=\pi k \frac{\partial\omega}{\partial y}(y_*)
\frac{1}{|k U'(y_*)|} |G(k,y_*,y,\sigma)|^2.
\label{out2}
\end{eqnarray}
Similarly,  Eq. (\ref{ham33}) can be written as
\begin{eqnarray}
P(k,y,\sigma)=\sum_b \gamma_b |G(k,y,y_*,\sigma)|^2\frac{\omega_b(y_*)}{|k
U'(y_*)|} .
\label{out3}
\end{eqnarray}
For a
single species system of point vortices with circulation $\gamma_b$, combining
Eqs. (\ref{out2}) and
(\ref{out3}), we obtain the out-of-equilibrium fluctuation-dissipation theorem
\begin{eqnarray}
{\rm Im}\,  G(k,y,y,\sigma)=\frac{\pi k}{\gamma_b}
\frac{\partial\ln|\omega_b|}{\partial y}(y_*)P(k,y,\sigma).
\label{out4}
\end{eqnarray}
In the case
where the field
vortices are at statistical equilibrium with the Boltzmann distribution
(\ref{ham52}) we recover the usual fluctuation-dissipation theorem
(\ref{ham50}).

\section{A simplified kinetic equation}
\label{sec_sim}

In this Appendix, we propose a simplified kinetic equation that may
approximately describe the dynamical evolution of a Hamiltonian system of 2D
point vortices in certain cases. To obtain this equation, we make the thermal
bath approximation in the Lenard-Balescu equation 
(\ref{aham105}), leading to Eq.
(\ref{tb6}), but we assume that $\psi(y,t)$ evolves self-consistently
with time, being
determined by the total vorticity $\omega(y,t)=\sum_a\omega_a(y,t)$ through the
Poisson equation
(\ref{ham2bb}) instead of being prescribed as in Sec.
\ref{sec_tbaq}.\footnote{In Sec.
\ref{sec_tbaq}, $\psi(y)$ is determined by the vorticity $\sum_b\omega_b(y)$ of
the field vortices assumed to be independent of time.} This gives
\begin{eqnarray}
\frac{\partial \omega_a}{\partial t}=\frac{\partial}{\partial y}\left \lbrack
D\left (\frac{\partial \omega_a}{\partial y}+\beta \gamma_a
\omega_a\frac{\partial \psi}{\partial y} \right )\right\rbrack,
\label{sim1}
\end{eqnarray}
\begin{eqnarray}
\Delta\psi=-\sum_a \omega_a,
\label{sim2}
\end{eqnarray}
where $D$ is given by Eq. (\ref{tb3}). This equation does not conserve the
energy
contrary to the Lenard-Balescu equation (\ref{aham105}). However, following
\cite{rsprl,lastepjb}, we can enforce the
energy conservation by allowing $\beta$ to depend on time in such
a way that $\dot E=\int\psi \frac{\partial\omega}{\partial t}\, dy=0$. This
yields
\begin{eqnarray}
\beta(t)=-\frac{\int D \frac{\partial\omega}{\partial
y}\frac{\partial\psi}{\partial y}\, dy}{\int D\omega_2 \left
(\frac{\partial\psi}{\partial
y}\right )^2\, dy}.
\label{sim3}
\end{eqnarray}
Equation (\ref{sim1}) with Eqs. (\ref{sim2}) and (\ref{sim3}) conserves the
circulations of each species, the energy, and increases the entropy
($H$-theorem) \cite{lastepjb}.\footnote{We can also conserve
the linear impulse by introducing a relative stream function $\psi_{\rm
eff}=\psi-V(t)y$ instead of $\psi$ and proceed like in Ref. \cite{lastepjb}.}
It relaxes towards the Boltzmann
distribution of statistical
equilibrium on a timescale $Nt_D$. This equation is well-posed mathematically
and interesting in
its own right. It can be seen as a heuristic approximation of the
Lenard-Balescu equation
(\ref{aham105})
providing a simplified kinetic equation for a Hamiltonian system of 2D point
vortices. However, since the approximation leading to Eq. (\ref{sim1}) is
uncontrolled, the solution of this equation may substantially differ from the
solution of the Lenard-Balescu equation (\ref{aham105}). For example, in
the case where there is no
resonance, Eq. (\ref{sim1}) gives a non-vanishing flux 
($\partial\omega/\partial t\neq 0$) driving the system towards the Boltzmann
distribution on a timescale $Nt_D$ while the Lenard-Balescu flux
vanishes ($\partial\omega_{\rm LB}/\partial t=0$) and the Boltzmann
distribution is reached on a longer timescale $N^2t_D$.\footnote{This
timescale discrepency could be corrected by empirically changing the value of
$D$ in Eq. (\ref{sim1}) to make it of order $1/N^2$.} More
generally, the relevance of Eq. (\ref{sim1}) should be determined
case by case by
solving this equation numerically and comparing its solution with the solution
of the Lenard-Balescu equation (\ref{aham105}) or with direct numerical
simulations of the $N$-point vortex system.

In principle, the diffusion coefficient $D$ is a functional of $\omega_a$ but we
shall assume $D={\rm
cst}$ for simplicity. We also take $\beta={\rm cst}$ in Eq. (\ref{sim1}) like in
the case of 2D Brownian vortices described by the canonical
ensemble. In that case, Eqs. (\ref{sim1}) and (\ref{sim2})
conserve the circulations of the different species of point vortices and
decrease the free energy
$F=E-TS$. With this setting, we can take fluctuations due to finite $N$
effects into account
by adding a noise term in the kinetic equation like in Sec. \ref{sec_tdbv}.
This leads
to\footnote{It is also possible to take into account
fluctuations in the more general equations (\ref{sim1})-(\ref{sim3}) and in the
Lenard-Balescu equation (\ref{aham105})
but the expression of the noise is more
complicated \cite{bouchetld}.}
\begin{eqnarray}
\frac{\partial \omega_a}{\partial t}+{\bf
u}\cdot\nabla\omega_a=\nabla\cdot \left \lbrack
D\left (\nabla\omega_a+\beta \gamma_a
\omega_a\nabla\psi \right
)\right\rbrack+\nabla\cdot \left\lbrack
\sqrt{2D\gamma_a\omega_a} {\bf R}_a({\bf r},t)\right\rbrack,
\label{sim4}
\end{eqnarray}
\begin{eqnarray}
\Delta\psi=-\sum_a \omega_a.
\label{sim5}
\end{eqnarray}
Interestingly, Eqs. (\ref{sim4}) and (\ref{sim5}) apply to arbitrary flows
(with the limitations
about their validity mentioned previously). These equations are interesting in
their own right. They could be used to describe random transitions
between different equilibrium states as discussed in Sec. \ref{sec_tdbv}.

\section{Lenard-Balescu equation for 2D point vortices}
\label{sec_j}

We consider an isolated system of $N$ point vortices with identical circulation
$\gamma$. We assume that the mean flow is unidirectional. We want to determine
the kinetic equation of 2D point vortices due to finite $N$ effects by using the
Klimontovich approach. The
derivation is similar to the one given for axisymmetric flows
in Refs. \cite{dubin,klim}. We start from the quasilinear equations
(\ref{ham12}) and (\ref{ham13}) without the external
potential ($\psi_e=0$) that we rewrite as
\begin{eqnarray}
\frac{\partial\omega}{\partial t}=\frac{\partial}{\partial y} \left\langle
\delta\omega \frac{\partial\delta\psi}{\partial x}\right\rangle,
\label{j1}
\end{eqnarray}
\begin{eqnarray}
\frac{\partial\delta\omega}{\partial t}+U\frac{\partial\delta\omega}{\partial
x}- \frac{\partial\delta\psi}{\partial x}
\frac{\partial\omega}{\partial y}=0.
\label{j2}
\end{eqnarray}

Taking the Fourier-Laplace transform of Eq. (\ref{j2}), we find that
\begin{eqnarray}
\label{j3}
\delta{\tilde\omega}(k,y,\sigma)=\frac{k\frac{\partial\omega}{\partial
y}}{kU-\sigma}\delta{\tilde\psi}(k,y,\sigma)+\frac{\delta{\hat\omega}(k,y,0)}{i(
kU-\sigma)},
\end{eqnarray}
where $\delta{\hat\omega}(k,y,0)$ is the Fourier transform of the initial
vorticity fluctuation caused by finite $N$ effects. Combining this
relation with the Poisson equation $\Delta\delta\psi=-\delta\omega$ written in
Fourier space [see Eq. (\ref{hamg2})], we get
\begin{eqnarray}
\left\lbrack \frac{d^2}{dy^2}-k^2+\frac{k\frac{\partial\omega}{\partial
y}}{kU(y)-\sigma}\right\rbrack
\delta{\tilde\psi}=-\frac{\delta{\hat\omega}(k,y,0)}{i(
kU-\sigma)}.
\label{j4}
\end{eqnarray}
The formal solution of this differential equation is
\begin{eqnarray}
\delta{\tilde\psi}(k,y,\sigma)=\int
G(k,y,y',\sigma)\frac{\delta{\hat\omega}(k,y',0)}{i(
kU'-\sigma)}\, dy',
\label{j5}
\end{eqnarray}
where the Green function is defined in Eq. (\ref{ham21}) and $U'$ stands for
$U(y')$. Taking the inverse
Laplace transform of this equation, using the Cauchy residue theorem, and
neglecting the contribution of the damped
modes for sufficiently
late times,\footnote{We only consider the contribution of the
pole $\sigma-kU'$ and ignore the contribution of the proper modes of the
flow which are the solutions of the Rayleigh equation (\ref{ham23}). See
Ref. \cite{linres} for general considerations about the linear response theory
of
systems with long-range interactions.} we obtain 
\begin{eqnarray}
\label{j6}
\delta{\hat\psi}(k,y,t)=\int dy'\,
G(k,y,y',kU') 
\delta{\hat \omega}(k,y',0)e^{-i k U't}.
\end{eqnarray}

On the other hand, taking the Fourier transform of Eq. (\ref{j2}), we find that
\begin{eqnarray}
\label{j7}
\frac{\partial \delta{\hat \omega}}{\partial t}+ikU\delta{\hat
\omega}=ik\frac{\partial \omega}{\partial y}\delta{\hat\psi}.
\end{eqnarray}
This first order differential equation in time can be solved with the method of
the variation of the constant, giving
\begin{eqnarray}
\label{j8}
\delta{\hat \omega}(k,y,t)=\delta{\hat\omega}(k,y,0)e^{-i kU
t}+ik
\frac{\partial \omega}{\partial y}\int_0^t dt'\, \delta{\hat
\psi}(k,y,t')e^{ikU(t'-t)}.
\end{eqnarray}
Substituting Eq. (\ref{j6}) into Eq. (\ref{j8}), we obtain 
\begin{eqnarray}
\label{j9}
\delta{\hat \omega}(k,y,t)=\delta{\hat\omega}(k,y,0)e^{-i k U
t}+ik\frac{\partial \omega}{\partial y}\int dy'\,
G(k,y,y',k U') \delta{\hat \omega}(k,y',0) e^{-i k U t}\int_0^t dt'\,
e^{ik(U-U')t'}.
\end{eqnarray}
Eqs.
(\ref{j6}) and
(\ref{j9}) relate $\delta{\hat\psi}(k,y,t)$ and $\delta{\hat \omega}(k,y,t)$ to
the
initial fluctuation $\delta{\hat \omega}(k,y,0)$.

We can now compute the flux
\begin{eqnarray}
\label{j10}
\left\langle \delta
\omega\frac{\partial\delta\psi}{\partial x}\right\rangle=\int dk dk'\, 
i k' e^{ikx}e^{ik'x}\langle \delta{\hat \omega}(k,y,t)\delta{\hat
\psi}(k',y,t)\rangle.
\end{eqnarray}
From Eqs. (\ref{j6}) and (\ref{j9}) we get
\begin{eqnarray}
\label{j11}
\langle \delta{\hat \omega}(k,y,t)\delta{\hat
\psi}(k',y,t)\rangle=\int dy'\, G(k',y,y',k'U')
e^{-i k' U't}e^{-i k U t}\langle \delta{\hat \omega}(k,y,0)\delta{\hat
\omega}(k',y',0)\rangle\nonumber\\
+\int dy'\, G(k',y,y',k'U')
e^{-i k' U't} ik\frac{\partial\omega}{\partial y}
\int dy''\, G(k,y,y'',kU'')
\langle\delta{\hat \omega}(k,y'',0)\delta{\hat \omega}(k',y',0)\rangle
e^{-i kU t}
\int_0^t dt'\, e^{ik(U-U'')t'}.\nonumber\\
\end{eqnarray}
The correlation function of the initial fluctuations in Fourier space is
given by (see, e.g., Appendix D of \cite{klim}) 
\begin{eqnarray}
\label{j12}
\langle \delta{\hat
\omega}(k,y,0)\delta{\hat\omega}(k',y',
0)\rangle=\frac{\gamma}{2\pi}\delta(k+k')\delta(y-y')\omega(y).
\end{eqnarray}
Eq. (\ref{j11}) then reduces to
\begin{eqnarray}
\label{j13}
\langle \delta{\hat \omega}(k,y,t)\delta{\hat
\psi}(k',y,t)\rangle=\frac{1}{2\pi} G(-k,y,y,
-kU)\gamma\omega(y)\delta(k+k')\nonumber\\
+\frac{1}{2\pi}\int dy'\, G(-k,y,y',-kU')ik\frac{\partial\omega}{\partial y}
G(k,y,y',kU')\delta(k+k')\gamma\omega(y')
\int_0^t ds\, e^{-ik(U-U')s},
\end{eqnarray}
where we have set $s=t-t'$. Substituting this relation into Eq. (\ref{j10}),
and taking the limit $t\rightarrow +\infty$, we obtain
\begin{eqnarray}
\label{j14}
\left\langle\delta\omega\frac{\partial\delta\psi}{\partial
x}\right\rangle=-\frac{1}{2\pi}\int
dk\, i k G(-k,y,y,-kU)\gamma\omega(y)
\nonumber\\
-\frac{1}{2\pi}\int dk\, i k \int dy'\,
G(-k,y,y',-kU')ik\frac{\partial\omega}{\partial y}
G(k,y,y',kU')\gamma\omega(y')\int_0^{+\infty} ds\, e^{-ik(U-U')s}.
\end{eqnarray}
Making the transformations $s\rightarrow -s$ and $k\rightarrow -k$ we see that
we can replace $\int_0^{+\infty} ds$ by $\frac{1}{2}\int_{-\infty}^{+\infty}
ds$. We then get 
\begin{eqnarray}
\label{j15}
\left\langle\delta\omega\frac{\partial\delta\psi}{\partial x}
\right\rangle=-\frac{1}{2\pi}\int dk\, i
k G(-k,y,y,-kU)\gamma\omega(y)
\nonumber\\
-\frac{1}{2\pi}\int dk\, i k \int dy'\, G(-k,y,y',-kU')
ik\frac{\partial\omega}{\partial y}
G(k,y,y',kU')\gamma\omega(y')
\frac{1}{2}\int_{-\infty}^{+\infty} dt'\, e^{ik(U-U')t'}.
\end{eqnarray}
Using the identities (\ref{delta}) and (\ref{gei7b}), we obtain
\begin{eqnarray}
\label{j17}
\left\langle\delta\omega\frac{\partial\delta\psi}{\partial x}
\right\rangle=-\frac{1}{2\pi}\int dk\, i
k G(-k,y,y,-kU)\gamma\omega(y)
\nonumber\\
+\frac{1}{2}\int dk\, |k| \int dy'\,
G(-k,y,y',-kU')\frac{\partial\omega}{\partial y}
G(k,y,y',kU')\gamma\omega(y')\delta(U-U').
\end{eqnarray}
Finally, using Eq. (\ref{obv}), we can rewrite the foregoing
equation as 
\begin{eqnarray}
\label{j18}
\left\langle\delta\omega\frac{\partial\delta\psi}{\partial x}
\right\rangle=-\frac{1}{2\pi}\int dk\, 
k\, {\rm Im}\, G(k,y,y,kU)\, \gamma\omega(y)
+\frac{1}{2}\int dk\, |k| \int dy'\, \frac{\partial\omega}{\partial y}
|G(k,y,y',kU')|^2\gamma\omega(y')\delta(U-U').
\end{eqnarray}
The first term is the drift term and the second term is the diffusion term.
Using the identity (\ref{hami4}) and substituting the flux from Eq.
(\ref{j18}) into Eq. (\ref{j1}), we obtain
the Lenard-Balescu-like equation 
\begin{eqnarray}
\frac{\partial\omega}{\partial t}=\frac{\gamma}{2}\frac{\partial}{\partial y} 
\int dy'\int dk\, |k| |G(k,y,y',kU(y))|^2   \delta(U(y')-U(y))\left
(\omega' \frac{\partial \omega}{\partial
y}-\omega\frac{\partial \omega'}{\partial y'}\right ).
\label{j19}
\end{eqnarray}
We recall that for a unidirectional flow made of a single species system of
point vortices, the Lenard-Balescu flux vanishes (see Sec. \ref{sec_mono}).
We can easily extend the derivation of the Lenard-Balescu equation to the
multispecies case, leading to Eq. (\ref{ham105}).

\end{document}